\documentclass[11pt]{article}

\usepackage[margin=1in]{geometry}
\usepackage{amsmath, amssymb, amsfonts, bbm}
\usepackage{xr-hyper} 
\usepackage[dvipsnames]{xcolor}
\usepackage{url}
\usepackage{soul}
\usepackage{color}
\usepackage{cancel}
\usepackage{natbib}
\usepackage{float}
\usepackage{mathrsfs}
\usepackage{enumitem}
\usepackage{changepage}       
\usepackage{xr}
\usepackage{mathtools}
\usepackage{parskip}

\usepackage{makecell}
\usepackage{caption}
\usepackage{colortbl}
\usepackage{hhline}
\usepackage{calc}
\usepackage{tabularx}

\usepackage[normalem]{ulem}
\newcommand{\stkout}[1]{\ifmmode\text{\sout{\ensuremath{#1}}}\else\sout{#1}\fi}

\usepackage{multirow}
\usepackage{arydshln}

\usepackage{calrsfs} 
\DeclareMathAlphabet{\pazocal}{OMS}{zplm}{m}{n}
\usepackage[mathscr]{euscript}

\usepackage{tikz}
\usepackage{tikz-qtree}
\usetikzlibrary{trees}
\usetikzlibrary{automata,positioning}

\usepackage{booktabs} 

\usepackage{algorithm}
\usepackage{algpseudocode}

\allowdisplaybreaks

\DeclareMathOperator{\dis}{dis}

\DeclareMathOperator{\pa}{pa}

\DeclareMathOperator{\ch}{ch}

\DeclareMathOperator{\mb}{mb}

\DeclareMathOperator*{\argmin}{arg\,min}

\newcommand{\red}[1]{\begingroup\color{red}#1\endgroup}

\newcommand{\blue}[1]{\begingroup\color{blue}#1\endgroup}

\newcommand{\E}{\mathbb{E}} 

\newcommand{\G}{\pazocal{G}}

\newcommand{\N}{\mathcal{N}}
\newcommand{\I}{\mathbb{I}}
 
\newcommand{\M}{\mathcal{M}}
\newcommand{\Z}{\mathscr{Z}}

\newcommand{\R}{\mathbb{R}}
\newcommand{\bbL}{\mathcal{L}}
\newcommand{\X}{\mathcal{X}}

\newcommand{\mpi}{\mathrm{mp}_{\G}}
\newcommand{\uniform}{\text{Uniform}}
\newcommand{\bin}{\text{Binomial}}
\newcommand{\expit}{\text{expit}}
\newcommand{\anna}[1]{\textcolor{Salmon}}
\newcommand{\smallO}{o_P}

\usepackage{xparse}

\newcommand{\mR}[1]{\mathscr{R}_{#1}}
\newcommand{\mB}[1]{\mathscr{B}_{#1}}
\newcommand{\f}[1]{f_{#1}}
\newcommand{\fr}[1]{f^r_{#1}}
\newcommand{\li}[1]{{h}_{#1}}
\newcommand{\lex}[1]{g_{#1}}

\NewDocumentCommand{\hmR}{ m g }{\IfNoValueTF{#2}{\hat{\mathscr{R}}_{#1}}{\hat{\mathscr{R}}_{#1}^{#2}}}

\NewDocumentCommand{\HmR}{ m g }{\IfNoValueTF{#2}{\hat{\mathscr{R}}_{#1}}{\hat{\mathscr{R}}_{#1}^{#2}}}

\newcommand{\tmR}[1]{\tilde{\mathscr{R}}_{#1}}
\NewDocumentCommand{\hmB}{ m g }{\IfNoValueTF{#2}{ \hat{\mathscr{B}}_{#1}}{\hat{\mathscr{B}}_{#1}^{#2}}}

\NewDocumentCommand{\HmB}{ m g }{\IfNoValueTF{#2}{ \hat{\mathscr{B}}_{#1}}{\hat{\mathscr{B}}_{#1}^{#2}}}

\newcommand{\tmB}[1]{\tilde{\mathscr{B}}_{#1}}

\NewDocumentCommand{\hf}{ m g}{\IfNoValueTF{#2}{\hat{f}_{#1}}{\hat{f}^{#2}_{#1}}}
\NewDocumentCommand{\Hf}{ m g}{\IfNoValueTF{#2}{\hat{f}_{#1}}{\hat{f}^{#2}_{#1}}}
\NewDocumentCommand{\tf}{ m g}{\IfNoValueTF{#2}{\tilde{f}_{#1}}{\tilde{f}^{#2}_{#1}}}

\NewDocumentCommand{\hfr}{ m g }{\IfNoValueTF{#2}{\hat{f}^r_{#1}}{\hat{f}^{r{#2}}_{#1}}}
\NewDocumentCommand{\Hfr}{ m g }{\IfNoValueTF{#2}{\hat{f}^r_{#1}}{\hat{f}^{r{#2}}_{#1}}}
\NewDocumentCommand{\tfr}{ m g }{\IfNoValueTF{#2}{\tilde{f}^r_{#1}}{\tilde{f}^{r{#2}}_{#1}}}

\NewDocumentCommand{\hli}{ m g }{ \IfNoValueTF{#2}{\hat{h}_{#1} }{\hat{h}^{#2}_{#1}}}
\NewDocumentCommand{\Hli}{ m g }{ \IfNoValueTF{#2}{\hat{h}_{#1} }{\hat{h}^{#2}_{#1}}}
\NewDocumentCommand{\tli}{ m g }{ \IfNoValueTF{#2}{\tilde{h}_{#1} }{\tilde{h}^{#2}_{#1}}}

\NewDocumentCommand{\hlex}{ m g }{\IfNoValueTF{#2}{\hat{g}_{#1}}{\hat{g}^{#2}_{#1}}}
\NewDocumentCommand{\Hlex}{ m g }{\IfNoValueTF{#2}{\hat{g}_{#1}}{\hat{g}^{#2}_{#1}}}
\NewDocumentCommand{\tlex}{ m g }{\IfNoValueTF{#2}{\tilde{g}_{#1}}{\tilde{g}^{#2}_{#1}}}

\DeclareMathOperator{\mpg}{mp_\G}
\newcommand{\mpgA}[1]{\mathrm{mp}^{-A}_\G(#1)}
\newcommand{\mpga}[1]{\mathrm{mp}^{-a}_\G(#1)}
\newcommand{\mBarg}[1]{\mathrm{mp}^{-A}_{\G_\prec}(\succeq #1)}

\newcommand{\mBarga}[1]{\mathrm{mp}^{-a}_{\G_\prec}(\succeq #1)}

\newcommand{\tmledens}{\psi_\text{densratio}(\hat{Q}^*)}
\newcommand{\tmlebayes}{\psi_\text{bayes}(\hat{Q}^*)}
\newcommand{\tmlednorm}{\psi_\text{dnorm}(\hat{Q}^*)}
\newcommand{\onedens}{\psi^+_\text{densratio}(\hat{Q})}
\newcommand{\onebayes}{\psi^+_\text{bayes}(\hat{Q})}
\newcommand{\onednorm}{\psi^+_\text{dnorm}(\hat{Q})}

\usepackage{pifont}

\usepackage{titletoc}
\titlecontents{section}
  [0em]
  {}
  {\contentslabel{2.3em}}
  {}
  {\titlerule*[0.5pc]{.}\contentspage}

\titlecontents{subsection}
  [2.3em]
  {}
  {\contentslabel{3.2em}}
  {}
  {\titlerule*[0.5pc]{.}\contentspage}

\usepackage{cancel}
\usepackage[hidelinks]{hyperref}

\newtheorem{theorem}{Theorem}     
\newtheorem{lemma}[theorem]{Lemma}           
\newtheorem{corollary}[theorem]{Corollary}   

\newtheorem{definition}{Definition} 
\newtheorem{example}{Example}       

\newtheorem{remark}{Remark}         

\title{Average Causal Effect Estimation in DAGs with Hidden Variables: Beyond Back-Door and Front-Door Criteria}

\date{}

\usepackage{authblk}

\author[]{Anna Guo} 
\author[]{Razieh Nabi} 
\affil[]{Department of Biostatistics and Bioinformatics, Emory University, Atlanta, GA, USA}

\begin{document}

\maketitle

\begin{abstract}
The identification theory for causal effects in directed acyclic graphs (DAGs) with hidden variables is well established, but methods for estimation and inference beyond the g-formula remain less developed. Previous studies have introduced influence-function-based estimators for broad classes of identifiable effects. When implemented through the observed-data density factorization, these estimators can require conditional density estimation and numerical integration, which can be statistically challenging and computationally demanding for continuous or multivariate variables. Prior work on related causal functionals has used sequential regressions and Bayes-rule representations of density ratios to avoid these requirements. Building on established identification and influence-function results, we make these reparameterizations explicit for effects identified under \textit{treatment primal fixability}, Tian and Pearl's necessary and sufficient condition for identifying the joint post-intervention distribution of all remaining variables, encompassing settings beyond classical back-door and front-door identification. We construct one-step estimators under alternative nuisance parameterizations and characterize how these choices affect robustness properties and nuisance-specific $L^2(P)$ product-rate conditions for asymptotic linearity and root-$n$ consistency. The estimators accommodate data-adaptive machine learning for nuisance estimation. To address potential violations of the target parameter bounds by one-step estimators, we also develop targeted minimum loss-based estimators (TMLEs) that respect these bounds. Our robustness characterizations and sufficient rate conditions apply to both the one-step estimators and TMLEs. Building on prior efficiency theory for hidden-variable DAGs, we further develop modifications of our estimators that exploit ordinary independence restrictions. To ensure our estimation methods are accessible in practice, we provide the \href{https://github.com/annaguo-bios/flexCausal}{\texttt{flexCausal}} package in \textsf{R}. 
\end{abstract}

\section{Introduction}
\label{sec:intro}

Controlling for treatment-outcome confounders is a  widely adopted approach for inferring the average causal effect (ACE). The presence of unmeasured (or hidden/latent) confounders in observational studies, however, presents challenges for causal identification and may invalidate causal inference. The literature studies this issue from multiple angles.   

In directed acyclic graphs (DAGs) with hidden variables, the two simplest criteria for identifying ACE are known as the back-door \citep{pearl2009causality, robins86new} and front-door \citep{pearl1995causal} criteria. The back-door criterion involves identifying a set of variables that blocks all confounding paths between the treatment and the outcome. In contrast, the front-door criterion assumes the existence of one or more mediating variables that fully mediate the treatment's effect on the outcome, while not being subject to the same unmeasured confounding between treatment and outcome. Estimating the back-door functional (also known as the adjustment formula or g-formula) has been extensively explored in the literature, with methods ranging from straightforward plug-in and inverse probability weighting estimators \citep{robins86new, hahn1998role, hirano2003efficient} to more involved influence function-based estimators, such as one-step corrected plug-in and targeted minimum loss-based estimators \citep{bickel1993efficient, van2000asymptotic, bang05doubly, tsiatis2007semiparametric, van2011targeted, double17chernozhukov}. Estimating the front-door functional has also been thoroughly investigated \citep{fulcher19robust, guo2026flexible}. 

In settings where neither the back-door nor the front-door criteria apply, identifying causal effects becomes considerably more complex. Sound and complete algorithms for identifying effects in hidden-variable DAGs have been well developed \citep{tian02general, shpitser06id, huang06do, bhattacharya2022semiparametric, richardson2023nested}. We focus on a broad class of such DAGs where the treatment satisfies a simple graphical criterion known as primal fixability, first established by \cite{tian02general} and further developed by \cite{bhattacharya2022semiparametric}. This criterion requires the absence of any confounding paths composed solely of hidden variables between the treatment and any of its children; equivalently, that the treatment and its children do not belong to the same latent component, referred to as a c-component by \citet{tian02general} and as a district by \citet{richardson2023nested}. The back-door and front-door criteria can be understood as instances of primal fixability, either in the original graph or in an appropriate latent projection where certain variables are marginalized out. \citet{tian02general} demonstrated that treatment primal fixability is necessary and sufficient for identifying the joint post-intervention distribution of all remaining variables. Consequently, the effect of a primal fixable treatment on any outcome is identifiable, although particular marginal effects may remain identifiable when primal fixability fails. Beyond its graphical simplicity, this class admits a common identifying functional that facilitates a unified treatment of estimation and inference, motivating our focus on causal effect identified under treatment primal fixability.

A lack of flexible and computationally feasible framework for estimating causal effects identifiable in hidden-variable DAGs has been a major barrier to their widespread application. Extending effect estimation beyond the back-door and front-door adjustments is therefore a key step toward bridging the gap between identification and estimation. Building on advances in causal nonparametric identification theory in hidden-variable DAGs, prior work has made important contributions to estimation. \citet{bhattacharya2022semiparametric} proposed one-step corrected plug-in estimators for causal effects identifiable under the primal fixability criterion, and established their double robustness properties. Their construction uses the observed-data density factorization, making its implementation more feasible in discrete settings or under parametric nuisance models, but potentially computationally intensive and statistically challenging in more complex continuous-variable settings, particularly due to the difficulties of conditional density estimation and numerical integration. They also characterized semiparametric efficiency bounds and efficient influence functions for a class of hidden-variable DAGs in which the model imposes only ordinary conditional independence restrictions. Focusing on discrete variables, \citet{jung2021estimating} developed estimators beyond primal fixability by combining influence functions with machine learning, and showed asymptotic linearity under the  $\smallO(n^{-1/4})$ rate for nuisance estimation. Most recently, \citet{jung2024unified} extended these methods to accommodate mixed discrete and continuous variables while avoiding direct density estimation and improving computational efficiency, and provided finite-sample analyses and asymptotic results under comparable rate conditions for nuisance estimates.

Building on these important contributions in identification and influence-function-based estimation, we address several remaining challenges in estimation and inference for primal-fixable causal effects. 
\textit{First}, one-step estimators do not generally guarantee that estimates lie within the target parameter space. This issue is particularly relevant for binary or bounded outcomes, where estimates of potential outcome means may fall outside the unit interval or the known outcome bounds, complicating interpretation and motivating parameter-space-preserving procedures, namely targeted minimum loss-based estimators (TMLEs). 
\textit{Second}, density-based implementations of existing influence-function estimators can require conditional density estimation and numerical integration. Sequential-regression and density-ratio approaches offer alternatives, but their use requires explicit nuisance representations and corresponding analyses of robustness and estimation error. 
\textit{Third}, prior work has established efficiency theory under ordinary independence restrictions for classes of hidden-variable DAGs; incorporating these restrictions into sequential-regression and targeting procedures requires corresponding estimator constructions.

To address these challenges, we develop new one-step corrected plug-in and targeted minimum loss-based estimators for primal-fixable causal effects, encompassing the back-door and front-door adjustment functionals as special cases. We systematically develop three complementary nuisance parameterizations for primal-fixable causal effects, two of which avoid direct conditional mediator density estimation and numerical integration through sequential regressions and density-ratio representations. These constructions accommodate flexible nuisance estimation with multivariate mediators of mixed types. 
We characterize double robustness and additional sufficient consistency regimes under the different parameterizations, and derive explicit remainder analyses and nuisance-specific $L^2(P)$ product-rate conditions sufficient for asymptotic linearity and root-$n$ consistency of both the one-step estimators and TMLEs. These conditions allow trade-offs between nuisance convergence rates rather than requiring every nuisance estimator to converge at rate $o_P(n^{-1/4})$. The TMLEs additionally preserve the target parameter space, and we investigate modifications that exploit ordinary independence restrictions. By incorporating cross-fitting and sample-splitting, an alternative to Donsker conditions when using flexible statistical and machine learning methods for nuisance estimation, our estimators meet the criteria of double-debiased estimators—a concept well-studied in statistics and popularized in machine learning following \citet{double17chernozhukov}. The explicit and transparent structure of our estimators has also enabled the development of the \href{https://github.com/annaguo-bios/flexCausal}{\texttt{flexCausal}} \textsf{R} package, making advanced causal inference tools accessible for applied research. 

The remainder of the paper is organized as follows. Section~\ref{sec:prelim} reviews the graphical framework, and Sections~\ref{sec:est}–\ref{sec:inference} develop our estimators and establish their theoretical properties. Section~\ref{sec:semiparam} considers efficiency improvements under additional ordinary independence restrictions. Sections~\ref{sec:sims} and \ref{sec:realdata} present simulation studies and a real-data application, respectively. Section~\ref{sec:discussion} concludes. All proofs are provided in the supplementary materials. 

\section{Causal models with hidden variables: \small{From DAGs to ADMGs}}
\label{sec:prelim}

Let $V$ denote the union set of observed variables $O$ and unmeasured (or hidden/latent) variables $U$. Assume that $V$ is sampled from a distribution $P$ within a, potentially unrestricted (including infinite-dimensional), statistical model $\pazocal{M}$. We assume the joint distribution $P(V)$ factorizes according to an \textit{acyclic directed graph} (DAG) $\G(V)$, that is $P(V) = \prod_{V_i \in V} P(V_i \mid \pa_\G(V_i)),$ where $\pa_\G(V_i)$ denotes the parents of $V_i$ in $\G(V)$. Throughout the manuscript, we refer to $V$ as variables and vertices interchangeably. The independence restrictions between variables in $V$ can be read-off from $\G(V)$ using d-separation rules \citep{pearl2009causality}. According to the \textit{global Markov property}, for disjoint sets of variables $X, Y, Z \subset V$ if $X$ and $Y$ are d-separated given $Z$ in $\G(V)$, then $X$ and $Y$ are independent given $Z$ in $P(V)$, denoted by $X \perp Y \mid Z$. 

Our target of inference is the average causal effect (ACE) of an observed binary treatment $A$ on an observed outcome $Y$, which is defined as $\E[Y(1) - Y(0)]$. Here, $Y(a)$ denotes the potential outcome $Y$ when treatment is assigned to $a \in \{0, 1\}$. For simplicity, we focus primarily on the potential outcome mean $\mathbb{E}[Y(a_0)]$, where $a_0$ represents a fixed treatment assignment with a value of either one or zero. The alternative treatment assignment is denoted by $a_1$. 

Assuming all parents of $A$ in $\G(V)$ are observed, meaning $\pa_\G(A) \cap U = \emptyset$, we can treat $\pa_\G(A)$ as a sufficient set for blocking all confounding pathways between $A$ and $Y$. Consequently, we can identify $\E[Y(a_0)]$ as a function of the observed data distribution via the \textit{back-door formula} \citep{pearl2009causality}. Figure~\ref{fig:graphs}(a) illustrates the simplest such DAG, where $U = \emptyset$ and $\pa_\G(A) = X$. The back-door formula is given by $\int y dP(y \mid a_0, x)dP(x)$. The validity of this identification relies on three key assumptions: \textit{ignorability}, which states that treatment is independent of the potential outcome given $X$; \textit{consistency}, which states that the potential outcome equals the observed outcome when the assigned treatment matches the observed treatment; and \textit{positivity}, which requires that $P(A=1 \mid \pa_\G(A)) > 0$ for every level of the treatment's parental set. It is worth noting that, beyond $\pa_\G(A)$, other valid adjustment sets may exist; for a detailed discussion on multiple valid adjustment sets, see \citep{rotnitzky2020efficient, runge2021necessary, henckel2022graphical}.

If not all parents of $A$ in $\G(V)$ are observed, meaning $\pa_\G(A) \cap U \neq \emptyset$, the process of identifying the counterfactual mean, and consequently the average causal effect, becomes more complex, and it may not be possible to identify it at all. The simplest hidden-variable DAG where the ACE is identified through a different functional than the back-door is the \textit{front-door} model \citep{pearl1995causal}, shown in Figure~\ref{fig:graphs}(b). Through assuming a set of measured variables $M$ that mediate all the effect of $A$ on $Y$ with all backdoor paths from $A$ to $M$ blocked by $X$ and all backdoor paths from $M$ to $Y$ blocked by $(A,X)$, the counterfactual mean $\E[Y(a_0)]$ can be identified as $\int y dP(y \mid m, a, x)dP(m \mid a_0, x) dP(a,x)$. 

\begin{figure}[t] 
	\begin{center}
    \scalebox{0.7}{
    \begin{tikzpicture}[>=stealth, node distance=1.7cm]
        \tikzstyle{format} = [thick, circle, minimum size=1.0mm, inner sep=2pt]
        \tikzstyle{square} = [draw, thick, minimum size=4.5mm, inner sep=2pt]
    
    \begin{scope}[xshift=0cm, yshift=0cm]
		\path[->, thick]
		
		node[] (a) {$A$}
		node[above right of=a, yshift=0.cm] (x) {$X$}
		node[right of=a, xshift=0.75cm] (y) {$Y$}
		 
		(x) edge[blue] (a) 
		(x) edge[blue] (y) 
		(a) edge[blue] (y)

        node[below of=a, xshift=1.25cm, yshift=0.85cm] (t1) {(a)} ;
		
	\end{scope}
    
	\begin{scope}[xshift=5cm, yshift=0cm]
		\path[->, thick]
		
		node[] (a) {$A$}
		node[right of=a, xshift=0.75cm] (m) {$M$}
		node[above right of=m, xshift=0cm, yshift=0.cm] (x) {$X$}
		node[above left of=m, yshift=0.cm] (u) {\red{$U$}}
		node[right of=m, xshift=0.75cm] (y) {$Y$}
		
		(x) edge[blue, bend left=0] (a) 
		(x) edge[blue] (m) 
		(x) edge[blue, bend right=0] (y) 
		(a) edge[blue] (m) 
		(m) edge[blue] (y) 
		(u) edge[blue] (y) 
		(u) edge[blue] (a)

        node[below of=a, xshift=2.5cm, yshift=0.85cm] (t3) {(b)} ;
		
	\end{scope}
    \begin{scope}[xshift=13cm, yshift=0cm]
		\path[->, thick]
		
		node[] (a) {$A$}
		node[right of=a, xshift=0.75cm] (m) {$M$}
		node[above right of=m,  xshift=0cm, yshift=0.0cm] (x) {$X$}
		node[above left of=m, yshift=0.cm] (u) {}
		node[right of=m, xshift=0.75cm] (y) {$Y$}
		
		(x) edge[blue, bend left=0] (a) 
		(x) edge[blue] (m) 
		(x) edge[blue, bend right=0] (y) 
		(a) edge[blue] (m) 
		(m) edge[blue] (y) 
        (a) edge[red, <->, out=60, in=155] (y)  

        node[below of=a, xshift=2.5cm, yshift=0.85cm] (t3) {(c)} ;
		
	\end{scope}
	\end{tikzpicture}
	}
    \vspace{-0.2cm}
	\caption{(a) A DAG where $\pa_\G(A)$ is fully observed, blocking the back-door path between $A$ and $Y$; (b) A front-door DAG where $\pa_\G(A)$ is partially unobserved, and thus not all the back-door paths can be blocked; (c) The latent projected ADMG of the DAG in (b).} 
	\label{fig:graphs}
	\end{center}
\end{figure}

Instead of directly analyzing a DAG with unmeasured/hidden variables $\G(O \cup U)$, the focus often shifts to its \textit{latent projection} onto the observed data $O$. This projection results in an acyclic directed mixed graph (ADMG), denoted by $\G(O)$. The latent projection, or marginalization, is governed by two primary rules: (i) $O_i \blue{\rightarrow} O_j$ is included in the ADMG $\G(O)$ if $O_i \blue{\rightarrow} O_j$ exists in $\G(O \cup U)$ or if a directed path from $O_i$ to $O_j$ exists that passes through only unmeasured variables in $\G(O \cup U)$; and (ii) $V_i \red{\leftrightarrow} V_j$ is included in the ADMG $\G(O)$ if a collider-free path, like $V_i \blue{\leftarrow} \cdots \blue{\rightarrow} V_j$, exists in $\G(O \cup U)$ where all the intermediate variables belong to $U$ \citep{verma90equiv}. For example, the ADMG in Figure~\ref{fig:graphs}(c) is the latent projection of the front-door DAG in Figure~\ref{fig:graphs}(b). Using an analogue of d-separation, known as \textit{m-separation}, we can also read off conditional independence restrictions from an ADMG \citep{richardson03markov}. 

Marginalization of hidden-variable DAGs primarily addresses challenges of global non-identifiability and singularities frequently encountered in models with latent variables \citep{drton2008lectures, allman2009identifiability}. Additionally, ADMG $\G(O)$ has been proven to effectively act as a smooth supermodel of all hidden-variable DAGs $\G(V)$ that share the same latent projection as $\G(O)$, maintaining the same theory for identifying causal effects and implying identical equality constraints on the observed data marginal distribution $P(O)$ \citep{evans2019smooth, richardson2023nested}. Consequently, utilizing ADMGs significantly streamlines the causal analysis process. 

Nonparametric identification theories for causal effects typically begin with a latent projected ADMG and determine whether the target parameter can be expressed using the joint distribution of observed variables. When identifiability is achievable, these theories then provide an identifying functional. A range of \textit{sound} and \textit{complete} identification algorithms, which embody the necessary and sufficient conditions for the nonparametric identification of the causal target parameter within the class of hidden-variable DAGs or ADMGs, are documented in the contributions by \cite{tian02general, shpitser06id, huang06do, bhattacharya2022semiparametric, richardson2023nested}. The terms ``sound'' and ``complete'' indicate that these algorithms reliably ascertain (non)identifiability based on the graphical model's structure. This aspect of identification theory not only highlights the utility of graphical models in causal inference but also underscores the importance of algorithmic approaches in navigating the complexities introduced by latent variables. 

\subsection{Factorization of observed data distributions in ADMGs}
\label{sec:factorization}

Although the latent projected ADMG and the original hidden-variable DAG impose the same equality constraints on observed variables, as per the Markov properties of the hidden-variable DAG $\G(O \cup U)$, formulating a factorization for $P(O)$ that fully encodes these constraints proves more complex than the DAG factorization initially introduced. 

To capture the structure of $P(O)$ in hidden variable models, \cite{tian02general} considered a factorization, referred to as the \textit{c-component factorization}, which in the language of ADMGs corresponds to the \textit{district factorization}. A district (or c-component) in an ADMG is defined as a maximal set of variables that are connected via bidirected paths, effectively partitioning variables into disjoint sets. Let $\cal{D}(\G)$ denote the set of all districts in $\G(O).$ \cite{tian02general} suggested to write $P(O) = \prod_{D \in \cal{D}(\G)} q_{D}(D \mid \pa_{\G}(D))$, where $q_{D}(D \mid \pa_{\G}(D))$, commonly known as a \textit{kernel} or \textit{c-factor}, associated with district $D$, maps values of $\pa_{\G}(D)$ to normalized densities over variables in district $D$. These kernels, despite facilitating probabilistic operations such as marginalization and conditioning, diverge from typical conditional distributions. Specifically, $q_{D}(D \mid \pa_{\G}(D))$ represents an intervention distribution that assumes variables outside of district $D$ are subjected to intervention and held at fixed values. 

Given a valid topological order $\tau$ over variables in $O$, where no variable is an ancestor of those preceding it in the order, the kernel for any district $D$ is linked to the observed data distribution $P(O)$ by $q_{D}(D \mid \pa_{\G}(D)) = \prod_{D_i \in D} P(D_i \mid \mpg(D_i))$. Here,  $\mpg(D_i)$ denotes the \textit{Markov pillow} of variable $D_i$, defined as the union of its district and the parental set of its district in a subgraph restricted to the predecessors of $D_i$ according to the topological order $\tau$. For simplicity, we omit $\tau$'s role in defining Markov pillows. This setup enables the district factorization of $P(O)$ to be expressed via a \textit{topological factorization} \citep{tian02general, bhattacharya2022semiparametric} as follows:
\begin{align}
    P(O) = \prod_{O_i \in O} P(O_i \mid \mpg(O_i)) \ . 
    \label{eq:top_fact}
\end{align}

This topological factorization simplifies the representation of $P(O)$ but may not capture all the equality constraints inherent in $P(O)$, as implied by the Markov properties of the hidden-variable DAG $\G(O \cup U)$. \cite{tian02on} provided a sound and complete algorithm for characterizing all such constraints in $P(O)$, while \cite{richardson2023nested} introduced the \textit{nested Markov factorization}, which encodes these constraints through a refined graphical factorization. For details on the nested Markov factorization, see \cite{richardson2023nested}, with a brief summary in Appendix~E of \cite{bhattacharya2022testability}. Our subsequent discussion in Section~\ref{sec:semiparam} delves into the implications of these constraints on the statistical efficiency of our proposed estimators. 

\subsection{A class of ADMGs defined by treatment primal fixability criterion}
\label{sec:class}

The intricate complexities associated with the use of ADMGs, ranging from statistical modeling and factorization to identification theories, have often deterred researchers and practitioners from leveraging this powerful graphical modeling framework, despite its effectiveness in modeling the complex dynamics between observed and unobserved variables in practice. In this paper, we narrow our focus down to a special class of ADMGs, characterized by the notion of \textit{primal fixability}, a criterion initially conceptualized by \cite{tian02general} and further developed by \cite{bhattacharya2022semiparametric}. This criterion examines the relationship between a treatment variable $A$, its immediate successors (or children), denoted as $\ch_\G(A)$, and its bidirectionally connected component, referred to as the treatment's district and denoted by $\dis_\G(A)$.
\begin{definition}\label{def:primal}
    In an ADMG $\G(O)$, a variable $O_i \in O$ is considered primal fixable if its children and its district do not overlap, that is, $\ch_\G(O_i) \cap \dis_\G(O_i) = \emptyset$.
\end{definition}

\begin{example}
Figure~\ref{fig:graphs_primal} illustrates examples where $A$ is primal fixable. For example, in Figure~\ref{fig:graphs_primal}(d), $\ch_\G(A) = \{M, Y\}$ and $\dis_\G(A) = \{A, L\}$, so the intersection is empty, and $A$ is primal fixable.
\label{ex:primal}
\end{example}

\begin{figure}[t] 
	\begin{center}
    \scalebox{0.65}{
    \begin{tikzpicture}[>=stealth, node distance=1.3cm]
        \tikzstyle{format} = [thick, circle, minimum size=1.0mm, inner sep=2pt]
        \tikzstyle{square} = [draw, thick, minimum size=4.5mm, inner sep=2pt]

        \begin{scope}[xshift=0cm, yshift=0cm]
		\path[->, thick]
		
		node[] (a) {$A$}
		node[right of=a, xshift=0.75cm] (m) {$M$}
		node[above of=m,  xshift=0cm, yshift=0cm] (x) {$X$}
		node[above left of=m, yshift=0.cm] (u) {}
		node[right of=m, xshift=0.75cm] (y) {$Y$}
		
		(x) edge[blue, bend left=0] (a) 
		(x) edge[blue, bend right=0] (y) 
		(a) edge[blue] (m) 
		(m) edge[blue] (y) 
        (a) edge[red, <->, out=340, in=200] (y)  
        (x) edge[red, <->, out=180, in=90] (a)  
        (x) edge[red, <->, out=0, in=90] (y)  

        node[below right of=m, xshift=-0.75cm, yshift=-0.2cm] (t3) {(a)} ;
		
	\end{scope}
    
    \begin{scope}[xshift=5.25cm, yshift=0cm]
		\path[->, thick]
		
		node[] (a) {$A$}
		node[right of=a, xshift=0.5cm] (m) {$M$}
        node[right of=m, xshift=0.5cm] (l) {$L$}
		node[above right of=m, yshift=0.2cm] (x) {$X$}
		node[right of=l, xshift=0.5cm] (y) {$Y$}
		
		(x) edge[blue, bend left=0] (a) 
		(x) edge[blue] (m) 
        (x) edge[blue] (l) 
		(x) edge[blue, bend right=0] (y) 
		(a) edge[blue] (m) 
		(m) edge[blue] (l) 
        (l) edge[blue] (y)    
        (a) edge[blue, bend right=25] (l) 
        (m) edge[blue, bend right=25] (y)

        (a) edge[red, <->, looseness=0.75, out=90, in=70] (y) 

        node[below right of=m, xshift=0.cm, yshift=-0.2cm] (t1) {(b)} ;
		
	\end{scope}
    \begin{scope}[xshift=11.85cm, yshift=0cm]
		\path[->, thick]
		
		node[] (a) {$A$}
		node[right of=a, xshift=0.5cm] (m) {$M$}
        node[right of=m, xshift=0.5cm] (l) {$L$}
		node[above right of=m, yshift=0.2cm] (x) {$X$}
		node[right of=l, xshift=0.5cm] (y) {$Y$}
		
		(x) edge[blue, bend left=0] (a) 
		(x) edge[blue] (m) 
        (x) edge[blue] (l) 
		(x) edge[blue, bend right=0] (y) 
		(a) edge[blue] (m) 
		(m) edge[blue] (l) 
        (l) edge[blue] (y)    
        (a) edge[red, <->, out=90, in=155] (l) 
        (l) edge[red, <->, out=25, in=90] (y) 

        node[below right of=m, xshift=0.cm, yshift=-0.2cm] (t1) {(c)} ;
		
	\end{scope}
    \begin{scope}[xshift=18.5cm, yshift=0cm]
		\path[->, thick]
		
		node[] (a) {$A$}
		node[right of=a, xshift=0.5cm] (m) {$M$}
        node[right of=m, xshift=0.5cm] (l) {$L$}
		node[above right of=m, yshift=0.2cm] (x) {$X$}
		node[above left of=m, yshift=0.cm] (u) {}
		node[right of=l, xshift=0.5cm] (y) {$Y$}
		
		(x) edge[blue, bend left=0] (a) 
		(x) edge[blue] (m) 
        (x) edge[blue] (l) 
		(x) edge[blue, bend right=0] (y) 
		(a) edge[blue] (m) 
		(m) edge[blue] (l) 
        (l) edge[blue] (y)    
        (a) edge[blue, bend right=15] (y) 
        (a) edge[red, <->, out=90, in=155] (l) 
        (m) edge[red, <->, out=25, in=90] (y) 

        node[below right of=m, xshift=0.cm, yshift=-0.2cm] (t2) {(d)} ;
		
	\end{scope}
	\end{tikzpicture}
	}
    \vspace{-0.2cm}
	\caption{(a) An example of a front-door ADMG containing ordinary independence constraint. (b-d) Examples of ADMGs that extend the front-door model; (a-c) $A$ and $Y$ share the same district; (d) $A$ and $Y$ belong to different districts.} 
	\label{fig:graphs_primal}
	\end{center}
\end{figure}

Our focus on the treatment primal fixability class is motivated by three main considerations. Graphically, primal fixability is straightforward to verify: the treatment must not share a district with any of its children. Conceptually, it provides an exact characterization of identification: the joint post-intervention distribution of all remaining variables is identifiable if and only if the treatment is primal fixable \citep{tian02general}, although causal effects on particular outcomes may remain identifiable even when this condition fails. Theoretically, the resulting identifying functional admits a unified representation across the entire class, providing a common foundation for developing flexible estimators and systematically studying their robustness properties, higher-order remainders, and nuisance-specific convergence-rate conditions. 

As we show below, the associated identifying functional under treatment primal fixability corresponds to intuitive extensions of the back-door and front-door criteria, although these connections often emerge only after marginalizing to a suitable subgraph. In other words, if there exists a valid back-door adjustment set $Z$ for the effect of $A$ on $Y$ in an ADMG $\G$, then $A$ is primal fixable either in $\G$ or in the ADMG derived from $\G$ by marginalizing all variables outside $\{A,Y,Z\}$. If $(A,Y)$ satisfy the front-door criterion with confounders $Z$ and mediators $M$, then $A$ is primal fixable either in $\G$ or in the ADMG derived from $\G$ by marginalizing all variables outside $\{A,Y,Z,M\}$. We illustrate these connections via the following example. 

\begin{example}
Figure~\ref{fig:graph_prunning} illustrates these connections. (a) The effect of $A$ on $Y$ is identifiable by the back-door criterion with an empty adjustment set (since $L$ is a collider), but $A$ is not primal fixable. (b–c) Marginalizing $M$ and then $L$ yields a graph where the effect is identifiable without adjustment and $A$ becomes primal fixable. (d) Another empty-adjustment case, where marginalizing $M$ (in $A$’s district but not a child) makes $A$ primal fixable. (e) A front-door example where $A$ becomes primal fixable only after marginalizing $L$. (f) A case where no marginalization yields primal fixability. 
\end{example} 

These examples highlight a general principle: a treatment not primal fixable in the original ADMG may become so after marginalizing certain children or district members that violate the criterion. In some cases multiple projections are needed. Thus, the relevant class of ADMGs includes not only those where the treatment is directly primal fixable, but also those where marginalization induces primal fixability. Developing efficient pruning algorithms and examining their impact on estimation efficiency remain important directions for future work.

\begin{figure}[t] 
	\begin{center}
    \scalebox{0.7}{
    \begin{tikzpicture}[>=stealth, node distance=1.4cm]
        \tikzstyle{format} = [thick, circle, minimum size=1.0mm, inner sep=2pt]
        \tikzstyle{square} = [draw, thick, minimum size=4.5mm, inner sep=2pt]
    \begin{scope}[xshift=0cm, yshift=0cm]
		\path[->, thick]
		
		node[] (a) {$A$}
		node[above right of=a, xshift=-0.25cm, yshift=0.cm] (m) {$M$}
        node[right of=m, xshift=-0.25cm, yshift=0.cm] (l) {$L$}
		node[below right of=l, xshift=-0.25cm] (y) {$Y$}
		 
		(a) edge[blue] (m) 
        (a) edge[red,<->,out=90,in=180] (m)
        (a) edge[blue] (y) 
		(m) edge[blue] (l) 
		(l) edge[red,<->,out=0, in=90] (y)

        node[below of=a, xshift=1.5cm, yshift=0.85cm] (t1) {(a)} ;
		
	\end{scope}
    \begin{scope}[xshift=3.75cm, yshift=0cm]
		\path[->, thick]
		
		node[] (a) {$A$}
		node[above right of=a, xshift=-0.25cm, yshift=0.cm] (m) {}
        node[right of=m, xshift=-0.25cm, yshift=0.cm] (l) {$L$}
		node[below right of=l, xshift=-0.25cm] (y) {$Y$}
		 
        (a) edge[blue] (y) 
		(a) edge[blue] (l) 
        (a) edge[red,<->, bend left] (l)
		(l) edge[red,<->,out=0, in=90] (y)

        node[below of=a, xshift=1.5cm, yshift=0.85cm] (t1) {(b)} ;
		
	\end{scope}
    \begin{scope}[xshift=7.5cm, yshift=0cm]
		\path[->, thick]
		
		node[] (a) {$A$}
		node[above right of=a, xshift=-0.25cm, yshift=0.cm] (m) {}
        node[right of=m, xshift=-0.25cm, yshift=0.cm] (l) {}
		node[below right of=l, xshift=-0.5cm] (y) {$Y$}
		 
        (a) edge[blue] (y) 

        node[below of=a, xshift=1.25cm, yshift=0.85cm] (t1) {(c)} ;
		
	\end{scope}
    \begin{scope}[xshift=11.cm, yshift=0cm]
		\path[->, thick]
		
		node[] (a) {$A$}
		node[above right of=a, xshift=0.5cm, yshift=0.cm] (m) {$M$}
		node[below right of=m, xshift=0.5cm] (y) {$Y$}
		 
		(a) edge[red, <->] (m) 
        (a) edge[blue] (y) 
		(m) edge[red, <->] (y) 

        node[below of=a, xshift=1.5cm, yshift=0.85cm] (t1) {(d)} ;
		
	\end{scope}
    \begin{scope}[xshift=15.cm, yshift=0cm]
		\path[->, thick]
		
		node[] (a) {$A$}
		node[above right of=a, xshift=0.5cm, yshift=0.cm] (l) {$L$}
        node[right of=a, xshift=0.1cm, yshift=0.cm] (m) {$M$}
		node[right of=m, xshift=0.cm] (y) {$Y$}
		 
		(a) edge[blue] (m) 
        (a) edge[red, <->, bend left] (y) 
		(m) edge[blue] (y) 
		(a) edge[red,<->, bend left] (l)
        (l) edge[red,<->, out=-20, in=20] (m)

        node[below of=a, xshift=1.5cm, yshift=0.85cm] (t1) {(e)} ;
		
	\end{scope}
    \begin{scope}[xshift=19cm, yshift=0cm]
		\path[->, thick]
		
		node[] (a) {$A$}
		node[above right of=a, xshift=0.5cm, yshift=0.cm] (m) {$M$}
		node[below right of=m, xshift=0.5cm] (y) {$Y$}
		 
		(a) edge[blue] (m) 
        (a) edge[red,<->,bend left] (m)
        (a) edge[blue] (y) 
		(m) edge[blue] (y) 

        node[below of=a, xshift=1.5cm, yshift=0.85cm] (t1) {(f)} ;
		
	\end{scope}
	\end{tikzpicture}
	}
    \vspace{-0.2cm}
	\caption{(a) An ADMG where $A$ is not primal fixable, but becomes so after marginalization; (b) Projection of (a) after marginalizing $M$; (c) Projection of (b) after marginalizing $L$; (d) An ADMG where marginalizing a district member ($M$) makes $A$ primal fixable; (e) An ADMG where marginalizing $L$ yields treatment primal fixability and front-door identification; (f) An ADMG where no marginalization leads to treatment primal fixability.} 
	\label{fig:graph_prunning}
	\end{center}
\end{figure}

Assuming, without any loss of generality, that our outcome variable $Y$ has no descendants in ADMG $\G(O)$, we establish a fixed valid topological order $\tau$ for all vertices in $\G(O)$. In this topological order, the treatment variable $A$ follows all its non-descendants, and $Y$ is the last element. Under this specified $\tau$, we partition $O$ into three disjoint subsets: $\X \cup \M \cup \bbL$, with $\X$ comprising all variables that precede the treatment, $\bbL$ consisting of all variables that are post-treatment and within $A$'s district, and $\M$ encompassing the rest of the post-treatment variables that lie outside of $A$'s district. We let $\bbL$ include the treatment $A$. The outcome variable may either be in $\M$ or $\bbL$; see Figure~\ref{fig:graphs_primal} for examples.  
For any $O_i \in O$, let $\mpgA{O_i}$ denote the Markov pillow of $O_i$ excluding treatment $A$. As shown by \cite{tian02general}, in an ADMG where $A$ is primal fixable, we can identify $\E[Y(a_0)]$ via the following functional of the observed data $P$, denoted by $\psi_{a_0}(P)$: 
\begin{equation}\label{eq:target_tian} 
\begin{aligned}
    \psi_{a_0}(P) 
    &=  \int  y  \ \Big\{ \prod_{L_i \in \bbL} dP(l_i \mid \mpga{l_i}, a_0) + \prod_{L_i \in \bbL} dP(l_i \mid \mpga{l_i}, a_1) \Big\}  \\ 
    &\hspace{3.5cm} \times \prod_{M_i \in  \M} \ dP(m_i \mid \mpga{m_i}, a_0) \  \prod_{X_i \in \X} dP(x_i \mid \mpg (x_i)) \ , 
\end{aligned}
\end{equation}
where a lower case letter $o_i$ denotes the realization of the variable $O_i$. To simplify notation yet maintain clarity, $\mpg(o_i)$ indicates the realization of the Markov pillow for $O_i$, and $\mpga{o_i}$ excludes the treatment realization. 

Let $\Z = \M \cup \bbL \setminus \{A, Y\}$ denote the collection of post-treatment pre-outcome variables. We index variables in $\Z$ according to $\tau$, as $Z_1, \ldots, Z_K$.  For any $Z_k \in \Z$, let $a_{Z_k} = a_0$ if $Z_k \in \M$ and $a_{Z_k} = a_1$ if $Z_k \in \bbL$. For outcome $Y$, we define $a_Y$ similarly. 

The following lemma re-expresses Tian and Pearl's identifying functional $\psi_{a_0}(P)$ in \eqref{eq:target_tian} in an algebraically equivalent form that facilitates our subsequent influence-function-based estimation procedures.

\begin{lemma}
    In an ADMG $\G(O)$ where $A$ is primal fixable, Tian and Pearl's identifying functional for $\E[Y(a_0)]$ in \eqref{eq:target_tian} admits the following representation:
    \begin{equation}
    \begin{aligned}
        \psi_{a_0}(P) 
        &=  \E\Big[ P(a_1 \mid \mpg(A))  \int  y \, dP(y \mid \mpga{y}, a_{Y})  \ \prod_{Z_k \in \Z} dP(z_k \mid \mpga{z_k}, a_{Z_k})  \Big] \\
        &\hspace{0.5cm} + \E\left[\I(A=a_0)Y\right]  \ , 
        \label{eq:target_par}
    \end{aligned}
    \end{equation}%
    where the first expectation is taken with respect to the joint distribution of covariates in $\X$. 
    \label{lem:reID}
\end{lemma}
See Appendix~\ref{app:proofs_id} for a proof of this equivalence.

\begin{example}
Since $A$ is primal fixable in Figure~\ref{fig:graphs_primal}(d) (see Example~\ref{ex:primal}), $\E[Y(a_0)]$ is identified by:
{\small 
\begin{align*}
    \text{(Fig.~\ref{fig:graphs_primal}(d))} \quad \psi_{a_0}(P) 
    &=  \int y \, dP(y \, | \, x, m, \ell, a_0) \, dP(m \, | \, x, a_0) \! \sum_{a \in \{0, 1\}} \! dP(\ell \, | \, x, m, a) \, dP(a \, | \, x) \, dP(x) \ , 
\end{align*}
}
and by Lemma~\ref{lem:reID}, it can be equivalently rewritten as
{\small 
\begin{align}
   \psi_{a_0}(P) 
     = \E\Big[ P(a_1 \, | \, X) \!\! \int \! y \,  dP(y \, | \, X, m, \ell, a_0) \, dP(\ell \, | \, X, m, a_1) \, dP(m \, | \, X, a_0)\Big] + \E\big[\I(A=a_0) \, Y \big] \, .  
   \label{eq:ex_reID}
\end{align}
}
\label{ex:psi_reID}
\end{example}
 
Our target parameter of inference is $\psi_{a_0}(P)$, as defined in \eqref{eq:target_par}. We first derive novel estimators, including both one-step corrected plug-in estimators and TMLEs, leveraging the efficient influence function within a nonparametric statistical framework, and discuss inference and asymptotic properties of our proposed estimators. Subsequently, we examine the relevance of these estimators within a semiparametric statistical model for an ADMG. 

\section{Influence function based estimation strategies}
\label{sec:est}

The parameter $\psi_{a_0}(P)$, defined in \eqref{eq:target_par}, depends on several key nuisance functional components: (i) the outcome regression $\E[Y \mid \mpg(Y)]$, represented as $\mu(\mpg(Y))$; (ii) the treatment propensity score $P(A=a \mid \mpg(A))$, noted as $\pi(a \mid \mpg(A))$; (iii) the conditional densities of mediators $P(Z_k \mid \mpg(Z_k))$, expressed as $f_{Z_k}(Z_k \mid \mpg(Z_k))$ for each $Z_k$ in the set $\Z$; and (iv) the distribution of covariates, $P_{X}$. Let $Q = \{ \mu, \pi, f_{Z_k} \ \forall Z_k \in \Z, P_{X}\}$ collect all the nuisances. With minor abuse of notation, we also write $\psi_{a_0}(Q)$ to refer to $\psi_{a_0}(P).$  

A plug-in estimate for $\psi_{a_0}(Q)$ in \eqref{eq:target_par} can be derived using the estimates for each nuisance in $Q$, namely $\hat{\mu}, \hat{\pi}, \hat{f}_{Z_k}$ for all $Z_k \in \Z$, while $P_{X}$ is approximated by the empirical distribution and $\E[\I(A=a_0)Y]$ is estimated by $n^{-1}\sum_{j=1}^n \I(A_j=a_0)Y_j$. Although the estimates for the outcome regression $\mu$ and the propensity score $\pi$ can be achieved through flexible regression models, the estimation of mediator densities beyond the scope of parametric models presents significant challenges, especially with continuous mediators. As an alternative, we adopt sequential regressions to propose a more flexible plug-in estimator.  

For a given mediator $Z_k \in \Z$, we define the following regression according to the conditional distribution of $Z_k$ evaluated at treatment level $A = a_{Z_k}$, namely, $f_{Z_k}(Z_k \mid \mpgA{Z_k}, a_{Z_k})$:
\begin{align}
    \mB{Z_k}\big(\mBarg{Z_k}, a_{Z_k}\big) = \E\big[ \mB{Z_{k+1}}(\mBarg{Z_{k+1}}, a_{Z_{k+1}}) \ \big| \ \mBarg{Z_k}, A=a_{Z_k} \big] \ ,
    \label{eq:B_k}
\end{align}
where $\mBarg{Z_k}$ denotes the union of the Markov pillows for $Z_k$ and its successors under the topological order $\tau$ (indicated by $\succeq Z_k$), excluding the treatment $A$ (indicated by the superscript $-A$), and retaining only variables that strictly precede $Z_k$ under $\tau$ (indicated by the subscript $\G_\prec$). This regression is obtained recursively for each $Z_k$ in $\Z$, with the convention that $Z_{K+1}$ coincides with $Y$ and $\mB{Z_{K+1}}(\mBarg{Z_{K+1}}, a_{Z_{K+1}})$ coincides with the outcome regression $\mu(\mpgA{Y}, a_Y)$. 

\begin{remark}
By the local Markov property, the regression representation in \eqref{eq:B_k} is equivalent to integration with respect to $f_{Z_k}(Z_k \mid \mpgA{Z_k}, a_{Z_k})$, as shown in Appendix~\ref{app:proofs_seqreg}. In nonparametrically saturated models (with no restriction), $\mBarg{Z_k}$ simplifies to the Markov pillow of $Z_k$, excluding $A$, denoted $\mpgA{Z_k}$. In contrast, in semiparametric models with independence restrictions (unsaturated), this simplification does not necessarily hold. For example, in Figure~\ref{fig:graphs}(c), where no restriction applies, $\mBarg{M} = \mpgA{M} = \{X\}$. In Figure~\ref{fig:graphs_primal}(a), under the constraint $M \perp X \mid A$, we have $\mBarg{M} \neq \mpgA{M}$ since $\mBarg{M} = \{\mpg(Y) \cup \mpg(M) \setminus A \} \cap \{X\} = \{X\}$ while $\mpgA{M} = \emptyset$. In this section, we develop estimation under an unrestricted statistical working model, without exploiting independence restrictions implied by the graph. We retain the conditioning sets $\mBarg{Z_k}$ in the regression notation throughout. Estimation that exploits ordinary conditional independence restrictions is discussed in Section~\ref{sec:semiparam}.
\end{remark}

Let $Q = \{\mu, \pi, \mB{Z_k} \ \forall Z_k \in \Z, P_X\}$ denote our new collection of nuisances. Given $n$ i.i.d. samples, our proposed plug-in estimator takes the following form: 
\begin{align}
    \psi^{\text{\tiny plug-in}}_{a_0}(\hat{Q}) = \frac{1}{n} \sum_{j = 1}^n \hat{\pi}( a_1 \mid \mpg(A_j)) \  \hmB{Z_1}(\mathrm{mp}^{-A}_{\G_\prec}(\succeq Z_{1j}), a_{Z_1}) + \frac{1}{n} \sum_{j = 1}^n \I(A_j=a_0)Y_j  \ , 
    \label{eq:plug-in}
\end{align}
where $\mathrm{mp}^{-A}_{\G_\prec}(\succeq Z_{1j})$ denotes the $j$-th instantiation of $\mathrm{mp}^{-A}_{\G_\prec}(\succeq Z_{1})$, $\hat{\pi}$ is an estimate of the propensity score $\pi$, and $\hmB{Z_1}$ is obtained by the following sequential regression procedure.

We begin by estimating the outcome regression and setting $\hmB{Z_{K+1}}(\mBarg{Z_{K+1}},a_{Z_{K+1}}) = \hat{\mu}(\mpgA{Y},A=a_Y)$. For $k=K,\ldots,1$, we treat the fitted values $\hmB{Z_{k+1}}(\mBarg{Z_{k+1}},a_{Z_{k+1}})$ as pseudo-outcomes, regress them on $(\mBarg{Z_k},A)$, and evaluate the fitted regression at $A=a_{Z_k}$ to obtain $\hmB{Z_k}(\mBarg{Z_k},a_{Z_k})$. Alternatively, the regression at step $k$ can be fitted using only observations with $A=a_{Z_k}$. This recursive procedure culminates in $\hmB{Z_1}(\mBarg{Z_1},a_{Z_1})$, which is substituted into \eqref{eq:plug-in}.

\begin{remark}
    Given the implications of the primal fixability criterion, $Z_1$, being the first mediator in the set $\Z$, is categorically part of the set $\M$. Consequently, the treatment level $a_{Z_1}$ attributed to $Z_1$ aligns with $a_0$. Accordingly, we can express $\hmB{Z_1}(\mBarg{Z_1}, a_{Z_1})$ as $\hmB{Z_1}(\mBarg{Z_1}, a_0)$, reflecting this alignment. 
\end{remark}

\begin{example}
    Continuing with the setup in Example~\ref{ex:psi_reID}, we can express $\psi_{a_0}(Q)$ in \eqref{eq:ex_reID} as: 
{\small
\begin{align*}
    \text{(Fig.~\ref{fig:graphs_primal}(d))} \qquad \psi_{a_0}(Q) 
    &= \E\big[ \pi(a_1 \mid X) \, \mB{Z_1}(X, a_{0}) \big] + \E\left[\I(A=a_0)Y\right] \ .
\end{align*}
}%
Here, $Z_1=M$ and $Z_2=L$, with $\mB{Z_1}(X,a_0) \coloneqq \E[\mB{Z_2}(M,X,a_1)\mid X,A=a_0]$, $\mB{Z_2}(M,X,a_1) \coloneqq \E[\mu_Y(L,M,X,a_0)\mid M,X,A=a_1]$, and $\mu_Y(L,M,X,a_0) \coloneqq \E[Y\mid L,M,X,A=a_0]$. The plug-in estimator of this estimand can be obtained by estimating $\mu_Y$ and sequentially regressing the resulting pseudo-outcomes to obtain $\hmB{Z_2}$ and $\hmB{Z_1}$, and substituting $\hat{\pi}$ and $\hmB{Z_1}$ into the expression above, with expectations evaluated empirically. 

\vspace{-0.15cm}
We note that a plug-in estimator based on the approach in \citet{bhattacharya2022semiparametric} requires estimating the conditional densities $P(M \! \mid \! X, A)$ and $P(L \! \mid \! M, X, A)$. For continuous mediators, this entails numerical integration (or approximation via Monte Carlo under a working model). Our parameterization avoids density estimation entirely. \citet{jung2024unified} also circumvent density estimation, but by using the empirical bifurcation method \citep{chernozhukov2023simple, xu2022neural}, which introduces an independent copy of the treatment variable $A' \sim P(A)$. Accordingly, their plug-in estimator relies on conditional regressions under an augmented law that involves $A'$, leading to terms like $\E[g(M)\mid A, A', X]$ and dependence on $P(M \! \mid \! A, A', X)$, rather than regressions that correspond directly to the joint factorization $P(M \! \mid \! A, X)$ that our approach uses. See Appendix~\ref{app:examples} for further discussion and concrete examples. 
\end{example}

We can examine the stochastic behavior of our plug-in estimator in \eqref{eq:plug-in}, through a linear expansion. Given an integrable function $f$ of the observed data $O$, let $P f \coloneqq \int f(o) \ dP(o)$ and $P_n f \coloneqq \frac{1}{n} \sum_{i = 1}^n f(O_i)$. The linear expansion is then given by 
\begin{align}
\label{eq:von_mise}
    \psi^{\text{plug-in}}_{a_0}(\hat{Q}) = \psi_{a_0}(Q) - P\Phi_{a_0}(\hat{Q}) + R_2(\hat{Q}, Q) \ ,
\end{align}
where $\Phi_{a_0}$ represents the gradient of the parameter, also known as the influence function, and $R_2(\hat{Q}, Q)$ encompasses the second-order remainder term of the linear approximation. In this context, the term $-P\Phi_{a_0}(\hat{Q})$ measures the first-order bias of the plug-in estimator, resulting from incorporating the estimates $\hat{Q}$ of $Q$. Notably, while the influence function $\Phi_{a_0}$ has a mean of zero under the probability distribution $P$, i.e., $P\Phi_{a_0} = 0$, the first-order bias $-P\Phi_{a_0}(\hat{Q})$ can still be significant. This observation leads us to introduce the one-step corrected estimator, designed to ``debias'' the initial plug-in estimates by incorporating an estimate of the first-order bias, specifically $-P_n\Phi_{a_0}(\hat{Q})$. We then adjust the plug-in estimator to $\psi^{\text{plug-in}}_{a_0}(\hat{Q}) + P_n\Phi_{a_0}(\hat{Q})$, which we will explore in further detail in the subsequent discussion. 

\subsection{One-step corrected plug-in estimator}
\label{subsec:one-step}

In an unrestricted nonparametric statistical model, the influence function for $\psi_{a_0}(P)$ is unique and is the nonparametric efficient influence function (EIF). \citet{bhattacharya2022semiparametric} derived this EIF for treatment primal-fixable effects using a parameterization based on the conditional densities and outcome regression in the observed-data factorization. Here, we re-express the same EIF in terms of sequential regressions and products of density ratios. This representation supports alternative nuisance estimation strategies and provides a basis for analyzing the nuisance-specific robustness properties and convergence-rate requirements. Discussions related to semiparametric statistical models are deferred to Section~\ref{sec:semiparam}. We first introduce the notation needed for this representation.

Given our established topological order, $\tau$, let $\M_{\prec Z_k}$ define the subset of variables in $\M$ that precede $Z_k$ according to $\tau$. Similarly, $\bbL_{\prec Z_k}$ refers to the subset of variables in $\bbL$ that come before $Z_k$ in the sequence. We introduce $\mathrm{mp}_{\G, \prec}(Z_k)$ to denote the collection of variables in $\M_{\prec Z_k}$ and their associated Markov pillows if  $Z_k \in \bbL$, or the collection of variables in $\bbL_{\prec Z_k}$ and their associated Markov pillows if $Z_k \in \M$. Additionally, $\mathrm{mp}^{-A}_{\G, \prec}(Z_k)$ signifies the same set as $\mathrm{mp}_{\G, \prec}(Z_k)$ but with the exclusion of the treatment variable $A$. 

For any $Z_k \in \Z$ and treatment $A$, we define the following ratios of conditional densities: 
\begin{align}
    \fr{Z_k}(Z_k, \mpgA{Z_k}) 
    = \frac{\f{Z_k}(Z_k \mid \mpgA{Z_k}, a_{Z_k})}{\f{Z_k}(Z_k \mid \mpgA{Z_k}, 1- a_{Z_k})} \ , \ \fr{A}(\mpg(A)) = \frac{\pi(a_1 \mid \mpg(A))}{\pi(a_0 \mid \mpg(A))} \ .  
    \label{eq:fr_density_ratio}
\end{align}
We denote $\mR{Z_k}$ as the product of these ratios of conditional mediator densities, defined as: 
{\small 
\begin{align}
    &\mR{Z_k}(\mathrm{mp}^{-A}_{\G, \prec}(Z_k))  \label{eq:R_prod_density_ratios} \\ 
    &\hspace{0.5cm}= \I(Z_k \in \bbL) \!\!\! \prod_{M_i \in \M_{\prec Z_k}} \fr{M_i}(M_i, \mpgA{M_i}) 
    + \I(Z_k \in \M) \ \fr{A}(\mpg(A)) \!\!\! \prod_{L_i \in \bbL_{\prec Z_k} \setminus A} \fr{L_i}(L_i, \mpgA{L_i})  \notag  \ ,
\end{align}%
}%
where the product calculation depends on whether $Z_k$ belongs to $\bbL$ or $\M$, using mediators from  $\M_{\prec Z_k}$ or $\bbL_{\prec Z_k}$, respectively. The term $\mR{Y}(\mathrm{mp}^{-A}_{\G, \prec}(Y))$ is defined similarly. 

\begin{example}
Consider the ADMG in Figure~\ref{fig:graphs_primal}(d), where $\M = \{M, Y\}$, $\bbL = \{A, L\}$, and $\Z = \{M, L\}$. The ratios of conditional densities $\fr{Z_k}$ are given by $\fr{L}(L, M, X) = f_L(L \! \mid \! M, X, a_1) / f_L(L \! \mid \! M, X, a_0)$, $\fr{M}(M, X) = f_M(M \! \mid \! X, a_0) / f_M(M \! \mid \! X, a_1)$, and $\fr{A}(X) = \pi(a_1 \! \mid \! X) / \pi(a_0 \! \mid \! X)$. The product of these ratios $\mR{Z_k}$ are defined as follows: for $M, Y \in \M$, we have $\mR{Y}(L,M,X)=\fr{A}(X)\, \fr{L}(L,M,X)$ and $\mR{M}(X)=\fr{A}(X)$; for $L \in \bbL$, $\mR{L}(M, X) = \fr{M}(M, X)$. 
\end{example}

Our collection of nuisances now expands to be $Q = \{\mu, \pi,\mR{Y}, \{\mB{Z_k}, \mR{Z_k} \ \forall Z_k \in \Z \}, P_X\}$. We have the following result. 

\begin{lemma}
The nonparametric efficient influence function derived by \citet{bhattacharya2022semiparametric} for our target parameter $\psi_{a_0}(Q)$ in \eqref{eq:target_par} admits the following representation in terms of the nuisance components in $Q$: 
\begin{subequations}
\begin{align}
    \Phi_{a_0}(Q)(O) 
    &=  \I(A = a_Y) \ \mR{Y}\big(\mathrm{mp}^{-A}_{\G, \prec}(Y)\big) \big\{ Y - \mu(\mpgA{Y}, a_Y) \big\} \label{eq:eif_y}
    \\
    \begin{split}
    &\hspace{0.5cm} + \sum_{Z_k \in \Z} \I(A = a_{Z_k}) \ \mR{Z_k}\big(\mathrm{mp}^{-A}_{\G, \prec}(Z_k)\big) \\
        &\hspace{2.5cm} \times \big\{  \mB{Z_{k+1}}(\mBarg{Z_{k+1}}, a_{Z_{k+1}}) - \mB{Z_{k}}(\mBarg{Z_k}, a_{Z_k}) \big\} 
    \end{split}
    \label{eq:eif_zk}
    \\
    &\hspace{0.5cm} + \left\{ \I(A = a_1) - \pi(a_1 \mid \mpg(A)) \right\} \ \mB{Z_1}(\mBarg{Z_1}, a_0) 
    \label{eq:eif_a}
    \\
    &\hspace{0.5cm} + \pi(a_1 \mid \mpg(A)) \ \mB{Z_1}(\mBarg{Z_1}, a_0) + \I(A = a_0) Y  - \psi_{a_0}(Q)  \ .
    \label{eq:eif_x}
\end{align}
\end{subequations}
\label{lem:eif}
\end{lemma}
See Appendix~\ref{app:proofs_eif} for a derivation of this representation. 

We note that although cancellation occurs between terms in \eqref{eq:eif_a} and \eqref{eq:eif_x}, we retain the current form of the EIF since \eqref{eq:eif_a} represents the component lying in the tangent space associated with the propensity score. This decomposition is particularly useful for constructing the targeted minimum loss-based estimator, as discussed in Section~\ref{subsec:tmle}, where \eqref{eq:eif_a} informs the targeting procedure that updates the initial estimate of the propensity score.

\begin{example}
In the ADMG of Figure~\ref{fig:graphs_primal}(d), the line \eqref{eq:eif_zk} in the EIF contains two terms corresponding to $M$ and $L$, namely $\I(A=a_0)\ \mR{M}(X) \ \{ \mB{L}(M,X, a_{1}) - \mB{M}(X, a_{0})\}$ and $\I(A=a_1) \ \mR{L}(M,X) \ \{ \mB{Y}(L,M,X, a_{0}) - \mB{L}(M,X, a_{1})\}$, respectively.
\end{example}

To construct our one-step corrected plug-in estimator, $\psi^\text{plug-in}_{a_0}(\hat{Q}) + P_n\Phi_{a_0}(\hat{Q})$, with $\Phi_{a_0}(Q)$ specified in Lemma~\ref{lem:eif}, we first need to obtain estimates for all nuisance components within $Q$. Estimates for the outcome regression and propensity scores, denoted as $\hat{\mu}$ and $\hat{\pi}$, can be acquired using flexible regression methods. The sequential regressions $\mB{Z_k}$ for each $Z_k$ in $\Z$ can be estimated flexibly through recursive pseudo-outcome regression methods previously described.

For estimating the product of density ratios, $\mR{Y}$ and $\mR{Z_k} \, \forall Z_k\in \Z$, we propose two flexible approaches. The first involves directly estimating the density ratios $\fr{Z_k}$ using methods highlighted by \cite{sugiyama2007direct, kanamori2009least, yamada2013relative, SUGIYAMA201044}. The estimates of $\mR{Y}$ and $\mR{Z_k}$ are then obtained by multiplying these density ratios as specified in \eqref{eq:R_prod_density_ratios}. Alternatively, individual density ratios $\fr{Z_k}$ can be estimated through regression methods, which often provide greater flexibility compared to direct estimation methods. For $Z_k\in\Z$, define
\begin{align}
    \li{Z_k}(Z_k,\mpgA{Z_k}) = P(a_{Z_k} \mid Z_k,\mpgA{Z_k}) \ , \ \lex{Z_k}(\mpgA{Z_k}) = P(a_{Z_k}\mid \mpgA{Z_k}) \ . 
    \label{eq:h_g_density_ratio}
\end{align}
This formulation aligns $\lex{Z_1}(\mpgA{Z_1})$ with the propensity score $\pi(a_0 \mid \mpg(A))$ as $\mpgA{Z_1} = \mpg(A)$ in an unrestricted nonparametric model. Using Bayes' theorem, we have:
\begin{align}
    \fr{Z_k}(Z_k, \mpgA{Z_k}) = \frac{\li{Z_k}(Z_k,\mpgA{Z_k})}{1-\li{Z_k}(Z_k,\mpgA{Z_k})} \times \frac{1-\lex{Z_k}(\mpgA{Z_k})}{\lex{Z_k}(\mpgA{Z_k})} \ . 
    \label{eq:fr_density_ratio_Bayes}
\end{align}
Subsequently, the estimate of $\mR{Z_k}(\mathrm{mp}^{-A}_{\G, \prec}(Z_k))$ for any $Z_k \in \Z$, is obtained through binary regressions. If $Z_k \in \bbL$, this involves estimating $g_{M_i}$ and $h_{M_i}$ for each $M_i \in \M_{\prec Z_k}$. Conversely, if $Z_k \in \M$, it requires estimating $g_{L_i}$ and $h_{L_i}$ for each $L_i \in \bbL_{\prec Z_k} \setminus A$. The estimate of $\mR{Y}(\mathrm{mp}^{-A}_{\G, \prec}(Y))$ can be obtained similarly.  

The use of Bayes' rule to express mediator density ratios through conditional treatment probabilities has precedents in causal mediation analysis \citep{tchetgen2013inverse, zheng2012targeted}, and in estimation of total effects under the front-door criterion and more general identifying functionals \citep{jung2024unified,guo2026flexible}. Here, we combine this established device with our sequential-regression representation to derive our estimators and subsequently analyze the resulting nuisance-specific robustness properties and convergence-rate requirements.

By aggregating all estimates into $\hat{Q}$ and applying the EIF from Lemma~\ref{lem:eif}, our proposed one-step corrected plug-in estimator, based on $n$ i.i.d. samples, denoted by $\psi^{+}_{a_0}(\hat{Q})$, is given by: 
\begin{align}
    \psi^{+}_{a_0}(\hat{Q}) 
    &= \frac{1}{n}\sum_{j = 1}^n \bigg\{ \I(A_j = a_Y) \  \hmR{Y}\big(\mathrm{mp}^{-A}_{\G, \prec}(Y_j)\big)  \big\{ Y_j - \hat{\mu}(\mpgA{Y_j}, a_Y) \big\} 
    \notag \\
    &\hspace{-0.95cm} + \!\!\! \sum_{Z_k \in \Z} \!\! \I(A_j = a_{Z_k}) \, \hmR{Z_k}\big(\mathrm{mp}^{-A}_{\G, \prec}(Z_{kj})\big)  \Big\{  \hmB{Z_{k+1}}(\mBarg{Z_{(k+1)j}}, a_{Z_{k+1}}) - \hmB{Z_{k}}(\mBarg{Z_{kj}}, a_{Z_k}) \Big\} \notag 
    \\
    &\hspace{-0.75cm} + \big\{ \I(A_j = a_1) - \hat{\pi}(a_1 \mid \mpg(A_j)) \big\} \, \hmB{Z_1}(\mBarg{Z_{1j}}, a_0) \notag 
    \\
    &\hspace{-0.75cm} + \hat{\pi}(a_1 \mid \mpg(A_j)) \ \hmB{Z_1}(\mBarg{Z_{1j}}, a_0) + \I(A_j = a_0) Y_j \bigg\}  \ . \qquad  \label{eq:one-step}  \textit{\small (one-step estimator)} 
\end{align}

Alternative one-step constructions for these identifying functionals have also been developed in the literature \citep{bhattacharya2022semiparametric, jung2024unified}. These constructions differ in their nuisance parameterizations and estimation strategies.

\citet{bhattacharya2022semiparametric} formulate their one-step estimator using the conditional-density representation of the identifying functional in \eqref{eq:target_tian} and its EIF. Their proposed implementation relies on estimating the conditional mediator densities, outcome regression, and propensity score, and evaluating integrals with respect to the fitted conditional mediator distributions. Our construction avoids mediator density estimation by first marginalizing over $A$ in the identifying functional, as shown in Lemma~1. This separates the factual contribution $\E[\I(A=a_0)Y]$ from the remaining term, in which each mediator integration conditions on a specified, fixed treatment value. Consequently, these integrations can be represented as sequential conditional expectations and estimated directly through pseudo-outcome regressions. Combined with direct density-ratio estimation or the Bayes-rule regressions in \eqref{eq:fr_density_ratio_Bayes}, this yields an implementation that requires neither direct mediator density estimation nor numerical integration. Thus, the treatment marginalization is the step that enables our particular sequential-regression construction; Lemma~\ref{lem:eif} then expresses the established EIF in terms of the resulting nuisance functions. Section~\ref{sec:inference} analyzes the population bias, robustness properties, and convergence-rate requirements of estimators constructed using these nuisance estimation strategies.

\citet{jung2024unified} develop one-step estimators for a broader class of sum-product identifying functionals through their unified covariate adjustment framework. Their construction uses empirical bifurcation \citep{chernozhukov2022automatic,xu2022neural}, which copies and randomly permutes selected variables. For the treatment variable $A$, this introduces an independent copy $A'\sim P(A)$, allowing its fixed and marginalized roles in the identifying functional to be represented separately. In the sequential regressions, the pseudo-outcomes are evaluated using either the original treatment $A$ or its permuted counterpart $A'$, while the predictors include one or both treatment variables, depending on the graph structure. They also use Bayes' rule to estimate density ratios through conditional treatment probabilities. Our construction instead starts from the treatment-level decomposition in Lemma~\ref{lem:reID} and uses the sequential regressions in \eqref{eq:B_k}, together with the density-ratio products in \eqref{eq:R_prod_density_ratios}. Comparing the resulting robustness guarantees and convergence-rate conditions requires matching the regression targets, weighting functions, and relationships among their estimates. Section~\ref{sec:inference} provides these analyses for our one-step estimators and TMLEs, and Appendix~\ref{app:examples} illustrates the nuisance representations in a common graphical example. 

While the one-step corrected plug-in estimator offers a flexible approach for estimating our target parameter $\psi_{a_0}(Q)$, it has a practical limitation in that it can generate estimates outside the intended parameter space, complicating interpretation---this is particularly problematic with binary or bounded continuous outcomes. Moreover, one-step estimators may lack stability under weak overlap conditions \citep{porter2011relative}. To address these challenges, we also propose TMLE as an alternative. TMLE not only ensures that estimates remain within the parameter space but also integrates with flexible modeling approaches to effectively handle complex data structures and exhibits favorable asymptotic behaviors similar to those of one-step estimators. 

\subsection{Targeted minimum loss based estimator}
\label{subsec:tmle}

Unlike the one-step estimator, which corrects for first-order bias by adding $-P_n\Phi_{a_0}(\hat{Q})$ to the plug-in estimator, the TMLE procedure aims to find estimates of $Q$ that make the first-order bias of the plug-in estimator negligible. Specifically, we seek to identify estimates of $Q$, denoted as $\hat{Q}^*$, such that $P_n\Phi_{a_0}(\hat{Q}^*) = o_P(n^{-1/2})$. Assuming such $\hat{Q}^*$ is given, our proposed TMLE is then given by: 
\begin{align}
    \psi_{a_0}(\hat{Q}^*) &= \frac{1}{n} \sum_{j = 1}^n \left\{ \hat{\pi}^*( a_1 \mid \mpg(A_j)) \  \hmB{Z_1}{*}(\mBarg{Z_{1j}}, a_{Z_1})  + \I(A_j=a_0)Y_j \right\} \ .   \qquad \textit{\small (TMLE)}
    \label{eq:tmle}
\end{align}%
Next, we outline how to obtain $\hat{Q}^*$. 

TMLE involves two primary steps, \textit{initialization} and \textit{targeting}. In the initialization step, we estimate $Q$ by individually estimating each nuisance parameter, as detailed in Section~\ref{subsec:one-step}. The targeting step then adjusts these estimates to ensure $P_n\Phi_{a_0}(\hat{Q}^*) = o_P(n^{-1/2})$. Below, we outline the key components and detailed procedure of the targeting step. 

We first decompose the EIF $\Phi_{a_0}(Q)$, as detailed in Lemma~\ref{lem:eif}, as follows:
\begin{align}
    \Phi_{a_0}(Q) = \Phi_{a_0, Y}(Q) + \sum_{Z_k \in \Z} \Phi_{a_0, Z_k}(Q)  + \Phi_{a_0, A}(Q) + \Phi_{a_0, \text{rem}}(Q) \ , 
    \label{eq:eif_decompose}
\end{align}
where each component is defined as:  
{\small 
\begin{equation*}
\begin{aligned}
    \Phi_{a_0, Y}(Q)(O) &=\I(A = a_Y) \ \mR{Y}\big(\mathrm{mp}^{-A}_{\G, \prec}(Y)\big) \left\{ Y - \mu(\mpgA{Y}, a_Y) \right\} \ , 
    \\
    \Phi_{a_0, Z_k}(Q)(O) &=\I(A = a_{Z_k}) \,\mR{Z_k}\big(\mathrm{mp}^{-A}_{\G, \prec}(Z_k)\big) \big\{  \mB{Z_{k+1}}(\mBarg{Z_{k+1}}, a_{Z_{k+1}}) - \mB{Z_{k}}(\mBarg{Z_k}, a_{Z_k}) \big\} \ , 
    \\
    \Phi_{a_0, A}(Q)(O) &= \left\{ \I(A = a_1) - \pi(a_1 \mid \mpg(A)) \right\} \ \mB{Z_1}(\mpgA{Z_1}, a_0) \ , 
    \\
    \Phi_{a_0, \text{rem}}(Q)(O) &=  \pi(a_1 \mid \mpg(A)) \ \mB{Z_1}(\mBarg{Z_1}, a_0)  +\I(A = a_0) Y  - \psi_{a_0}(Q)  \ . 
\end{aligned}
\end{equation*}
}%
This breakdown allows us to target nuisance components separately in the TMLE procedure. 

We begin by noting that using the empirical distribution for $P_X$ and sample mean for $\I(A=a_0)Y$ satisfies $P_n[\Phi_{a_0, \text{rem}}(\hat{Q}^*)] = o_P(n^{-1/2})$. Therefore, to ensure that $P_n[\Phi_{a_0}(\hat{Q}^*)] = o_P(n^{-1/2})$, it suffices to concurrently satisfy the following conditions: $P_n[\Phi_{a_0, Y}(\hat{Q}^*)] = o_P(n^{-1/2})$, $P_n[\Phi_{a_0, Z_k}(\hat{Q}^*)] = o_P(n^{-1/2})$ for each $Z_k$ in $\Z$, and $P_n[\Phi_{a_0, A}(\hat{Q}^*)] = o_P(n^{-1/2})$. Given the geometric interpretation of the EIF as a sum of terms in orthogonal subspaces that construct the model's tangent space, each condition effectively serves as a score equation. These score equations are solved by identifying the appropriate combination of a submodel and a loss function, aligning the derivatives of the chosen loss function with the EIF components, as detailed below. 

Let $\pazocal{M}_{Q_i}$ denote the functional space of nuisance component $Q_i \in Q$. For a given $\tilde{{\mu}}\in \pazocal{M}_{\mu}$, $\tilde{\pi}\in \pazocal{M}_{\pi}$,  $\tmR{Z_k}\in \pazocal{M}_{\mR{Z_k}}$, and $\tmB{Z_k}\in \pazocal{M}_{\mB{Z_k}}\ \forall Z_k\in\Z$, consider the following loss function $(L)$ and submodel (indexed by $\epsilon$) combinations: 
{\small 
\begin{equation}\label{display:loss_submodel}
\begin{aligned}
    &\begin{cases}
        L_Y(\tilde{{\mu}}; \tmR{Y}) = \tmR{Y} \{ Y - \tilde{\mu} \}^2  \\
        \tilde{\mu}(\varepsilon_Y) = \tilde{\mu} + \varepsilon_Y 
    \end{cases}
    \ , \qquad   
    \begin{cases}
        L_{A}(\tilde{\pi}) = - \log \tilde{\pi}   \\
        \tilde{\pi}(\varepsilon_A; \tmB{Z_1}) = \operatorname{expit} \Big[\operatorname{logit} \ \tilde{\pi} + \varepsilon_A\ \tmB{Z_1} \Big]  
    \end{cases} \ , 
    \\ 
    &\hspace{2cm}
    \begin{cases}
        L_{Z_k}(\tmB{Z_k}; \tmR{Z_k}, \tmB{Z_{k+1}}) = \tmR{Z_k}\{\tmB{Z_{k+1}}- \tmB{Z_k}\}^2  \\
        \tmB{Z_k}(\varepsilon_{Z_k}) = \tmB{Z_k} + \varepsilon_{Z_k} 
    \end{cases}  \quad \forall Z_k \in \Z \ . 
\end{aligned}
\end{equation}
}%

The loss functions and submodels are indexed by the targeted nuisance parameter and a univariate real-valued parameter $\epsilon$, respectively. In some cases, they are also indexed by additional nuisance components, as indicated by a semicolon in the function inputs, to emphasize the role of these components in solving a score equation when targeting a specific nuisance estimate. In Appendix~\ref{app:tmle_details}, we verify that these combinations yield score equations proportional to the corresponding components of the EIF in display \eqref{eq:eif_decompose}, as stated below: 
{\small 
\begin{align*}
    \frac{\partial L_Y(\tilde{\mu}(\epsilon_Y); \tmR{Y})}{\partial \epsilon_Y} \propto \Phi_{a_0, Y}(\tilde{Q}) 
    \ ,  \ 
    \frac{\partial L_{A}(\tilde{\pi}(\epsilon_A; \tmB{Z_1}))}{\partial \epsilon_{A}} \propto \Phi_{a_0, A}(\tilde{Q})
    \ , \
    \frac{\partial L_{Z_k}(\tmB{Z_k}(\epsilon_{Z_k}); \tmR{Z_k}, \tmB{Z_{k+1}} )}{\partial \epsilon_{Z_k}} \propto \Phi_{a_0, Z_k}(\tilde{Q}) \ . 
\end{align*}
}

Note that the submodel for $\tilde{\mu}$ in display~\eqref{display:loss_submodel} is applicable when $Y$ is continuous. Generalization to binary outcomes is straightforward and is elaborated in Appendix~\ref{app:tmle_binary_y}.

The nuisance estimates in $\hat{Q}$ are then updated through an iterative risk minimization process. We begin with initial estimates, denoted by $\hat{Q}^{(0)}$. At each iteration $t$, with the current estimates denoted by $\hat{Q}^{(t)}$, we follow three targeting steps (T1-T3) to update these estimates, and arrive at $\hat{Q}^{(t+1)}$.

\noindent \emph{(T1): Risk minimization for $\pi$.} 

Define $\hat{\pi}^{(t+1)}$ as $\operatorname{expit} [\operatorname{logit} \hat{\pi}^{(t)} + \hat{\varepsilon}_A\ \hmB{Z_1}{(t)} ]$, where $\hat{\varepsilon}_A$ is the minmizer of the following empirical risk: $\hat{\varepsilon}_A = \argmin_{\varepsilon_A \in \mathbb{R}} P_n L_A(\hat{\pi}^{(t)}(\varepsilon_A; \hmB{Z_1}{(t)}))$. This minimization is equivalent to finding the coefficient of $\hmB{Z_1}{(t)}$ in the following logistic regression model with no intercept: $\I(A=a_1)\sim\mathrm{offset}(\operatorname{logit}\ \hat{\pi}^{(t)}(\mpi(A)))+\hmB{Z_1}{(t)}$. 

Updating the estimate of $\pi$ implies updating the estimate of the density ratios $\hmR{Y}{(t)}, \hmR{Z_k}{(t)}$, denoted by $\hmR{Y}{(t+1)}, \hmR{Z_k}{(t+1)}$ for all $Z_k\in \Z$. 

Define $\hat{Q}^{(t, *)}=(\hat{\mu}^{(t)}, \hat{\pi}^{(t+1)}, \hmR{Y}{(t+1)}, \{ \hmR{Z_k}{(t+1)}, \hmB{Z_k}{(t)} \  \forall Z_k \in \Z\})$. It is noted that $P_n\Phi_{A}(\hat{Q}^{(t, *)}) = 0$. 

\vspace{0.25cm}
\noindent \emph{(T2): Risk minimization for $\mu$.} 

Define $\hat{\mu}^{(t+1)} = \hat{\mu}^{(t)} + \hat{\epsilon}_Y$, where $\hat{\epsilon}_Y$ is the minimizer of the empirical risk: \\ 
$\hat{\varepsilon}_Y = \argmin_{\varepsilon_Y \in \mathbb{R}} P_n L_Y(\hat{\mu}^{(t)}(\varepsilon_Y); \hmR{Y}{(t+1)})$. The solution $\hat{\varepsilon}_Y$ corresponds to the intercept in the following weighted regression model $Y \sim \mathrm{offset}(\hat{\mu}^{(t)})+1$ with weights $\hmR{Y}{(t+1)}$. 

Redefine $\hat{Q}^{(t, *)}$ as $\{\hat{\mu}^{(t+1)}, \hat{\pi}^{(t+1)}, \hmR{Y}{(t+1)}, \{ \hmR{Z_k}{(t+1)}, \hmB{Z_k}{(t)}, \  \forall Z_k \in \Z\} \}$. It holds that $P_n\Phi_{Y}(\hat{Q}^{(t, *)}) = 0$. 

\vspace{0.25cm}
\noindent \emph{(T3): Sequential risk minimization for $\mB{Z_k}\ , \forall Z_k\in\Z$.} 

We update the sequential regressions in reverse order, starting from $k=K$ down to $k=1$. We define $\mB{\succ Z_k}$ to include all sequential regressions subsequent to $\mB{Z_k}$: $\{\mB{Z_{k+1}}, \ldots, \mB{Z_K}\}$. Assuming $\hmB{\succ Z_k}{(t+1)}$ encompasses the updated estimates, which includes ${\hmB{Z_{k+1}}{(t+1)}, \ldots, \hmB{Z_K}{(t+1)}}$, the following are the steps for updating the estimate of the sequential regression related to the $k$-th mediator:

\hspace{0.25cm} \emph{(T3-a):} Given $\hmB{\succ Z_k}{(t+1)}$, we first revise the estimate $\hmB{Z_k}{(t)}$ using the previously described sequential regression scheme. This involves treating estimates of $\hmB{Z_{k+1}}{(t+1)}$ as realizations of a pseudo-outcome and regressing it onto $\mpg(Z_k)$, evaluated at $a_{Z_k}$. 

\hspace{0.25cm} \emph{(T3-b):} Define $\hmB{Z_k}{(t+1)} = \hmB{Z_k}{(t)} + \hat{\epsilon}_{Z_k}$, where $\hat{\epsilon}_{Z_k}$ is the minimizer to the empirical risk defined as $\hat{\varepsilon}_{Z_k} \ = \ \argmin_{\varepsilon_{Z_k} \in \mathbb{R}} \ P_n L_{Z_k}(\hmB{Z_k}{(t)}(\varepsilon_{Z_k}); \hmR{Z_k}{(t+1)},\hmB{Z_{k+1}}{(t+1)} )$. The solution corresponds to the intercept in the following weighted linear regression model $\hmB{Z_{k+1}}{(t+1)} \sim \mathrm{offset}(\hmB{Z_k}{(t)})+1$ with weights $\hmR{Z_k}^{(t+1)}$. 

\hspace{0.25cm} \emph{(T3-c):}  Update $\hmB{\succ Z_{k-1}}{(t+1)} = \{\hmB{Z_k}{(t+1)}, \hmB{\succ Z_k}{(t+1)}\}$, and decrement $k$ by one. If we redefine $\hat{Q}^{(t, *)}$ as $(\hat{Q}^{(t, *)} \setminus \hmB{Z_k}{(t)}, \hmB{Z_k}{(t+1)})$---replacing the old sequential regression estimate with the new one---it is noted that $P_n\Phi_{Z_k}(\hat{Q}^{(t, *)}) = 0$.

Repeat steps (T3-a) through (T3-c) while $k > 1$. Once $k=1$, all sequential regressions are updated. At the conclusion of the three targeting steps (T1-T3), we define $\hat{Q}^{(t+1)} = \hat{Q}^{(t, *)}$ which includes $\hat{\mu}^{(t+1)}, \hat{\pi}^{(t+1)}, \hmR{Y}{(t+1)}, \{ \hmR{Z_k}{(t+1)}, \hmB{Z_k}{(t+1)}, \  \forall Z_k \in \Z\}$.  

If at iteration $t^*$, the magnitude of $|P_n \Phi_{a_0}(\hat{Q}^{t^*})|$ is less than a pre-specified threshold $C_n = o_P(n^{-1/2})$, the TMLE procedure concludes, and $\hat{Q}^{*}=\hat{Q}^{(t^*)}$ becomes our final estimate of $Q$. 

For ease of implementation, we summarize this TMLE procedure in Algorithm~\ref{appalg:continuous} of Appendix~\ref{app:tmle_alg}.

\section{Asymptotic properties and robustness behaviors}
\label{sec:inference}

In this section, we examine the asymptotic behaviors of our proposed one-step estimator, as detailed in Section~\ref{subsec:one-step} by \eqref{eq:one-step}, and TMLE, as detailed in Section~\ref{subsec:tmle} by \eqref{eq:tmle}.

Let $\psi_{a_0}(\hat{Q})$ denote either the one-step estimator $\psi_{a_0}^+(\hat{Q})$ or the TMLE $\psi_{a_0}(\hat{Q}^*)$. To facilitate the analysis of the asymptotic behavior exhibited by these proposed estimators, we reformulate the linear expansion presented in \eqref{eq:von_mise} as follows, leveraging the fact that $P\Phi_{a_0}(Q) = 0$:
\begin{align}
    \psi_{a_0}(\hat{Q}) - \psi_{a_0}(Q) = P_n \Phi_{a_0}(Q) + (P_n - P) \{\Phi_{a_0}(\hat{Q}) - \Phi_{a_0}(Q)\} + R_2(\hat{Q}, Q) \ .
    \label{eq:expansion}
\end{align}%
The first term in expansion \eqref{eq:expansion} is a sample average of the mean-zero influence function and is  $O_P(n^{-1/2})$ by the central limit theorem, assuming finite variance. The second term is a centered empirical process term. Under the Donsker and influence-function convergence conditions detailed in Appendix~\ref{app:proofs_r2}, it satisfies $(P_n - P) \{\Phi_{a_0}(\hat{Q}) - \Phi_{a_0}(Q)\} = o_P(n^{-1/2})$. In Section~\ref{subsec:cross_fit}, we discuss cross-fitting as an alternative to the Donsker conditions \citep{kennedy2022semiparametric,double17chernozhukov}. The final term in this expansion, $R_2(\hat{Q}, Q)$, is the second-order remainder term. 
Its asymptotic behavior depends on the nuisance estimation errors and their relationships under the chosen estimation strategy. Controlling this remainder is crucial to establishing asymptotic linearity. For sequential regressions, we first clarify the regression targets relevant to this analysis.

\begin{remark}
\label{rem:seq_reg}
For the constructions based on directly fitted sequential regressions, accurate estimation at step $k$ means that $\hmB{Z_k}$ approximates the conditional expectation of its fitted pseudo-outcome, with the fitted downstream function held fixed. Thus, the relevant regression error is
\begin{align*}
    \hmB{Z_k}(\mBarg{Z_k},a_{Z_k})
    -
    \E_P\big[
        \hmB{Z_{k+1}}(\mBarg{Z_{k+1}},a_{Z_{k+1}})
        \mid \mBarg{Z_k},A=a_{Z_k}
    \big].
\end{align*}
Requiring this error to vanish does not require consistent estimation of the downstream regressions used to construct the pseudo-outcome. It also does not necessarily imply convergence of $\hmB{Z_k}$ to the fixed population regression $\mB{Z_k}$ in \eqref{eq:B_k}, which is defined using the true downstream function. For the directly fitted sequential-regression constructions, we henceforth use $\|\widehat B_{Z_k}-B_{Z_k}\|$ as shorthand for the norm of the regression error displayed above, with the fitted downstream function held fixed. The corresponding convergence rates and consistency conditions are understood in this sense.
\end{remark}

Let $\pazocal{Q}$ denote the space of nuisance collections under consideration. We now characterize the remainder for a generic collection $\tilde Q\in\pazocal{Q}$ relative to the true collection $Q$. For two functions $f(o)$ and $g(o)$, write $\{f-g\}(o)=f(o)-g(o)$. We also let $\{\prec z_k\}^{-a}$ denote the realization of the variables preceding $Z_k$, excluding $A$.
\begin{lemma}\label{lem:r2_compact}
    For any $\tilde{Q} = \{\tilde{\mu}, \tmR{Y}, \{\tmR{Z_k}, \tmB{Z_k} \forall Z_k\in \Z\}\} \in \pazocal{Q}$, we have: 
    \begin{align*}
        R_2(\tilde{Q}, Q)
        &=\sum_{Z_k \in \Z} \int \{\tmR{Z_k} -\mR{Z_k}\}(\mathrm{mp}^{-a}_{\G, \prec}(z_k)) \times \{\mB{Z_k} -\tmB{Z_k}\}(\mBarga{z_k}, a_{Z_k})\ dP(\{\prec z_k\}^{-a},a_{Z_k})
        \\
        &\hspace{0.25cm}+\int \{\tmR{Y} -\mR{Y} \}(\mathrm{mp}^{-a}_{\G, \prec}(y)) \times \{\mu -\tilde{\mu} \}(\mpga{y}, a_{Y})\ dP(\{\prec y\}^{-a},a_Y) \ . 
    \end{align*}
\end{lemma}
Recall that our nuisance components $Q$ include $\mu$, $\pi$, $\mR{Y},$ and $\{ \mR{Z_k}, \mB{Z_k} \ \forall Z_k \in \Z \}$. We can readily use flexible regression methods to estimate $\mu$ and $\pi$. However, estimation of the other components depends on their specific parameterization. In Section~\ref{sec:est}, we discussed methods for estimating $\mR{Y}$ and $\mR{Z_k}$, either through  density estimations, or direct density ratio estimations, or by employing a series of binary regressions using the Bayes' parameterization. Similarly, we discussed estimating $\mB{Z_k}$ by integrating over an estimate of the conditional densities or through sequential regressions. 
Below, we consider three combinations of these estimation strategies and establish sufficient conditions for consistency and asymptotic linearity of the resulting one-step estimators and TMLEs. We use $\|f\| = (Pf^2)^{1/2}$ to denote the $L^2(P)$-norm of $f$. Before stating these conditions, we clarify their treatment-specific interpretation.

\begin{remark}
\label{rem:trt-specific-nuisances}
The conditions below concern nuisance functions evaluated at the treatment levels at which they enter the estimator. In particular, the outcome-regression requirements concern $\mu(\mpgA{Y},a_Y)$, and the sequential-regression requirements concern the conditional expectations evaluated at $A=a_{Z_k}$, as specified above. Correct estimation of these functions at other treatment levels is not required unless those levels enter another component of the estimator. For density-based constructions, the required treatment levels depend on whether a conditional density enters an integration step or the numerator or denominator of a density ratio; a density ratio generally involves both treatment levels. These statements concern a single intervention mean $\psi_{a_0}(P)$. Estimating an average causal effect by combining two intervention means requires accounting for the nuisance components entering both estimators.
\end{remark}


\subsection{Sequential regressions and direct density ratio estimations} 
\label{subsec:inference_ratio}

For our first nuisance estimation method, we fit the sequential regressions $\mB{Z_k}$ for all $Z_k \in \Z$, directly using pseudo-outcomes. Further, we directly estimate the individual density ratios $\fr{Z_k}$ for all $Z_k \in \Z$. Their products yield estimates of $\mR{Y}$ and $\mR{Z_k}$ according to \eqref{eq:R_prod_density_ratios}. With this, we avoid the complexities associated with mediator density estimations. We denote the one-step estimator and TMLE that employ these nuisance estimates as  $\psi^+_{a_0, 1}(\hat{Q})$ and $\psi_{a_0, 1}(\hat{Q}^*)$, respectively. Here, $\hat{Q}$ includes $\hat{\mu}$, $\hat{\pi}$, $\hfr{Z_k}$, and $\hmB{Z_k}, \forall Z_k \in \Z$. 

Given the $R_2$ term in Lemma~\ref{lem:r2_compact}, we can express the difference $\tmR{Z_k} -\mR{Z_k}$ as follows, where the function inputs are removed for simplicity: 
{\small
\begin{equation}\label{eq:Rz_difference_ratio}
\begin{aligned}
    \tmR{Z_k} -\mR{Z_k} 
    &= \I(Z_k \in \bbL) \sum_{Z_i\in \M_{\prec Z_k}}\left(\tfr{Z_i}-\fr{Z_i}\right) \prod_{\{j<i:Z_j\in\M_{\prec Z_k}\}} \fr{Z_j} \prod_{\{h>i:Z_h\in\M_{\prec Z_k}\}} \tfr{Z_h} \\
    &\hspace{0.25cm} + \I(Z_k \in \mathcal{M}) \sum_{Z_i\in \bbL_{\prec Z_k}}\left(\tfr{Z_i}-\fr{Z_i}\right) \prod_{\{j<i:Z_j\in\bbL_{\prec Z_k}\}} \fr{Z_j} \prod_{\{h>i:Z_h\in\bbL_{\prec Z_k}\}} \tfr{Z_h} \ . 
\end{aligned}
\end{equation}
}%

We have the following theorem establishing the asymptotic linearity of $\psi^+_{a_0, 1}(\hat{Q})$ and $\psi_{a_0, 1}(\hat{Q}^*)$, where we define $\Z_{\prec Z_k}=\{Z_i: Z_i\in\Z, Z_i\prec Z_k\}$.

\begin{theorem}\label{thm:asymp_ratio}
Assume the $L^2(P)$ convergence behaviors of the nuisance estimates in $\hat{Q}$ are as follows: $|| \hat{\mu} - \mu || =\smallO(n^{-{1}/{b_Y}})$, $|| \hmB{Z_k} - \mB{Z_k} || = o_P(n^{-{1}/{b_{Z_k}}})$ and $|| \hfr{Z_k} - \fr{Z_k} || = o_P(n^{-{1}/{r_{Z_k}}})$ for all $Z_k \in \Z$, and $|| \hat{f}^r_{A} - \fr{A} || = o_P(n^{-{1}/{r_{A}}})$. 
Under bounded conditions discussed in Appendix~\ref{appsubsec:r2_ratio}, if 
\begin{enumerate}
    \item $\frac{1}{b_{Z_k}}+\frac{1}{r_{Z_i}}\geq \frac{1}{2} \ , \ \forall Z_k\in\Z$  and $\forall Z_i\in\Z_{\prec Z_k}$ s.t. $a_{Z_i} \neq a_{Z_k}$,  
    \item $\frac{1}{b_Y}+\frac{1}{r_{Z_i}}\geq \frac{1}{2} \ , \ \forall Z_i\in \Z$ s.t. $a_{Z_i} \neq a_Y$, 
    \item $\frac{1}{r_{A}}+\frac{1}{b_{Z_i}}\geq \frac{1}{2} \ , \ \forall Z_i \in \Z\cap\M$,
    \item $\frac{1}{b_Y}+\frac{1}{r_A}\geq \frac{1}{2} \ $, if $Y\in\M$,
\end{enumerate}
then the one-step estimator and TMLE are both asymptotically linear, that is $\psi_{a_0, 1}^+(\hat{Q}) - \psi_{a_0}(Q) = P_n\Phi_{a_0}(Q) + o_P(n^{-1/2})$ and $\psi_{a_0, 1}(\hat{Q}^*) - \psi_{a_0}(Q) = P_n\Phi_{a_0}(Q) + o_P(n^{-1/2})$, where $\Phi_{a_0}(Q)$ is the influence function, given in Lemma~\ref{lem:eif}.  
\end{theorem}

The conditions of Theorem~4 identify which nuisance estimation errors must have products of order $o_P(n^{-1/2})$, according to the graphical structure. They permit trade-offs between the convergence rates of the paired nuisance estimators, rather than requiring every nuisance estimator to satisfy the commonly used sufficient rate condition $o_P(n^{-1/4})$. Such rate trade-offs are a general feature of product-remainder analyses, including \citet{jung2024unified}; Theorem~4 makes the required pairings explicit for our sequential-regression and direct density-ratio construction across the treatment primal-fixable class. The following corollary states sufficient consistency conditions for this construction, expressed in terms of $L^2(P)$-consistency of the nuisance estimates.

\begin{corollary}\label{cor:robust_ratio}
$\psi^+_{a_0, 1}(\hat{Q})$ and $\psi_{a_0, 1}(\hat{Q}^*)$ are consistent estimators for $\psi(Q)$ if the following conditions are simultaneously true: 
\begin{enumerate}[label=(\roman*)]
    \item $|| \hat{\mathscr{B}}_{Z_k} - \mB{Z_k} || = o_P(1)$ or $||\hat{f}^r_{Z_i} - f^r_{Z_i} || = o_P(1)$, $\forall Z_k \in \Z$ and $\forall Z_i\in \Z_{\prec Z_k}$ s.t. $a_{Z_i} \not= a_{Z_k}$, 
    \item $|| \hat{\mu} - \mu || = o_P(1)$ or $|| \hat{f}^r_{Z_i} - f^r_{Z_i} || = o_P(1)$, $\forall Z_i \in \Z$ s.t. $a_{Z_i} \not= a_{Y}$, 
    \item $|| \hat{f}^r_{A} - f^r_{A} || = o_P(1)$ or $|| \hat{\mathscr{B}}_{Z_i} - \mB{Z_i} || = o_P(1)$, $\forall Z_i \in \Z\cap\mathcal{M}$,
    \item $|| \hat{\mu} - \mu || = o_P(1)$ or $|| \hat{f}^r_{A} - f^r_{A} || = o_P(1)$, if $Y\in\M$.
\end{enumerate}
\end{corollary}

The above conditions may overlap due to the hierarchical structure of the variables. For example, if $Y\in\M$, the requirement in Condition (ii) that $|| \hat{f}^r_{Z_i} - f^r_{Z_i} || = o_P(1)$ for all $Z_i \in \Z$ such that $a_{Z_i} \neq a_{Y}$  involves the consistent estimation of mediator density ratios $f^r_{Z_i}$ for all $Z_i\in\bbL$. Thus, Condition (i) for $Z_k \in \Z\cap\mathcal{M}$ is automatically assured. 

Corollary~\ref{cor:robust_ratio} provides a broad set of robustness results, allowing multiple ways to satisfy Conditions (i), (ii), and (iii). A notable scenario where these conditions are met results in the estimators $\psi^+_{a_0, 1}(\hat{Q})$ and $\psi_{a_0, 1}(\hat{Q}^*)$ being \textbf{doubly robust}. Specifically, the estimators are consistent if either
\begin{itemize}
    \item all sequential regression estimates $\hat{\mathscr{B}}_{Z_k}$ for all $Z_k \in \Z$, along with $\hat{\mu}$, are consistent, or 
    \item all density ratio estimates $\hat{f}^r_{A}$ and $\hat{f}^r_{Z_k}$ for all $Z_k \in \Z$ are consistent. 
\end{itemize}

\begin{remark}
    We refer to our estimators as ``doubly robust'' because they remain consistent if at least one of two \textit{disjoint} sets of nuisance functions is consistently estimated, even when each set includes multiple nuisance components. This usage extends the notion of ``double robustness'' from applying to the number of nuisance functions to instead applying to the number of disjoint nuisance sets, a generalization also considered in recent work \citep{bhattacharya2022semiparametric, jung2024unified}.
\end{remark}

\begin{example}
Consider ADMGs in Figures~\ref{fig:graphs_primal}(b-d). The above double robustness property ensures that $\psi^+_{a_0, 1}(\hat{Q})$ and $\psi_{a_0, 1}(\hat{Q}^*)$ remain consistent if either the sequential regression estimates ($\hat{\mu},\hmB{L}$, $\hmB{M}$) are consistent, or the density ratio estimates ($\hat{f}^r_{L}, \hat{f}^r_{M}$, $\hat{f}^r_{A}$) are consistent. By Corollary~\ref{cor:robust_ratio}, other combinations can also guarantee consistency. 
For example, in Figure~\ref{fig:graphs_primal}(b), where $\M=\{M,L\}$ and $\bbL=\{A,Y\}$, $\psi^+_{a_0, 1}(\hat{Q})$ and $\psi_{a_0, 1}(\hat{Q}^*)$ are consistent if $\fr{A}$ is consistently estimated, along with either $\mu$ or both $\fr{L}$ and $\fr{M}$.
In Figure~\ref{fig:graphs_primal}(c), where $\M = \{M\}$ and $\bbL = \{A, L, Y\}$, $\psi_{a_0, 1}(\hat{Q}^*)$ are consistent if $\hat{f}^r_{M}$ is consistent, together with either $\hmB{M}$ or $\hat{f}^r_{A}$.
Lastly, in Figure~\ref{fig:graphs_primal}(d), where $\M=\{M,Y\}$ and $\bbL=\{A,L\}$, consistency is achieved if $\hat{f}^r_{A}$ is consistent, together with either $\hmB{L}$ and $\hat{f}^r_{L}$, or $\hat{f}^r_{M}$ and $\hat{\mu}$. 
\end{example}

\subsection{Sequential regressions and Bayesian reparameterization of density ratios}
\label{subsec:inference_bayes}

For our second method of estimating nuisance components, we continue to use recursive sequential regressions to estimate $\mB{Z_k}$ for each mediator $Z_k \in \Z$. Additionally, we employ reparameterization based on Bayes' rule (detailed in \eqref{eq:fr_density_ratio_Bayes}) to estimate $\mR{Y}$ and $\mR{Z_k}$ for each $Z_k \in \Z$. This approach avoids the need for direct estimation of mediator density ratios. The one-step estimator and TMLE that employ these nuisance estimates are denoted as $\psi^+_{a_0, 2}(\hat{Q})$ and $\psi_{a_0, 2}(\hat{Q}^*)$, respectively. Here, $\hat{Q}$ includes $\hat{\mu}$, binary regression estimates $\hat{g}_{Z_k}$ and $\hat{h}_{Z_k}$, and the sequential regressions $\hmB{Z_k}$, $\forall Z_k \in \Z$.

In this subsection, the difference in the remainder term, compared to that in the previous subsection, is characterized by how we express the differences in the estimated density ratios, $\tfr{Z_i} - f^r_{Z_i}$. This is achieved using the formula in \eqref{eq:fr_density_ratio_Bayes}, and the derivation is as follows:
{\small 
\begin{align}
    & \{\tfr{Z_i} - \fr{Z_i}\}(Z_i,\mpgA{Z_i}) \label{eq:fr_difference} \\
    &\hspace{1.cm}= \frac{\tilde{h}_{Z_i}(Z_i,\mpgA{Z_i})}{1-\tilde{h}_{Z_i}(Z_i,\mpgA{Z_i})} \frac{1-\tilde{g}_{Z_i}(\mpgA{Z_i})}{\tilde{g}_{Z_i}(\mpgA{Z_i})} 
    - \frac{\li{Z_i}(Z_i,\mpgA{Z_i})}{1-\li{Z_i}(Z_i,\mpgA{Z_i})} \frac{1-\lex{Z_i}(\mpgA{Z_i})}{\lex{Z_i}(\mpgA{Z_i})} \ . \notag 
\end{align}
}

We have the following theorem establishing the asymptotic linearity of $\psi^+_{a_0, 2}(\hat{Q})$ and $\psi_{a_0, 2}(\hat{Q}^*)$. 

\begin{theorem}\label{thm:asymp_bayes}
Assume the $L^2(P)$ convergence behaviors of the nuisance estimates in $\hat{Q}$ are as follows: 
$|| \hat{\mu} - \mu || =\smallO(n^{-{1}/{b_Y}})$, 
$|| \hmB{Z_k} - \mB{Z_k} || = o_P(n^{-{1}/{b_{Z_k}}})$, 
$|| \hli{Z_k} - \li{Z_k} || = o_P(n^{-{1}/t_{Z_k}})$,  
$|| \hlex{Z_k} - \lex{Z_k} || = o_P(n^{-{1}/d_{Z_k}})$  $\forall Z_k \in \Z$. Under certain boundedness conditions discussed in Appendix~\ref{appsubsec:r2_bayes}, if 
\begin{enumerate}
    \item $\frac{1}{b_{Z_k}}+\frac{1}{t_{Z_i}}\geq \frac{1}{2}$, and $\frac{1}{b_{Z_k}}+\frac{1}{d_{Z_i}}\geq \frac{1}{2}$, 
    $\forall Z_k\in\Z$ and $\forall Z_i\in\Z_{\prec Z_k}$ s.t. $a_{Z_i} \neq a_{Z_k},$,  
    \item $\frac{1}{b_Y}+\frac{1}{t_{Z_i}}\geq \frac{1}{2}$, and $\frac{1}{b_Y}+\frac{1}{d_{Z_i}}\geq \frac{1}{2}$,
    $\ \forall Z_i\in \Z$ s.t. $a_{Z_i} \neq a_Y$,  
    \item $\frac{1}{d_{Z_1}}+\frac{1}{b_{Z_i}}\geq \frac{1}{2}$, 
    $\forall Z_i\in\Z\cap\M$, 
    \item $\frac{1}{b_Y}+\frac{1}{d_{Z_1}}\geq \frac{1}{2} \ $, if $Y\in\M$,
\end{enumerate}
then the one-step estimator and TMLE are both asymptotically linear, 
that is $\psi_2^+(\hat{Q}) - \psi_{a_0}(Q) = P_n\Phi_{a_0}(Q) + o_P(n^{-1/2})$ and $\psi_{a_0, 2}(\hat{Q}^*) - \psi_{a_0}(Q) = P_n\Phi_{a_0}(Q) + o_P(n^{-1/2})$, where $\Phi_{a_0}(Q)$ is the influence function, given in Lemma~\ref{lem:eif}. 
\end{theorem}

Similar to Theorem~\ref{thm:asymp_ratio}, Theorem~\ref{thm:asymp_bayes} allows nuisance estimators to converge to their respective truths at rates slower than $n^{-1/2}$. However, for Theorem~\ref{thm:asymp_bayes}, the convergence conditions for the density ratio $f^r_{Z_k}$ are subdivided into separate requirements for $h_{Z_k}$ and $g_{Z_k}$. Additionally, the multiple robustness behavior observed for $\psi^+_{a_0, 1}(\hat{Q})$ and $\psi_{a_0, 1}(\hat{Q}^*)$, as detailed in Corollary~\ref{cor:robust_ratio}, similarly applies to $\psi^+_{a_0, 2}(\hat{Q})$ and $\psi_{a_0, 2}(\hat{Q}^*)$.

\begin{corollary}\label{cor:robust_bayes}
$\psi^+_{a_0, 2}(\hat{Q})$ and $\psi_{a_0, 2}(\hat{Q}^*)$ are consistent estimators for $\psi_{a_0}(Q)$ if the following conditions are simultaneously true: 
\begin{enumerate}[label=(\roman*)]
    \item $|| \hat{\mathscr{B}}_{Z_k} - \mB{Z_k} || = o_P(1)$ or both $||\hat{h}_{Z_i} - h_{Z_i} || = o_P(1)$ and $||\hat{g}_{Z_i} - g_{Z_i} || = o_P(1)$, $\forall Z_k \in \Z$ and $Z_i\in\Z_{\prec Z_k}$ s.t. $a_{Z_i} \not= a_{Z_k}$, 
    \item $|| \hat{\mu} - \mu || = o_P(1)$ or both $||\hat{h}_{Z_i} - h_{Z_i} || = o_P(1)$ and $||\hat{g}_{Z_i} - g_{Z_i} || = o_P(1)$, $\forall Z_i \in \Z$ s.t. $a_{Z_i} \not= a_{Y}$,
    \item $||\hat{g}_{Z_1} - g_{Z_1} || = o_P(1)$ or $|| \hat{\mathscr{B}}_{Z_i} - \mB{Z_i} || = o_P(1)$, $\forall Z_i \in \Z\cap\M$,
    \item $|| \hat{\mu} - \mu || = o_P(1)$ or $||\hat{g}_{Z_1} - g_{Z_1} || = o_P(1)$, if $Y\in\M$. 
\end{enumerate}
\end{corollary}

Similar to Corollary~\ref{cor:robust_ratio}, Corollary~\ref{cor:robust_bayes} entails specific conditions under which the estimators $\psi^+_{a_0, 2}(\hat{Q})$ and $\psi_{a_0, 2}(\hat{Q}^*)$ are \textbf{doubly robust}. These estimators remain consistent if either 
\begin{itemize}
    \item all sequential regression estimates $\hmB{Z_k}$ for all $Z_k\in\Z$, along with $\hat{\mu}$, are consistent, or 
    \item all density ratio estimates $\hat{h}_{Z_k}$ and $\hat{g}_{Z_k}$ for all $Z_k\in\Z$ are consistent. 
\end{itemize}

\begin{example}
Consider the ADMGs in Figures~\ref{fig:graphs_primal} (b-d). The above double robustness property ensures that $\psi^+_{a_0, 2}(\hat{Q})$ and $\psi_{a_0, 2}(\hat{Q}^*)$ are consistent if either the sequential regressions ($\mB{M}, \mB{L},$ $\mu$), or the nuisance parameters used for density ratio estimation ($h_M, g_M$, $h_L$), are consistently estimated. Corollary~\ref{cor:robust_bayes} also offers alternative ways to achieve consistency. 
For instance, in Figure~\ref{fig:graphs_primal}(b) ($\M=\{M,L\}$ and $\bbL=\{A,Y\}$), the estimators are consistent if $g_M$ is consistently estimated, together with either $\mu$ or both $h_L$ and $h_M$. 
In Figure~\ref{fig:graphs_primal}(c) ($\M = \{M\}$ and $\bbL = \{A, L, Y\}$), the estimators are consistent if either the sequential regressions ($\mu,\mB{L}$, $\mB{M}$) are consistently estimated or both $h_M$ and $g_M$ are consistently estimated. This is a weaker requirement than that of the initially stated double robustness property, which also necessitates consistent estimates of $h_L$. 
Lastly, in Figure~\ref{fig:graphs_primal}(d) ($\M=\{M,Y\}$ and $\bbL=\{A,L\}$), the estimators are consistent if both $h_M$ and $g_M$ are consistently estimated, together with either $\mu$ or $h_L$. Note that in this example $g_L=h_M$. 
\end{example}

\subsection{Conditional density-based estimations} 
\label{subsec:inference_densities}

In our third method of estimation, we directly estimate the conditional densities of all mediators. The nuisance functional parameters in this approach are $Q = \{\mu, \pi, \{ f_{Z_k} \ \forall Z_k \in \Z \} \}$. The product of density ratios $\mR{Y}$ and $\mR{Z_k}$ for each mediator $Z_k \in \Z$ are estimated via functions of the estimated mediator densities. The sequential regressions $\mB{Z_k}$ are estimated by numerically integrating $\mu$ with respect to the product of mediator densities, $\prod_{i=k}^{K} f_{Z_i}$. The one-step estimator and TMLE that utilize these nuisance estimates are denoted by $\psi^+_{a_0, 3}(\hat{Q})$ and $\psi_{a_0, 3}(\hat{Q}^*)$, respectively. These estimators are particularly suitable for scenarios involving discrete-valued mediators, where estimating densities through regression and performing numerical integration via finite summation are both practical and reliable. Note that by using the density estimation, the TMLE procedure in Section~\ref{subsec:tmle} requires modified submodels and loss functions to target conditional densities and update density ratios accordingly. Further details are provided in Appendix~\ref{app:tmle_density}.

We have the following theorem establishing the asymptotic linearity of $\psi^+_{a_0, 3}(\hat{Q})$ and $\psi_{a_0, 3}(\hat{Q}^*)$. 

\begin{theorem}\label{thm:asymp_density}
Assume the $L^2(P)$ convergence behaviors of the nuisance estimates in $\hat{Q}$ are as follows: 
$|| \hat{\mu} - \mu || =\smallO(n^{-{1}/{b_Y}})$ \ , 
$|| \hat{\pi} - \pi || =\smallO(n^{-{1}/l_A})$, 
$|| \hat{f}_{Z_k} - {f}_{Z_k} || = o_P(n^{-{1}/{l_{Z_k}}})$, for all $Z_k$ in $\Z$.  
Under certain bounded conditions discussed in Appendix~\ref{appsubsec:r2_density}, if 
\begin{enumerate}
    \item $\frac{1}{l_{Z_k}}+\frac{1}{l_{Z_i}}\geq \frac{1}{2},\ \forall Z_k\in\Z,\ \forall Z_i\in \Z_{\prec Z_k}$ s.t. $a_{Z_i} \not= a_{Z_k}$,  
    \item $\frac{1}{b_Y}+\frac{1}{l_{Z_i}}\geq \frac{1}{2},\ \forall Z_i\in\Z$ s.t. $a_{Z_i} \not= a_{Y}$, 
    \item $\frac{1}{l_A}+\frac{1}{l_{Z_i}}\geq \frac{1}{2},\ \forall Z_i\in\Z\cap\M$,
    \item $\frac{1}{b_{Y}}+\frac{1}{l_A}\geq \frac{1}{2}$, if $Y\in\M$,
\end{enumerate}
then the one-step estimator and TMLE are both asymptotically linear, 
that is $\psi_3^+(\hat{Q}) - \psi_{a_0}(Q) = P_n\Phi(Q) + o_P(n^{-1/2})$ and $\psi_{a_0, 3}(\hat{Q}^*) - \psi_{a_0}(Q) = P_n\Phi_{a_0}(Q) + o_P(n^{-1/2})$, where $\Phi_{a_0}(Q)$ is the influence function, given in Lemma~\ref{lem:eif}. 
\end{theorem}

Similar to the previous estimators, $\psi^+_{a_0, 3}(\hat{Q})$ and $\psi_{a_0, 3}(\hat{Q}^*)$ enjoy slower nuisance convergence rates than $n^{-1/2}$. The robustness properties are formalized as follows. 

\begin{corollary}\label{cor:robust_density}
$\psi^+_{a_0, 3}(\hat{Q})$ and $\psi_{a_0, 3}(\hat{Q}^*)$ are consistent estimators for $\psi_{a_0}(Q)$ if the following conditions are simultaneously true: 
\begin{enumerate}[label=(\roman*)]
    \item $|| \hf{Z_k} - f_{Z_k} || = o_P(1)$ or $||\hf{Z_i} - f_{Z_i} || = o_P(1)$, $\forall Z_k \in \Z$ and $\forall Z_i\in \Z_{\prec Z_k}$ s.t. $a_{Z_i} \not= a_{Z_k}$, 
    \item $|| \hat{\mu} - \mu || = o_P(1)$ or $|| \hf{Z_i} - f_{Z_i} || = o_P(1)$, $\forall Z_i \in \Z$ s.t. $a_{Z_i} \not= a_{Y}$, 
    \item $|| \hat{\pi} - \pi || = o_P(1)$ or $|| \hf{Z_i} - f_{Z_i} || = o_P(1)$, $\forall Z_i \in \Z\cap\M$, 
    \item $|| \hat{\mu} - \mu || = o_P(1)$ or $|| \hat{\pi} - \pi || = o_P(1)$, if $Y\in\M$. 
\end{enumerate}
\end{corollary}
The most notable aspect of Corollary~\ref{cor:robust_density} is its demonstration that the estimators $\psi^+_{a_0, 3}(\hat{Q})$ and $\psi_{a_0, 3}(\hat{Q}^*)$ are \textbf{doubly robust}. These estimators remain consistent if either:
\begin{itemize}
    \item all $f_{Z_k}$ for all $Z_k \in \Z\cap\M$, along with $\mu$ if $Y \in \M$, are consistently estimated, or 
    \item all $f_{Z_k}$ for all $Z_k \in \Z\cap\bbL$, along with $\mu$ if $Y \in \bbL$ and $\pi$, are consistently estimated. 
\end{itemize}

While other conditions could also guarantee consistency, they tend to be overly conservative and less practical in application. 

\begin{example}
Consider the ADMGs in Figures~\ref{fig:graphs_primal}(b-d). 
By Corollary~\ref{cor:robust_density},  $\psi^+_{a_0, 3}(\hat{Q})$ and $\psi_{a_0, 3}(\hat{Q}^*)$ are consistent under the following conditions: 
In Figure~\ref{fig:graphs_primal}(b), consistency holds if either $\hat{\pi}$ and $\hat{\mu}$, or $\hat{f}_M$ and $\hat{f}_L$ are consistent. 
In Figure~\ref{fig:graphs_primal}(c), consistency holds if either $\hat{f}_M$ is consistent, or $\hat{\pi},\hat{\mu}$, and $\hat{f}_L$ are all consistent. 
In Figure~\ref{fig:graphs_primal}(d), consistency holds if either $\hat{\pi}$ and $\hat{f}_L$, or $\hat{f}_M$ and $\hat{\mu}$ are consistent.
\end{example}

\subsection{Special cases and connections to prior robustness results} 
\label{subsec:special_cases}

In this subsection, we demonstrate how our general framework for estimating causal effects and the associated robustness behaviors under the treatment primal fixability criterion encompasses and extends several well-known results from the causal inference literature. Specifically, we show that the asymptotic properties and robustness behaviors of our proposed estimators include the back-door criterion \citep{robins1994estimation} and the front-door criterion \citep{fulcher19robust, wang2024pac, guo2026flexible}, as well as results from previous work on the treatment primal fixability \citep{bhattacharya2022semiparametric, jung2024unified}.

The back-door criterion, illustrated in Figure~\ref{fig:graphs}(a), represents a simple special case within our general framework based on the primal fixability criterion. In this setting, there are no mediators ($\Z = \emptyset$), so any condition involving variables $Z_i \in \Z$ holds trivially. As a result, conditions (i-iii) in Corollaries~\ref{cor:robust_ratio}, \ref{cor:robust_bayes}, and \ref{cor:robust_density} are automatically satisfied. Corollary~\ref{cor:robust_ratio} reduces to requiring either $|| \hat{\mu} - \mu || = o_P(1)$ or $|| \hat{f}^r_{A} - f^r_{A} || = o_P(1)$. Corollary~\ref{cor:robust_bayes} and \ref{cor:robust_density} simplify to $|| \hat{\mu} - \mu || = o_P(1)$ or $|| \hat{\pi} - \pi || = o_P(1)$, given that $g_{Z_1} \equiv \pi$. These conditions mirror the standard double robustness property of the one-step estimator and TMLE (i.e., the well-known augmented inverse probability weighted estimator), which are consistent when either the outcome regression $\mu$ or the propensity score $\pi$ is correctly specified.

The front-door criterion, shown in Figures~\ref{fig:graphs}(c) and \ref{fig:graphs_primal}(a), represents another special case within our framework, with $\Z = \{M\}$. Employing direct density estimation, as detailed in Section~\ref{subsec:inference_densities}, condition (i) of Corollary~\ref{cor:robust_density} is automatically true given that $\Z_{\prec M}=\emptyset$. As a result, Corollary~\ref{cor:robust_density} reduces to conditions (ii-iv), which imply that our one-step estimator and TMLE are consistent if: either both the propensity score and outcome regression estimators are consistent, or the mediator density estimator is consistent. This implies a form of double robustness for the front-door model: the estimator remains consistent if either $|| \hf{M} - f_{M} || = o_P(1)$, or if both $|| \hat{\mu} - \mu || = o_P(1)$ and $|| \hat{\pi} - \pi || = o_P(1)$. These findings are consistent with the results reported by \cite{fulcher19robust, wang2024pac, guo2026flexible}. As clarified in Remark~\ref{rem:trt-specific-nuisances}, the outcome-regression requirement for our construction of $\psi_{a_0}(P)$ concerns only $\mu(M,X,a_1)$, because the factual contribution $\E[\I(A=a_0)Y]$ is estimated directly.

Directly fitting the sequential regression and density ratios, as described in Section~\ref{subsec:inference_ratio}, permits consistency conditions expressed in terms of those fitted functions. In the front-door model, Corollary~\ref{cor:robust_ratio} yields four sufficient regimes: correct estimation of $(\mu,\mB{M})$, $(\mu,\fr{A})$, $(\fr{M},\mB{M})$, or $(\fr{M},\fr{A})$, with regression accuracy interpreted as specified below. These regimes follow by controlling each product in the population bias through either its regression error or its density-ratio error. The regime $(\mu,\fr{A})$ recovers the outcome-regression and propensity-score condition in the preceding paragraph when $\fr{A}$ is constructed from $\pi$. As clarified in Remarks~\ref{rem:trt-specific-nuisances} and~\ref{rem:seq_reg}, the requirement on $\mu$ concerns only $A=a_1$, while accuracy of the fitted sequential regression at $A=a_0$ means $\|\hmB{M}(X,a_0)-\E_P[\hat\mu(M,X,a_1)\mid X,A=a_0]\|=o_P(1)$. This latter condition does not require correct estimation of the outcome regression used to construct the pseudo-outcome.

Under the Bayes-rule construction in Section~\ref{subsec:inference_bayes}, the ratios are estimated as $\hfr{A}=(1-\hat g_M)/\hat g_M$ and $\hfr{M}=\{\hat h_M/(1-\hat h_M)\}\{(1-\hat g_M)/\hat g_M\}$, using the fitted $\hat g_M=\hat\pi(a_0\mid X)$ and $\hat h_M=\hat P(A=a_0\mid M,X)$. Corollary~\ref{cor:robust_bayes} guarantees consistency under correct estimation of $(\mu,\mB{M})$, $(\mu,g_M)$, or $(h_M,g_M)$, with the treatment-specific and sequential-regression requirements interpreted as above. The regime $(\mu,g_M)$ recovers the outcome-regression and propensity-score condition discussed previously. In the regime $(h_M,g_M)$, both ratios are consistently estimated, without requiring correct estimation of either $\mu$ or the sequential regression. This construction complements direct density-ratio estimation by using binary regressions to estimate both $\fr{M}$ and $\fr{A}$. 

For the broad\sout{er} class of treatment primal-fixable effects, \citet{bhattacharya2022semiparametric} establish double robustness of their influence-function-based estimator under correctness of either of two collections of conditional models. Under our notation, Corollary~\ref{cor:robust_density} recovers these density-based regimes: consistency holds if either the mediator densities associated with $\Z\cap\M$, together with $\mu$ if $Y\in\M$, or the mediator densities associated with $\Z\cap\bbL$, together with $\pi$ and $\mu$ if $Y\in\bbL$, are consistently estimated. Our TMLE construction has the same sufficient consistency regimes when its updated nuisance estimates satisfy the corresponding conditions. Beyond this conditional-density construction, Corollaries~\ref{cor:robust_ratio} and~\ref{cor:robust_bayes} characterize sufficient consistency regimes for estimators constructed using directly fitted sequential regressions and density ratios, including ratios estimated through binary regressions.

Related regression-and-weighting robustness results are established by \citet{jung2024unified}. Their Theorem~3 gives a product-form decomposition involving regression and ratio errors, including conditional expectations of fitted pseudo-outcomes. Their robustness characterization is stated in terms of regression and ratio nuisances, while Bayes' rule is presented as a strategy for estimating the ratios. Our analyses distinguish these two levels explicitly. For directly estimated ratios, Corollary~\ref{cor:robust_ratio} states sufficient conditions in terms of which ratios and sequential regressions must be accurately estimated. For ratios constructed through Bayes' rule, Corollary~\ref{cor:robust_bayes} instead identifies which combinations of the constituent binary-regression nuisances $g_{Z_k}$ and $h_{Z_k}$ and the sequential regressions suffice for consistency. Comparing the resulting regimes requires matching the regression targets and weighting functions, including those defined under the augmented laws used in their empirical-bifurcation construction. Appendix~\ref{app:examples} illustrates the respective nuisance representations in a common graphical example.

\subsection{Cross fitting as an alternative to Donsker conditions}
\label{subsec:cross_fit} 

The Donsker conditions limit the complexity of the nuisance parameter space and are prerequisites for achieving asymptotic linearity of our one-step and TMLE estimators. However, these conditions can be relaxed by employing cross-fitting, an approach also referred to as double-debiased machine learning \citep{double17chernozhukov} and cross-validated TMLE \citep{zheng2010asymptotic}.

In a $K$-fold cross-fitting setup, the data are divided into $K$ non-overlapping, equally sized subsamples ${S_k,\ k=1,\cdots,K}$. For each fold, we train the nuisance parameters on the complement of the $k$-th subsample, $S^{(-k)}$, yielding estimates denoted by $\hat{Q}^{(-k)}$. Predictions are then made for the excluded subsample $S_k$, referred to as $\hat{Q}^{(-k)}(S_k)$. By aggregating predictions from all $K$ subsamples, we assemble the cross-fitted nuisance estimate $\hat{Q}^{\text{cf}}(O) = ({\hat{Q}^{(-1)}(S_1), \ldots, \hat{Q}^{(-K)}(S_K)})$. This aggregated estimate is used to compute the cross-fitted one-step estimator, $\psi^{+}_{a_0}(\hat{Q}^{\text{cf}})$ \citep{double17chernozhukov}. For the TMLEs, updating the initial nuisance parameter estimates $\hat{Q}^{\text{cf}}$ to $\hat{Q}^*$ completes the procedure \citep{zheng2010asymptotic}.

\section{On semiparametric efficiency bounds}
\label{sec:semiparam}

Thus far, we have focused on nonparametric statistical models, where the tangent space is $L^2_0(P)$, the space of square-integrable, mean-zero functions of the observed data. Imposing restrictions on the joint observed-data distribution can reduce the tangent space. In a restricted model, influence functions need not be unique: they differ by elements of the orthogonal complement of the tangent space. The efficient influence function is the unique influence function in the closure of the tangent space and attains the semiparametric efficiency bound. This bound is no larger than the nonparametric efficiency bound and can be strictly smaller when the restrictions provide information about the target parameter. Deriving the efficient influence function provides a basis for constructing regular and asymptotically linear (RAL) estimators that attain this bound under suitable conditions \citep{bickel1993efficient, van2000asymptotic, tsiatis2007semiparametric}.

In DAGs where all variables are observed, restrictions on the joint observed data distribution are directly linked to the structure of the graph: the absence of an edge between two variables implies ordinary independence restriction given an appropriate conditioning set. These restrictions are encoded in the DAG factorization and the equivalent Markov properties, and can be read off from DAGs via d-separation rules \citep{pearl2009causality}. In such settings, deriving the semiparametric efficient influence function is facilitated by the known tangent-space characterization \citep{van2000asymptotic, tsiatis2007semiparametric}. Specifically, the semiparametric EIF can be obtained by projecting any influence function for the target parameter onto the closure of the tangent space of the DAG's statistical model; see \citep{rotnitzky2020efficient, bhattacharya2022semiparametric} for details. 

In addition to ordinary independence restrictions, ADMGs encode a more complex type of constraints known as generalized independence restrictions, or Verma constraints \citep{verma90equiv}. While ordinary independence restrictions in ADMGs can be read off using m-separation rules \citep{richardson03markov}, determining generalized independence restrictions is less straightforward, and roughly translates to independence restrictions in truncated or intervention distributions (a.k.a. kernels)  obtained through fixing operations. An example of an ordinary independence restriction in an ADMG is shown in Figure~\ref{fig:graphs_semiparam}(a), where the absence of an edge between $A$ and $Y$ implies $A \perp Y \mid X, M, L$ according to m-separation rules. In contrast, the missing edge between $A$ and $Y$ in Figure~\ref{fig:graphs_semiparam}(b) does not imply any independence restriction between $A$ and $Y$ because there is an open path from $A$ to $Y$ through $M$ and $L$, and conditioning on either variable opens the collider path $A \ \blue{\rightarrow} \ M \ \red{\leftrightarrow} \ Y$. However, under the nested Markov factorization \citep{richardson2023nested}, this missing edge corresponds to a generalized independence restriction of the form $A \perp Y \mid X, L$ in a truncated distribution $P(V)/P(L \mid X, A, M)$. The nested Markov model of an ADMG encodes all equality (ordinary and generalized independence) constraints implied by the graph. For further details, refer to \citep{verma90equiv, richardson2023nested}. 

\begin{figure}[t] 
	\begin{center}
    \scalebox{0.75}{
    \begin{tikzpicture}[>=stealth, node distance=1.4cm]
        \tikzstyle{format} = [thick, circle, minimum size=1.0mm, inner sep=2pt]
        \tikzstyle{square} = [draw, thick, minimum size=4.5mm, inner sep=2pt]
    \begin{scope}[xshift=0.cm, yshift=0cm]
		\path[->, thick]
		
		node[] (a) {$A$}
		node[right of=a, xshift=0.5cm] (m) {$M$}
        node[right of=m, xshift=0.5cm] (l) {$L$}
		node[above right of=m, yshift=0.2cm] (x) {$X$}
		node[right of=l, xshift=0.5cm] (y) {$Y$}
		
		(x) edge[blue, bend left=0] (a) 
		(x) edge[blue] (m) 
        (x) edge[blue] (l) 
		(x) edge[blue, bend right=0] (y) 
		(a) edge[blue] (m) 
		(m) edge[blue] (l) 
        (l) edge[blue] (y)    

        (m) edge[blue, ->, out=25, in=90] (y)
        (a) edge[red, <->, out=90, in=155] (l) 

        node[below right of=m, xshift=0.cm, yshift=-0.cm] (t1) {(a)} ;
		
	\end{scope}
    \begin{scope}[xshift=7.5cm, yshift=0cm]
		\path[->, thick]
		
		node[] (a) {$A$}
		node[right of=a, xshift=0.5cm] (m) {$M$}
        node[right of=m, xshift=0.5cm] (l) {$L$}
		node[above right of=m, yshift=0.2cm] (x) {$X$}
		node[above left of=m, yshift=0.cm] (u) {}
		node[right of=l, xshift=0.5cm] (y) {$Y$}
		
		(x) edge[blue, bend left=0] (a) 
		(x) edge[blue] (m) 
        (x) edge[blue] (l) 
		(x) edge[blue, bend right=0] (y) 
		(a) edge[blue] (m) 
		(m) edge[blue] (l) 
        (l) edge[blue] (y)    
        (a) edge[red, <->, out=90, in=155] (l) 
        (m) edge[red, <->, out=25, in=90] (y) 

        node[below right of=m, xshift=0.cm, yshift=-0.cm] (t2) {(b)} ;
		
	\end{scope}
    \begin{scope}[xshift=15cm, yshift=0cm]
		\path[->, thick]
		
		node[] (a) {$A$}
		node[right of=a, xshift=0.5cm] (m) {$M$}
        node[right of=m, xshift=0.5cm] (l) {$L$}
		node[above right of=m, yshift=0.2cm] (x) {$X$}
		node[right of=l, xshift=0.5cm] (y) {$Y$}
		
		(x) edge[blue, bend left=0] (a) 
		(x) edge[blue] (m) 
        (x) edge[blue] (l) 
		(x) edge[blue, bend right=0] (y) 
		(a) edge[blue] (m) 
		(m) edge[blue] (l) 
        (l) edge[blue] (y)    
        (a) edge[red, <->, out=90, in=155] (l) 
        (l) edge[red, <->, out=35, in=110] (y) 

        (m) edge[blue, ->, out=25, in=90] (y) 
        
        node[below right of=m, xshift=0.cm, yshift=-0.cm] (t3) {(c)} ;
		
	\end{scope}
	\end{tikzpicture}
	}
    \vspace{-0.2cm}
	\caption{Example ADMGs illustrating different types of equality restrictions: (a) An ordinary independence restriction due to the missing edge between $A$ and $Y$; (b) A generalized independence (Verma) restriction; (c) No restriction despite the missing edge between $A$ and $Y$.} 
	\label{fig:graphs_semiparam}
	\end{center}
\end{figure}

\citet{bhattacharya2022semiparametric} proposed a sound and complete algorithm for determining whether an ADMG implies any equality constraints on the observed-data distribution when the hidden variables are unrestricted. If no such constraints exist, the corresponding nested Markov model is nonparametrically saturated. Consequently, the estimators proposed in Section~\ref{sec:est} attain the nonparametric efficiency bound under the conditions discussed in Section~\ref{sec:inference}. For example, in Figure~\ref{fig:graphs_semiparam}(c), the nested Markov model remains saturated despite the missing edge between $A$ and $Y$.

Although nested Markov models provide a graphical characterization of both ordinary independence and Verma constraints, translating these restrictions into tractable tangent-space characterizations is more challenging. Verma constraints can couple components of the observed-data likelihood, so independently perturbing its conditional factors need not preserve the model restrictions. Smooth parameterizations are available for discrete nested Markov models \citep{evans2019smooth} and efficiency results have been developed under specialized no-direct-effect restrictions in Robins' g-null graph and related models \citep{robins1999testing, liu2021efficient, guo2026flexible}. However, a complete, broadly applicable characterization of tangent spaces and efficient influence functions for general nested Markov models remains an open challenge.

For an important subclass of ADMGs, however, the tangent-space characterization is established. \cite{bhattacharya2022semiparametric} characterized a class of ADMGs, referred to as \textit{mb-shielded} ADMGs, in which all equality restrictions are implied by ordinary conditional independence restrictions. An ADMG is mb-shielded if an edge between two vertices $O_i$ and $O_j$ in $\G(O)$ is absent only if $O_i \notin \mb_\G(O_j)$ \emph{and} $O_j \not\in \mb_\G(O_i),$ where $\mb_\G(O_i)$ denotes the Markov blanket of a vertex $O_i$ as the district of $O_i$ and the parents of its district, excluding $O_i$ itself. 
This graphical characterization simplifies the analysis and allows the use of established results from semiparametric theory to derive the tangent space and, consequently, the semiparametric EIF \citep{bickel1993efficient, van2000asymptotic, rotnitzky2020efficient, bhattacharya2022semiparametric}. For this subclass, \citet{bhattacharya2022semiparametric} established tangent-space characterizations and efficient influence functions by applying standard semiparametric theory to the resulting conditional independence restrictions. 

Building on the tangent-space characterization and projection results of \citet{bhattacharya2022semiparametric}, we investigate how ordinary independence restrictions can be incorporated into our sequential-regression one-step and TMLE constructions. We focus on mb-shielded ADMGs and defer the incorporation of Verma constraints to future work. Deriving the semiparametric EIF under ordinary independence constraints can be understood as projecting the EIF from Lemma~\ref{lem:eif} onto the reduced tangent space of the ADMG's statistical model. In the following discussion, we simplify the projection of the EIF from Lemma~\ref{lem:eif}, by excluding the term $\I(A=a_0)Y - \E[\I(A=a_0)Y]$, for clarity. Given the substantial computational efficiency achieved by not  projecting the entire EIF onto the more restricted tangent space---particularly in settings where numerical integration is required---we suggest modifying the EIF selectively. 

Consider an independence constraint involving a variable $Z_k \in \Z$, stating that given $\mpg(Z_k)$, $Z_k$ is independent of $Z_j$, where $Z_j \prec Z_k$ and $Z_j \notin \mpg(Z_k)$. This independence constraint reduces the model space for the conditional density $f_{Z_k}$. Accordingly, the corresponding term in the EIF from Lemma~\ref{lem:eif}, i.e., $\Phi_{a_0,Z_k}$, is adjusted to reflect the independence restriction, resulting in $ \Phi^\text{eff}_{a_0,Z_k}$:

\vspace{-0.25cm}
{\small 
\begin{equation}
\label{eq:eif_k_semi}
    \begin{aligned}
    \Phi^\text{eff}_{a_0,Z_k}(Z_k,\mpg(Z_k)) 
    &= \E\Big[ \I(A=a_{Z_k}) \mR{Z_k}\big(\mathrm{mp}^{-A}_{\G, \prec}(Z_k)\big)  \times \Big\{  \mB{Z_{k+1}}(\mBarg{Z_{k+1}}, a_{Z_{k+1}}) \\
    &\hspace{1.5cm} - \mB{Z_{k}}(\mBarg{Z_k}, a_{Z_k}) \Big\} \ \Big| \  Z_k,\mpg(Z_k)\Big] \ . 
\end{aligned}
\end{equation}
}%

Unlike in the nonparametric setting, where $\mB{Z_k}$ is defined on $\mpgA{Z_k}$ by leveraging the simplification that $\mBarg{Z_k}=\mpgA{Z_k}$, in the semiparametric context, $\mB{Z_k}$ is defined as a function of the union of the Markov pillows of $Z_k$ and all variables succeeding $Z_k$, restricted to those that precede $Z_k$. This distinction arises because the embedding relationship $\mpg(Z_{k}) \subset \mpg(Z_{k+1})$, which holds in the nonparametric statistical model, need not hold in the semiparametric setting. 

To construct the semiparametric efficient one-step estimator, we modify the one-step procedure outlined in Section~\ref{subsec:one-step}, by adding an additional regression step to estimate $\Phi^{\text{eff}}_{a_0,Z_k}$. This estimate is obtained from the predictions of the following regression model, evaluated at $A = a_{Z_k}$: 
{\small 
\begin{align*} 
\mathbb{I}(A = a_{Z_k}) \hmR{Z_k}\big(\mathrm{mp}^{-A}_{\G, \prec}(Z_k)\big) \big( \hmB{Z_{k+1}}(\mBarg{Z_{k+1}}, a_{Z_{k+1}}) - \hmB{Z_{k}}(\mBarg{Z_k}, a_{Z_k}) \big) \sim \{Z_k, \mpg(Z_k)\} \ . 
\end{align*}
}%
The semiparametric efficient one-step estimator is then constructed by replacing the estimate of $\Phi^{\text{eff}}_{a_0,Z_k}$ in the nonparametric one-step estimator \eqref{eq:one-step}. 

Constructing the semiparametric efficient TMLE requires modifications to our TMLE procedure in Section~\ref{subsec:tmle}, particularly in updating the estimate of $\mB{Z_k}$. Rather than targeting $\mB{Z_k}$ directly, the focus shifts to the conditional density $f_{Z_k}$. To this end, we introduce the following submodel and loss function pair, defined for any valid $\tf{Z_k}$ within the model space:
\begin{align}
L_{Z_k}(\tf{Z_k}) &= -\mathbb{I}(A = a_{Z_k}) \log \tf{Z_k} \ , \label{eq:semi_f_loss} \\
\tf{Z_k}(\varepsilon_{Z_k}; \tmB{Z_k}, \tmR{Z_k}, \tmB{Z_{k+1}}) &= \tf{Z_k} \times \left(1 + \varepsilon_{Z_k} \tilde{\Phi}^{\text{eff}}_{a_0,Z_k}\right), \quad \forall \varepsilon_{Z_k} \in (-\delta_{Z_k}, \delta_{Z_k}) \ .  \label{eq:semi_f_submodel1} 
\end{align}
The submodel given by \eqref{eq:semi_f_submodel1} is simple but has limitations. It requires finding a small real number, $\delta_{Z_k}$, such that for all $\varepsilon_{Z_k} \in (-\delta_{Z_k}, \delta_{Z_k})$, the submodel $\tf{Z_k}(\varepsilon_{Z_k})$ remains positive. This constraint can complicate the optimization process, making it less robust in practice. To address these shortcomings, an alternative submodel to consider is as follows: 
\begin{align}
\tf{Z_k}(\varepsilon_{Z_k}; \tmB{Z_k}, \tmR{Z_k}, \tmB{Z_{k+1}}) &= \frac{\tf{Z_k} \times  \exp{\left(\varepsilon_{Z_k} \tilde{\Phi}^{\text{eff}}_{a_0,Z_k}\right)}}{\int \exp{\big(\varepsilon_{Z_k} \tilde{\Phi}^{\text{eff}}_{a_0,Z_k}(z_k, \mpg(Z_k))\big)} \tf{Z_k}(z_k\mid \mpgA{Z_k},a_{Z_k}) dz_k} \ . 
\label{eq:semi_f_submodel2}
\end{align}
The submodel shown in \eqref{eq:semi_f_submodel2} guarantees a valid submodel for all $\varepsilon_{Z_k} \in \mathbb{R}$, thus offering greater robustness. However, it may involve increased computational complexity during risk minimization. Both submodels have their respective advantages and disadvantages, and the choice between them depends on the specific practical requirements. Unlike regression-based methods, risk minimization for these submodels with the specified loss function requires numerical optimization techniques. The validity of these submodel-loss function pairs can be demonstrated using methods similar to those outlined in Appendix~\ref{app:tmle_details}. 

\section{Simulations}
\label{sec:sims}

In this section, we examine the theoretical properties and practical performance of our proposed one-step estimators and TMLEs across eight simulation studies. We investigate asymptotic behavior and computational runtime (Simulation~1), performance under weak overlap (Simulation~2), misspecified parametric models versus flexible nuisance estimation (Simulation~3), the impact of cross-fitting (Simulation~4), the impact of graph pruning (Simulation~5), parameter-space preservation (Simulation~6), behavior under unequal nuisance convergence rates (Simulation~7), and the incorporation of ordinary independence restrictions (Simulation~8). The implementation code is accessible through the Github repository: \href{https://github.com/annaguo-bios/ADMGs-Estimation-paper}{annaguo-bios/ADMGs-Estimation-paper}.


\subsection*{Simulation 1: Asymptotic behavior and computational runtime}

Our first set of simulations is designed to examine the asymptotic properties of the proposed estimators under conditions that favor achieving asymptotic linearity. We focus on two key  aspects: (i) $\sqrt{n}$-scaled bias, which is expected to converge to zero when the discussed conditions are satisfied, and (ii) $n$-scaled variance, which should converge to the variance of the EIF, $\E[\Phi^2(Q)]$, under the same conditions. We examine two representative scenarios: one where the outcome $Y$ is within the district of the treatment $A$, as shown in Figure~\ref{fig:graphs_primal}(b), and another where $Y$ is outside the district of $A$, as shown in Figure~\ref{fig:graphs_primal}(d). In both scenarios, the joint distribution of the variables is Markov relative to the depicted graphs, with the mediator $M$ generated as a bivariate normal variable and the mediator $L$ as a univariate normal variable. Nuisance estimations were performed using fully parametric models and a combination of parametric models with nonparametric kernel methods for density ratio estimation. 

We implemented three TMLE estimators, namely $\tmledens$, $\tmlebayes$, and $\tmlednorm$, along with their corresponding one-step estimators, $\onedens$, $\onebayes$, and $\onednorm$. The subscripts `densratio', `bayes', and `dnorm' denote the methods used to estimate the product of density ratios $\mR{Y}$ and $\mR{Z_k}$, as detailed in Section~\ref{sec:inference}. $\mB{Z_k}$ is estimated via sequential regression technique. Simulations were conducted at sample sizes of 250, 500, 1000, 2000, 4000, and 8000, with each sample replicated 1000 times. 
The results in Appendix Figures~\ref{fig:YinL} and~\ref{fig:YnotL} show behavior consistent with the asymptotic theory across the settings considered. These figures and detailed descriptions of the data generating processes (DGPs) are provided in Appendix~\ref{app:sims:consistency}.

We also report the combined runtime for computing each one-step estimator and its corresponding TMLE (Appendix Table~\ref{table:runtime-yinl}). Averaged over 1000 replications, combined runtimes ranged from 0.111 to 0.652 seconds for the Bayes-rule strategy, 0.347 to 7.974 seconds for the parametric density strategy, and 5.519 to 364.589 seconds for the kernel density-ratio strategy, across the sample sizes considered.

\subsection*{Simulation 2: TMLEs versus one-step estimators under weak overlap}

This simulation examined the finite-sample performance of our estimators under weak overlap conditions, where the treatment effect was unevenly distributed across the sample. Weak overlap posed substantial challenges to estimator performance, making TMLE particularly appealing because of its anticipated robustness in such settings \citep{porter2011relative}.

We considered scenarios in which the outcome $Y$ was either within or outside the district of the treatment $A$. The DGPs mirrored those in Simulation 1, with the modification that $A$ was generated from a $\operatorname{Binomial}(0.001 + 0.998X)$ distribution, resulting in a propensity score range of $[0.001, 0.999]$. This setup enforced the weak overlap condition. We used the same three TMLEs and their corresponding one-step counterparts as in the previous simulation. Further details were provided in Appendix~\ref{app:sims:overlap}.

Data were generated for $n$ equals 500, 1000, and 2000, with each sample undergoing 1000 replications. We evaluate the estimators based on estimation bias, sample standard deviation (SD), mean squared error (MSE), 95\% confidence interval (CI) coverage, and average CI width. The 95\% CI coverage is calculated as the proportion of simulations where the interval, $\hat{\psi} \pm z_{0.975} n^{-1/2} \hat{\sigma}$, contains the true estimand. Here, $z_{0.975}$ is the 97.5\% quantile of the standard normal distribution, and $\hat{\sigma}$ is the estimated standard deviation from $\Phi^2(\hat{Q})$ for one-step estimators and from $\Phi^2(\hat{Q}^*)$ for TMLEs.

The findings for scenarios illustrated in Figures~\ref{fig:graphs_primal}(b) and \ref{fig:graphs_primal}(d) are summarized in Tables~\ref{table:weakoverlap-YinL} and \ref{table:weakoverlap-YnotL}, respectively. In both cases, the one-step estimators and TMLEs show minimal bias, while TMLE estimators consistently demonstrating a smaller SD, leading to lower MSE across all sample sizes.

\providecommand{\huxb}[2]{\arrayrulecolor[RGB]{#1}\global\arrayrulewidth=#2pt}
\providecommand{\huxvb}[2]{\color[RGB]{#1}\vrule width #2pt}
\providecommand{\huxtpad}[1]{\rule{0pt}{#1}}
\providecommand{\huxbpad}[1]{\rule[-#1]{0pt}{#1}}

\begin{table}[t]
\centering
\captionsetup{justification=centering,singlelinecheck=off}
\caption{\texttt{[Simulation~2]} Comparative performance of TMLEs and one-step estimators under weak overlap, with data generated according to Figure~\ref{fig:graphs_primal}(b).}
\setlength{\tabcolsep}{0pt}
\resizebox{0.85\textwidth}{!}{
\begin{tabular}{l l l l l l l}

\Xhline{1pt}

\multicolumn{1}{!{\huxvb{0, 0, 0}{0}}l!{\huxvb{0, 0, 0}{0}}}{\huxtpad{1pt + 1em}\raggedright \hspace{1pt} \textbf{{\fontsize{8pt}{9.6pt}\selectfont }} \hspace{1pt}\huxbpad{1pt}} &
\multicolumn{1}{c!{\huxvb{0, 0, 0}{0}}}{\huxtpad{1pt + 1em}\centering \hspace{1pt} {{\fontsize{8pt}{9.6pt}\selectfont \(\psi_\text{dnorm}^{+}(\hat{Q})\)}} \hspace{1pt}\huxbpad{1pt}} &
\multicolumn{1}{c!{\huxvb{0, 0, 0}{0}}}{\huxtpad{1pt + 1em}\centering \hspace{1pt} {{\fontsize{8pt}{9.6pt}\selectfont \(\psi_\text{dnorm}(\hat{Q}^\star)\)}} \hspace{1pt}\huxbpad{1pt}} &
\multicolumn{1}{c!{\huxvb{0, 0, 0}{0}}}{\huxtpad{1pt + 1em}\centering \hspace{1pt} {{\fontsize{8pt}{9.6pt}\selectfont \(\psi_\text{densratio}^{+}(\hat{Q})\)}} \hspace{1pt}\huxbpad{1pt}} &
\multicolumn{1}{c!{\huxvb{0, 0, 0}{0}}}{\huxtpad{1pt + 1em}\centering \hspace{1pt} {{\fontsize{8pt}{9.6pt}\selectfont \(\psi_\text{densratio}(\hat{Q}^\star)\)}} \hspace{1pt}\huxbpad{1pt}} &
\multicolumn{1}{c!{\huxvb{0, 0, 0}{0}}}{\huxtpad{1pt + 1em}\centering \hspace{1pt} {{\fontsize{8pt}{9.6pt}\selectfont \(\psi_\text{bayes}^{+}(\hat{Q})\)}} \hspace{1pt}\huxbpad{1pt}} &
\multicolumn{1}{c!{\huxvb{0, 0, 0}{0}}}{\huxtpad{1pt + 1em}\centering \hspace{1pt} {{\fontsize{8pt}{9.6pt}\selectfont \(\psi_\text{bayes}(\hat{Q}^\star)\)}} \hspace{1pt}\huxbpad{1pt}} \tabularnewline
\end{tabular}}\label{table:weakoverlap-YinL}

\end{table}

  \providecommand{\huxb}[2]{\arrayrulecolor[RGB]{#1}\global\arrayrulewidth=#2pt}
  \providecommand{\huxvb}[2]{\color[RGB]{#1}\vrule width #2pt}
  \providecommand{\huxtpad}[1]{\rule{0pt}{#1}}
  \providecommand{\huxbpad}[1]{\rule[-#1]{0pt}{#1}}

\begin{table}[t]
\centering
\captionsetup{justification=centering,singlelinecheck=off}
\caption{\texttt{[Simulation~2]} Comparative performance of TMLEs and one-step estimators under weak overlap, with data generated according to Figure~\ref{fig:graphs_primal}(d).}
 \setlength{\tabcolsep}{0pt}
\resizebox{0.85\textwidth}{!}{
\begin{tabular}{l l l l l l l}

\Xhline{1pt}

\multicolumn{1}{!{\huxvb{0, 0, 0}{0}}l!{\huxvb{0, 0, 0}{0}}}{\huxtpad{1pt + 1em}\raggedright \hspace{1pt} \textbf{{\fontsize{8pt}{9.6pt}\selectfont }} \hspace{1pt}\huxbpad{1pt}} &
\multicolumn{1}{c!{\huxvb{0, 0, 0}{0}}}{\huxtpad{1pt + 1em}\centering \hspace{1pt} {\fontsize{8pt}{9.6pt}\selectfont \(\psi_\text{dnorm}^{+}(\hat{Q})\)} \hspace{1pt}\huxbpad{1pt}} &
\multicolumn{1}{c!{\huxvb{0, 0, 0}{0}}}{\huxtpad{1pt + 1em}\centering \hspace{1pt} {\fontsize{8pt}{9.6pt}\selectfont \(\psi_\text{dnorm}(\hat{Q}^\star)\)} \hspace{1pt}\huxbpad{1pt}} &
\multicolumn{1}{c!{\huxvb{0, 0, 0}{0}}}{\huxtpad{1pt + 1em}\centering \hspace{1pt} {\fontsize{8pt}{9.6pt}\selectfont \(\psi_\text{densratio}^{+}(\hat{Q})\)} \hspace{1pt}\huxbpad{1pt}} &
\multicolumn{1}{c!{\huxvb{0, 0, 0}{0}}}{\huxtpad{1pt + 1em}\centering \hspace{1pt} {\fontsize{8pt}{9.6pt}\selectfont \(\psi_\text{densratio}(\hat{Q}^\star)\)} \hspace{1pt}\huxbpad{1pt}} &
\multicolumn{1}{c!{\huxvb{0, 0, 0}{0}}}{\huxtpad{1pt + 1em}\centering \hspace{1pt} {\fontsize{8pt}{9.6pt}\selectfont \(\psi_\text{bayes}^{+}(\hat{Q})\)} \hspace{1pt}\huxbpad{1pt}} &
\multicolumn{1}{c!{\huxvb{0, 0, 0}{0}}}{\huxtpad{1pt + 1em}\centering \hspace{1pt} {\fontsize{8pt}{9.6pt}\selectfont \(\psi_\text{bayes}(\hat{Q}^\star)\)} \hspace{1pt}\huxbpad{1pt}} \tabularnewline
\end{tabular}}\label{table:weakoverlap-YnotL}

\end{table}

\subsection*{Simulation 3: Misspecified parametric models vs. flexible nuisance estimation}

We evaluated the performance of our estimators under conditions of nuisance model misspecification, incorporating interaction terms as detailed in Appendix~\ref{app:sims:misspecification} and using the ADMGs depicted in Figures~\ref{fig:graphs_primal}(b) and (d). These simulations, conducted with sample sizes of 500, 1000, and 2000, aimed to assess the robustness of the estimators when faced with either misspecified models or models enhanced through flexible machine learning techniques.

For these simulations, we employed the same three TMLEs and their corresponding one-step versions used in previous settings. We adopted three approaches for nuisance estimation: (i) Main term linear regressions that excluded interaction terms, resulting in model misspecification; (ii) The Super Learner, an ensemble method detailed by \citep{van2007super}, which combines estimates from a diverse set of models including intercept-only regression, generalized linear models, Bayesian generalized linear models, multivariate adaptive regression splines, generalized additive models, random forests, SVM, and XGBoost, designed to capture complex interactions in the DGP; and (iii) A combination of the Super Learner with five-fold cross-fitting to mitigate potential violations of the Donsker conditions that might occur with complex ML algorithms.

  \providecommand{\huxb}[2]{\arrayrulecolor[RGB]{#1}\global\arrayrulewidth=#2pt}
  \providecommand{\huxvb}[2]{\color[RGB]{#1}\vrule width #2pt}
  \providecommand{\huxtpad}[1]{\rule{0pt}{#1}}
  \providecommand{\huxbpad}[1]{\rule[-#1]{0pt}{#1}}

\begin{table}[htbp]
\captionsetup{justification=centering,singlelinecheck=off}
\caption{\texttt{[Simulation~3]} Comparative performance of TMLEs and one-step estimators under model misspecification, with data generated according to Figure~\ref{fig:graphs_primal}(b).}
 \setlength{\tabcolsep}{0pt}
\resizebox{\textwidth}{!}{
\begin{tabular}{l l l l l l l l l l l l l l l l l l l}

\Xhline{1pt}

\multicolumn{1}{!{\huxvb{0, 0, 0}{0}}l!{\huxvb{0, 0, 0}{0}}}{\huxtpad{-1pt + 1em}\raggedright \hspace{-1pt} \textbf{{\fontsize{7pt}{8.4pt}\selectfont }} \hspace{-1pt}\huxbpad{-1pt}} &
\multicolumn{9}{c!{\huxvb{0, 0, 0}{0}}}{\huxtpad{-1pt + 1em}\centering \hspace{-1pt} \textbf{{\fontsize{7pt}{8.4pt}\selectfont TMLEs}} \hspace{-1pt}\huxbpad{-1pt}} &
\multicolumn{9}{c!{\huxvb{0, 0, 0}{0}}}{\huxtpad{-1pt + 1em}\centering \hspace{-1pt} \textbf{{\fontsize{7pt}{8.4pt}\selectfont One-step estimators}} \hspace{-1pt}\huxbpad{-1pt}} \tabularnewline
\end{tabular}}\label{table:misspecification-YinL}

\end{table}

  \providecommand{\huxb}[2]{\arrayrulecolor[RGB]{#1}\global\arrayrulewidth=#2pt}
  \providecommand{\huxvb}[2]{\color[RGB]{#1}\vrule width #2pt}
  \providecommand{\huxtpad}[1]{\rule{0pt}{#1}}
  \providecommand{\huxbpad}[1]{\rule[-#1]{0pt}{#1}}

\begin{table}[htbp]
\captionsetup{justification=centering,singlelinecheck=off}
\caption{\texttt{[Simulation~3]} Comparative performance of TMLEs and one-step estimators under model misspecification, with data generated according to Figure~\ref{fig:graphs_primal}(d).}
 \setlength{\tabcolsep}{0pt}
\resizebox{\textwidth}{!}{
\begin{tabular}{l l l l l l l l l l l l l l l l l l l}

\Xhline{1pt}

\multicolumn{1}{!{\huxvb{0, 0, 0}{0}}l!{\huxvb{0, 0, 0}{0}}}{\huxtpad{-1pt + 1em}\raggedright \hspace{-1pt} \textbf{{\fontsize{7pt}{8.4pt}\selectfont }} \hspace{-1pt}\huxbpad{-1pt}} &
\multicolumn{9}{c!{\huxvb{0, 0, 0}{0}}}{\huxtpad{-1pt + 1em}\centering \hspace{-1pt} \textbf{{\fontsize{7pt}{8.4pt}\selectfont TMLEs}} \hspace{-1pt}\huxbpad{-1pt}} &
\multicolumn{9}{c!{\huxvb{0, 0, 0}{0}}}{\huxtpad{-1pt + 1em}\centering \hspace{-1pt} \textbf{{\fontsize{7pt}{8.4pt}\selectfont One-step estimators}} \hspace{-1pt}\huxbpad{-1pt}} \tabularnewline
\end{tabular}}\label{table:misspecification-YnotL}

\end{table}

The results, detailed in Tables~\ref{table:misspecification-YinL} and \ref{table:misspecification-YnotL}, indicate that using misspecified linear models leads to significant bias and poor CI coverage, even as sample size increases. In contrast, estimators that incorporate the Super Learner show markedly improved performance, with reduced bias, better CI coverage, and narrower confidence intervals. Additionally, incorporating cross-fitting with the Super Learner did not significantly affect bias or CI coverage in our settings. These findings underscore the importance of proper model specification and advocate the use of flexible modeling approaches like the Super Learner in scenarios involving complex data structures. 

\subsection*{Simulation 4: Impact of cross-fitting}

In this simulation, we examined the impact of cross-fitting, particularly when used with the random forest algorithm, which is known to provide poor estimates if cross-fitting is not applied \citep{double17chernozhukov, biau2012analysis}. We configured the random forest with standard settings: 500 trees, a minimum node size of 5 for continuous variables, and 1 for binary variables. For the scenario with $Y\in\bbL$, the covariates $X$ were ten-dimensional, with each element independently drawn from a uniform distribution over the range $[0,1]$. The remaining variables were generated with a complex structure, including interaction terms and higher-order terms, as detailed in Appendix~\ref{app:sims:cross-fit}. For the scenario with $Y\in\M$, the DGP mirrors that used in Simulation 3. We conducted 1000 simulations for each sample size of 500, 1000, and 2000, evaluating scenarios where the outcome $Y$ is either within or outside the district of the treatment $A$.

  \providecommand{\huxb}[2]{\arrayrulecolor[RGB]{#1}\global\arrayrulewidth=#2pt}
  \providecommand{\huxvb}[2]{\color[RGB]{#1}\vrule width #2pt}
  \providecommand{\huxtpad}[1]{\rule{0pt}{#1}}
  \providecommand{\huxbpad}[1]{\rule[-#1]{0pt}{#1}}

\begin{table}[t]
\captionsetup{justification=centering,singlelinecheck=off}
\caption{\texttt{[Simulation~4]} Comparative performance of TMLEs and one-step estimators with and without cross-fitting, using random forests. `RF' denotes random forests with 500 trees, a minimum node size of 5 for continuous variables, and 1 for binary variables. `CF' denotes random forests with 5-fold cross-fitting. Data generated according to the model shown in Figure~\ref{fig:graphs_primal}(b).}
 \setlength{\tabcolsep}{0pt}
\resizebox{\textwidth}{!}{
\begin{tabular}{l l l l l l l l l l l l l}

\Xhline{1pt}

\multicolumn{1}{!{\huxvb{0, 0, 0}{0}}l!{\huxvb{0, 0, 0}{0}}}{\huxtpad{0pt + 1em}\raggedright \hspace{0pt} \textbf{{\fontsize{6.5pt}{7.8pt}\selectfont }} \hspace{0pt}\huxbpad{0pt}} &
\multicolumn{6}{c!{\huxvb{0, 0, 0}{0}}}{\huxtpad{0pt + 1em}\centering \hspace{0pt} \textbf{{\fontsize{6.5pt}{7.8pt}\selectfont TMLEs}} \hspace{0pt}\huxbpad{0pt}} &
\multicolumn{6}{c!{\huxvb{0, 0, 0}{0}}}{\huxtpad{0pt + 1em}\centering \hspace{0pt} \textbf{{\fontsize{6.5pt}{7.8pt}\selectfont One-step estimators}} \hspace{0pt}\huxbpad{0pt}} \tabularnewline
\end{tabular}}\label{table:crossfit-YinL}

\end{table}

  \providecommand{\huxb}[2]{\arrayrulecolor[RGB]{#1}\global\arrayrulewidth=#2pt}
  \providecommand{\huxvb}[2]{\color[RGB]{#1}\vrule width #2pt}
  \providecommand{\huxtpad}[1]{\rule{0pt}{#1}}
  \providecommand{\huxbpad}[1]{\rule[-#1]{0pt}{#1}}

\begin{table}[t]
\captionsetup{justification=centering,singlelinecheck=off}
\caption{\texttt{[Simulation~4]} Comparative performance of TMLEs and one-step estimators with and without cross-fitting, using random forests. `RF' denotes random forests with 500 trees, a minimum node size of 5 for continuous variables, and 1 for binary variables. `CF' denotes random forests with 5-fold cross-fitting. Data generated according to the model shown in Figure~\ref{fig:graphs_primal}(d).}
 \setlength{\tabcolsep}{0pt}
\resizebox{\textwidth}{!}{
\begin{tabular}{l l l l l l l l l l l l l}

\Xhline{1pt}
\multicolumn{1}{!{\huxvb{0, 0, 0}{0}}l!{\huxvb{0, 0, 0}{0}}}{\huxtpad{0pt + 1em}\raggedright \hspace{0pt} \textbf{{\fontsize{6.5pt}{7.8pt}\selectfont }} \hspace{0pt}\huxbpad{0pt}} &
\multicolumn{6}{c!{\huxvb{0, 0, 0}{0}}}{\huxtpad{0pt + 1em}\centering \hspace{0pt} \textbf{{\fontsize{6.5pt}{7.8pt}\selectfont TMLEs}} \hspace{0pt}\huxbpad{0pt}} &
\multicolumn{6}{c!{\huxvb{0, 0, 0}{0}}}{\huxtpad{0pt + 1em}\centering \hspace{0pt} \textbf{{\fontsize{6.5pt}{7.8pt}\selectfont One-step estimators}} \hspace{0pt}\huxbpad{0pt}} \tabularnewline
\end{tabular}}\label{table:crossfit-YnotL}

\end{table}

The results for the two scenarios are presented in Tables~\ref{table:crossfit-YinL} and \ref{table:crossfit-YnotL}. 
In both cases, employing cross-fitting along with random forest for nuisance estimation led to a noticeable reduction in bias and a slight increase in SD compared to using random forest without cross-fitting. This different is also reflected in the larger CI coverage achieved with cross-fitting. These findings are consistent across different sample sizes and scenarios. 

Moreover, using the same DGP as in the simulation setting described in Simulation 3 allows a direct comparison between the results in Table~\ref{table:misspecification-YnotL}, which presents estimation outputs using Super Learner with a range of machine learning algorithms, and those in Table~\ref{table:crossfit-YnotL}, which shows estimation outputs primarily using random forests. Under the same DGP, the results in Table~\ref{table:misspecification-YnotL} exhibit smaller bias and better CI coverage than those in Table~\ref{table:crossfit-YnotL}, highlighting the importance of incorporating flexible nuisance estimation methods when dealing with complex data generating processes. 

\subsection*{Simulation 5: Impact of graph pruning on finite-sample performances}

In this simulation, we focused on the scenario depicted in Figure~\ref{fig:graphs_primal}(b), for which two distinct interpretations of the DGP were possible. One interpretation strictly followed the DGP shown in Figure~\ref{fig:graphs_primal}(b), whereas the other treated the mediators $M$ and $L$ as a single multivariate variable, reducing the DGP to the front-door model shown in Figure~\ref{fig:graphs}(c). These two interpretations led to different constructions of the TMLE and one-step estimators. The objective was to determine whether the resulting estimators exhibited different asymptotic efficiencies.

The data-generating mechanism mirrored that of Simulation~1, with additional details provided in Appendix~\ref{app:sims:consistency}. For both the TMLE and one-step estimator, we considered three approaches for estimating the mediator densities, consistent with the previous simulations. For each sample size of $500$, $1000$, $2000$, and $8000$, we generated $1000$ simulated datasets.

  \providecommand{\huxb}[2]{\arrayrulecolor[RGB]{#1}\global\arrayrulewidth=#2pt}
  \providecommand{\huxvb}[2]{\color[RGB]{#1}\vrule width #2pt}
  \providecommand{\huxtpad}[1]{\rule{0pt}{#1}}
  \providecommand{\huxbpad}[1]{\rule[-#1]{0pt}{#1}}

\begin{table}[!t]
\captionsetup{justification=centering,singlelinecheck=off}
\caption{\texttt{[Simulation~5]} Comparative performance of TMLEs and one-step estimators when variables M and L are treated separately verse combined.}
 \setlength{\tabcolsep}{0pt}
\resizebox{\textwidth}{!}{
\begin{tabular}{l l l l l l l l l l l l l}

\Xhline{1pt}

\multicolumn{1}{!{\huxvb{0, 0, 0}{0}}l!{\huxvb{0, 0, 0}{0}}}{\huxtpad{-1pt + 1em}\raggedright \hspace{-1pt} \textbf{{\fontsize{7pt}{8.4pt}\selectfont }} \hspace{-1pt}\huxbpad{-1pt}} &
\multicolumn{6}{c!{\huxvb{0, 0, 0}{0}}}{\huxtpad{-1pt + 1em}\centering \hspace{-1pt} \textbf{{\fontsize{7pt}{8.4pt}\selectfont TMLEs}} \hspace{-1pt}\huxbpad{-1pt}} &
\multicolumn{6}{c!{\huxvb{0, 0, 0}{0}}}{\huxtpad{-1pt + 1em}\centering \hspace{-1pt} \textbf{{\fontsize{7pt}{8.4pt}\selectfont One-step estimators}} \hspace{-1pt}\huxbpad{-1pt}} \tabularnewline
\end{tabular}}\label{table:combine_mediators}

\end{table}

The findings in Table~\ref{table:combine_mediators} show that both approaches to interpreting the DGP yield highly consistent estimation results across all evaluation metrics, especially at larger sample sizes. This consistency is observed for both TMLE and one-step estimator. Thus, the choice of interpretation depends on the specific objectives of the study.

\subsection*{Simulation 6: Parameter-space preservation}

This simulation investigated when one-step estimators produce out-of-bounds (OOB) estimates and compared their behavior with the corresponding TMLEs. We considered the ADMG in Figure~\ref{fig:graphs_primal}(b) and targeted $\E(Y^1)$ with a binary outcome. Four data-generating scenarios, labeled ``Favorable,'' ``Moderate,'' ``Severe,'' and ``Extreme,'' were constructed with progressively weaker overlap. The DGPs and details on nuisance-estimation procedures are described in Appendix~\ref{app:sims:oob}.

Across the four scenarios, the treatment mechanism and the conditional distributions of $M$ and $L$ were varied to induce increasingly extreme density ratios. Under the ``Favorable'' scenario, these ratios remained relatively stable, with ranges $[0.92,1.08]$, $[0.77,1.32]$, and $[0.76,1.37]$, respectively. Under the ``Extreme'' scenario, the corresponding ranges were $[65.67,99.00]$, $[0.06,7.75]$, and $[0.05,9.49]$. The ``Moderate'' and ``Severe'' scenarios represented intermediate levels of weak overlap. Detailed summaries of the conditional distributions and density-ratio ranges are provided in Appendix Table~\ref{table:weak-overlap-severity}. The outcome mechanism also varied across scenarios, with the true intervention mean  $\E(Y^1)$ increasing from approximately $0.753$ in the Favorable'' scenario to $0.926$ in the Extreme'' scenario. Thus, these scenarios examine parameter-space preservation under changing overlap and proximity of the target to the upper boundary.

\begin{figure}[t]
    \centering
    \includegraphics[width=0.8\linewidth]{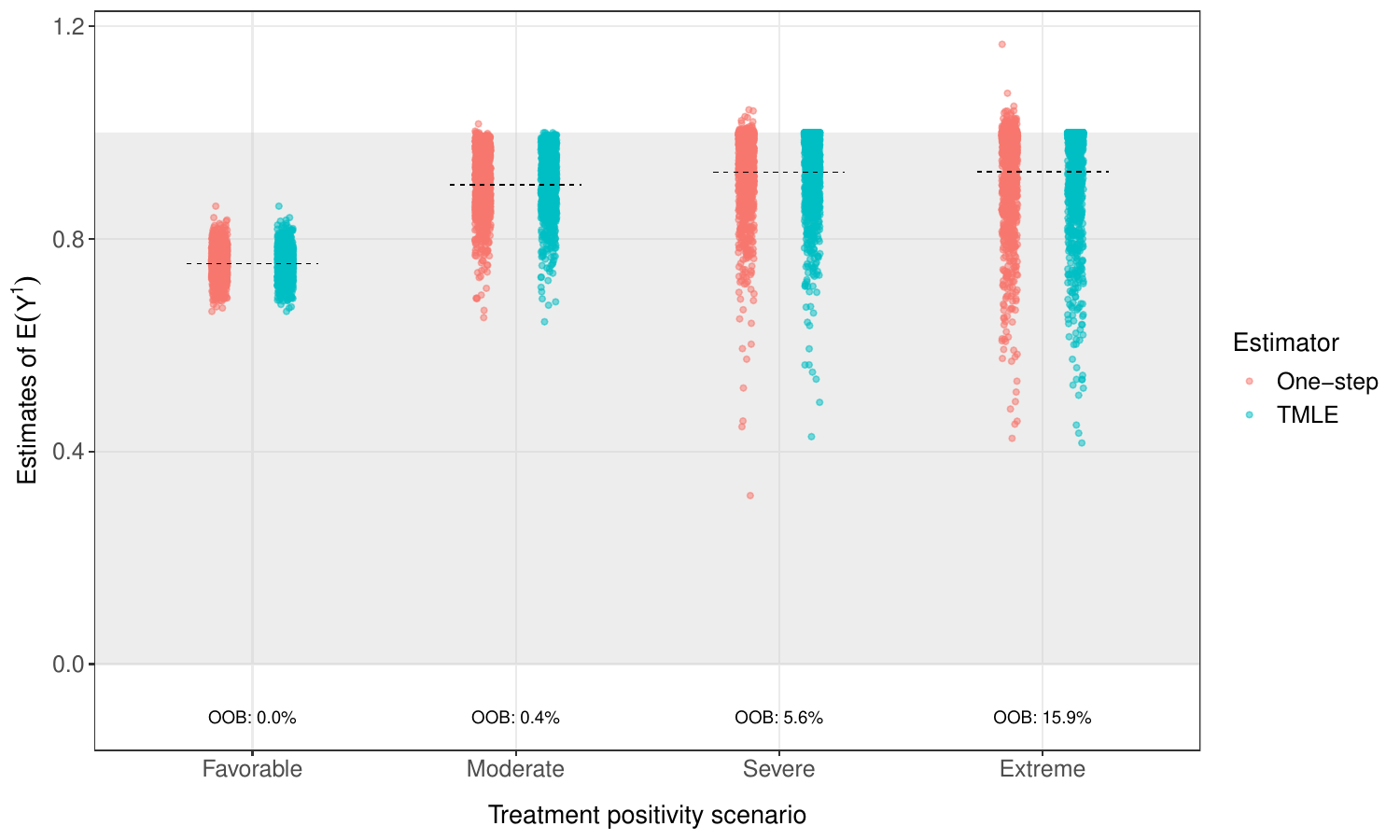}
    \caption{\texttt{[Simulation~6]} Distributions of one-step and TMLE estimates of $\E(Y^1)$ across four data-generating scenarios with progressively weaker overlap, based on $1000$ replications with sample size $250$ per scenario. Dashed lines indicate the scenario-specific true values. Percentages at the bottom indicate the frequency of one-step estimates outside the parameter space $[0,1]$. No out-of-bounds TMLE estimates were observed in any scenario.}
    \label{fig:oob}
\end{figure}

For each scenario, we generated $1000$ datasets of size $250$ and estimated $\E(Y^1)$ using the one-step estimator and TMLE. We compared the distributions of the estimates and recorded the frequency of estimates outside $[0,1]$. As shown in Figure~\ref{fig:oob}, neither estimator produced OOB estimates in the ``Favorable'' scenario. The frequency of OOB one-step estimates increased from $0.4\%$ in the ``Moderate'' scenario to $15.9\%$ in the ``Extreme'' scenario. In contrast, the TMLE estimates remained within the target parameter space across all scenarios.

These results illustrate that one-step estimates can fall outside the parameter space in finite samples, particularly in the scenarios with weaker overlap and intervention means closer to the upper boundary. The TMLE estimates remained within $[0,1]$ throughout this experiment. This study complements Simulation~2 by examining parameter-space violations for a binary-outcome intervention mean, rather than estimation performance for a continuous outcome under weak overlap.

\subsection*{Simulation 7: Asymptotic behavior under unequal nuisance convergence rates}

This simulation examines the behavior of the one-step estimator and TMLE when nuisance parameters are estimated at rates slower than $\sqrt{n}$. We considered the ADMG in Figure~\ref{fig:graphs_primal}(b), with the DGP specified in Appendix~\ref{app:sims:rate}. We used the same sample sizes and number of replications as in Simulation~1. Both estimators were constructed using sequential regressions and density ratios reparameterized via Bayes' rule. Motivated by the simulation designs of \citet{kennedy2023towards}, we introduced perturbations to the regression and density-ratio estimates at different sample-size-dependent rates.

We considered five perturbation scenarios. In the first two scenarios, the sequential regressions were constructed to converge at rates of $o_p(n^{-2/5})$ and $o_p(n^{-1/3})$, respectively, while the corresponding density ratios converged at rates of $o_p(n^{-1/5})$ and $o_p(n^{-1/6})$. Both scenarios satisfy the required product-rate condition of $o_p(n^{-1/2})$. In the remaining three scenarios, the convergence rates for the sequential regressions were $o_p(n^{-1/6})$, $o_p(n^{-1/9})$, and $o_p(n^{-1/10})$, respectively, with corresponding rates of $o_p(n^{-1/8})$, $o_p(n^{-1/10})$, and $o_p(n^{-1/15})$ for the density ratios. In these scenarios, the product-rate condition is not guaranteed. This design illustrates the product-rate trade-offs in our theoretical conditions: slower convergence of one nuisance component can be offset by faster convergence of another, rather than requiring both components to satisfy an $o_p(n^{-1/4})$ condition.

The perturbation mechanisms are described in Appendix~\ref{app:sims:rate}. The results, summarized in Appendix Figure~\ref{fig:rates}, are consistent with the asymptotic theory for both the one-step estimator and TMLE. In the first two scenarios, the $\sqrt{n}$-scaled empirical bias approaches zero, and the $n$-scaled empirical variance approaches the corresponding theoretical asymptotic variance as the sample size increases. In contrast, in the remaining scenarios, the $\sqrt{n}$-scaled empirical bias fails to approach zero, and the $n$-scaled empirical variance fails to converge to the theoretical asymptotic variance, with different degrees of deviation across scenarios.

\subsection*{Simulation 8: Leveraging ordinary independence restrictions for efficiency gains}

In this simulation, we consider an ADMG obtained from Figure~\ref{fig:graphs_primal}(b) by removing the direct edge from $A$ to $L$. This modification introduces the ordinary independence restriction $L \perp A \mid M,X$. We use this setting to examine the efficiency gains from incorporating this restriction into estimating the causal effect using the construction described in Section~\ref{sec:semiparam}, compared with estimators based on the nonparametric EIF in Section~\ref{sec:est}.

\begin{figure}[!t]
    \centering
    \includegraphics[width=1\linewidth]{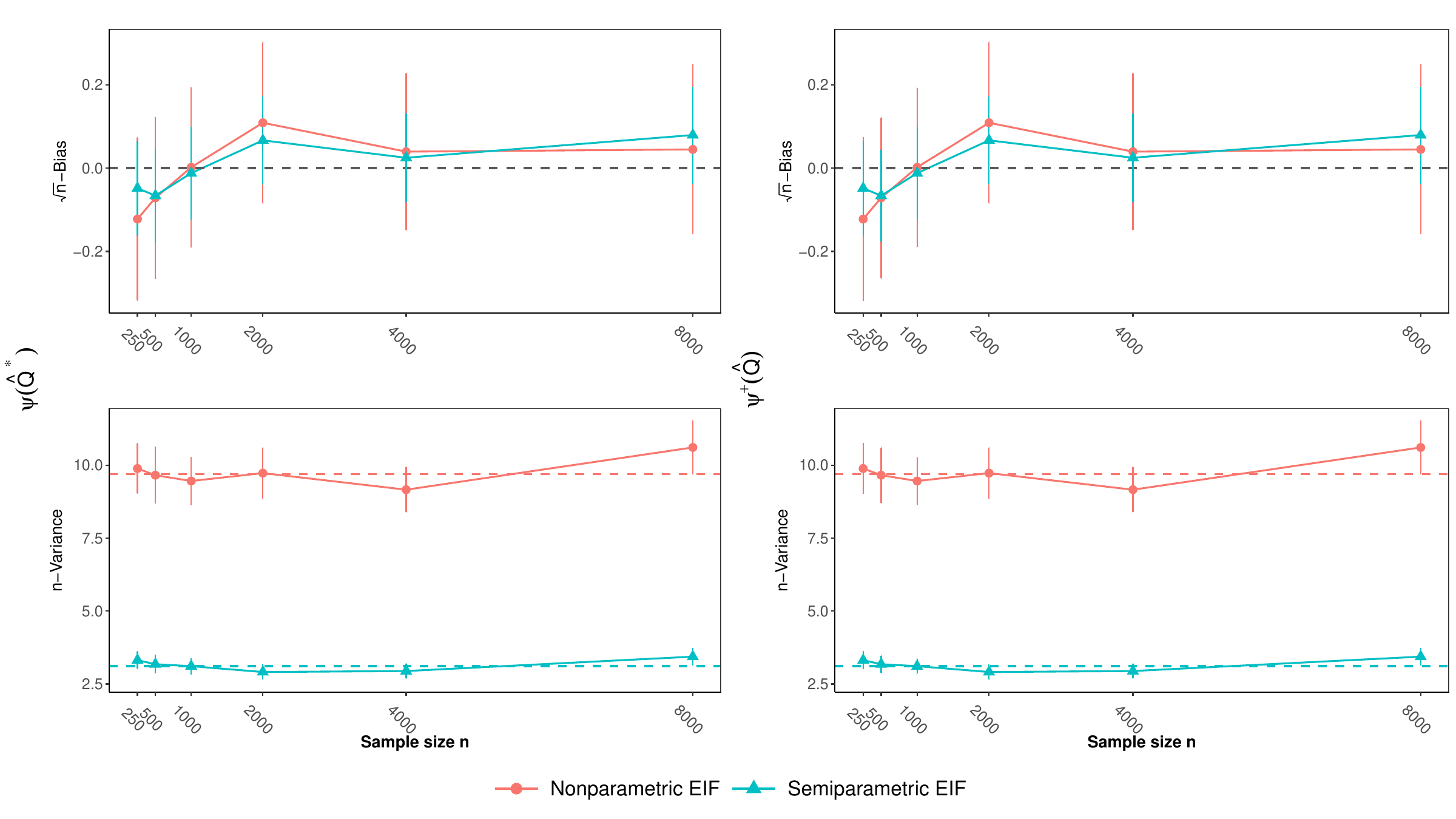}
    \caption{\texttt{[Simulation~8]} 
    Simulation results illustrating the efficiency gain achieved by the semiparametrically efficient estimators introduced in Section~\ref{sec:semiparam}, compared with estimators based on the nonparametric EIF in Section~\ref{sec:est}, under an ADMG obtained by removing the $A\rightarrow L$ edge from Figure~\ref{fig:graphs_primal}(b). The left and right columns correspond to the TMLE and one-step estimators, respectively.
    }
    \label{fig:semieff}
\end{figure}

We used the same sample sizes and number of replications as in Simulation~1. The DGP and details of nuisance-estimation procedures are provided in Appendix~\ref{app:sims:eff}. We implemented one-step estimators and TMLEs, both with and without incorporating the independence restriction. Figure~\ref{fig:semieff} shows negligible $\sqrt{n}$-scaled empirical bias for both constructions, but substantial differences in their $n$-scaled empirical variances. At the larger sample sizes, these variances were approximately $3$ for the semiparametrically efficient estimators incorporating the restriction, highlighted in pink, compared with approximately $10$ for those based on the nonparametric EIF, highlighted in blue. These results illustrate a more than three-fold reduction in variance from incorporating the independence restriction in this setting.

\section{Real data application}
\label{sec:realdata}

We applied our proposed estimation methods to investigate the causal effect of parental socioeconomic status (SES) on children's future annual income, using the Life Course 1971–2002 dataset from the Finnish Social Science Data Archive \citep{fsd}. This dataset originates from a longitudinal study that tracked 634 children born between 1964 and 1968 in Jyväskylä, Finland. In the early 1970s, when the children were between the ages of 3 and 9, verbal intelligence was assessed using the Illinois Test of Psycholinguistic Abilities (ITPA). Follow-up data were collected in 1984, 1991, and 2002, capturing a broad range of life events throughout early adulthood.

Our analysis adopts the same causal graph structure as the one used in \cite{helske2021estimation}, although we investigate a different causal relationship. Their study focused on estimating the causal effect of educational attainment ($X$) on income in adulthood ($Y$), measured in 2000 euros. Educational attainment was categorized into three levels: secondary or less, lower tertiary, and higher tertiary. The analysis included parental SES ($W$, categorized as low, middle, or high), primary school GPA ($Z$), ITPA score ($S$), and gender ($G$) as relevant covariates. The assumed causal relationships among these variables are represented by the ADMG shown in Figure~\ref{fig:graph_realdata}.

Under this graphical model, we shift the focus to the causal effect of parental SES on children's future income. Specifically, we estimate two average causal effects: the effect of high vs. middle parental SES and the effect of middle vs. low parental SES on children's future income. To establish the primal fixability of $W$, it is essential to assume that $W$ has no direct effect on either $X$ or $Y$. This assumption is supported by prior works \citep{acacio2018socioeconomic, osterbacka2001family}, as discussed in \cite{helske2021estimation}. Specifically, \cite{acacio2018socioeconomic} suggested that in the Finnish context, the influence of parental SES on educational attainment is largely mediated by the child’s academic performance, while \cite{osterbacka2001family} argued that the direct effect of family income on children’s future income is minimal after accounting for educational attainment.

\begin{figure}[t] 
	\begin{center}
    \scalebox{0.85}{
    \begin{tikzpicture}[>=stealth, node distance=2.5cm]
        \tikzstyle{format} = [thick, circle, minimum size=1.0mm, inner sep=2pt]
        \tikzstyle{square} = [draw, thick, minimum size=4.5mm, inner sep=2pt]
        
	\begin{scope}[xshift=0cm, yshift=0cm]
		\path[->, thick]
		node[] (w) {$\underset{\substack{\text{Parental SES}}}W$}
  
        node[right of=w, xshift=1cm] (z) {$\underset{\substack{\text{Primary school GPA} }}Z$}
        
        node[right of=z, xshift=1.75cm] (x) {$\underset{\text{Educational attainment}}{X}$}

        node[right of=x, xshift=1cm] (y) {$\underset{\text{Income}}{Y}$}
        
        node[above of=x, xshift=-1.25cm, yshift=-1.1cm] (s) {$\underset{\text{ITPA score}}{S}$}

        node[above of=x, xshift=1.25cm, yshift=-1.1cm] (g) {$\underset{\text{Gender}}{G}$}
		 
        (w) edge[blue] (z)
		(z) edge[blue] (x) 
		(x) edge[blue] (y) 
		(s) edge[blue] (z)
        (s) edge[blue] (x)
        (s) edge[blue] (y)
        (g) edge[blue] (z)
        (g) edge[blue] (x)
        (g) edge[blue] (y)
        (w) edge[red, <->, out=-15, in=195] (x)  
        (w) edge[red, <->, out=-15, in=195] (y)  
        (w) edge[red, <->, out=30, in=180] (s)  
        ;
	\end{scope}
	\end{tikzpicture}}
    \vspace{-0.25cm}
	\caption{Causal relationships among variables affecting parental SES and annual income.} 
	\label{fig:graph_realdata}
	\end{center}
\end{figure} 

We estimated the causal effects using the one-step estimator $\psi^+_{a_0, 2}(\hat{Q})$ and its TMLE counterpart $\psi_{a_0, 2}(\hat{Q}^*)$. To account for potential nonlinearities and interactions among variables, we used Super Learner with five-fold cross-fitting. The ensemble library included an intercept-only model, generalized linear models, multivariate adaptive regression splines, random forests, and XGBoost.  

The one-step estimator and TMLE produced highly similar results. For the high vs. middle SES comparison, the one-step estimator yielded an ACE of $673.72$ (95\% CI: $-1845.84$, $3193.29$), while the TMLE yielded $674.28$ (95\% CI: $-1845.48$, $3194.04$). For the middle vs. low SES comparison, the one-step estimator gave an ACE of $100.01$ (95\% CI: $-1616.89$, $1816.91$), and the TMLE produced $100.34$ (95\% CI: $-1607.32$, $1807.99$). The point estimates suggest that children from families with higher parental SES tend to earn more as adults, with a larger estimated income gap between the high and middle SES groups than between the middle and low groups. However, the confidence intervals include zero for both comparisons, indicating substantial uncertainty.

As a benchmark, we also computed plug-in estimators using simple linear nuisance models, with 95\% confidence intervals based on 1000 bootstrap replications. For the high vs. middle SES comparison, the plug-in estimator yielded a larger ACE of $899.82$ (95\% CI: $186.34$, $1613.30$). For the middle vs. low SES comparison, the plug-in estimator yielded an ACE of $198.92$ (95\% CI: $-93.31$, $491.16$).  The plug-in estimators yielded larger point estimates and narrower confidence intervals than the proposed estimators, illustrating that the estimation approach can affect both effect estimates and inferences. Because the true data-generating process is unknown, restrictive parametric nuisance models may miss important nonlinearities or interactions. We therefore favor flexible ensemble learning combined with influence-function-based correction, which can capture such complexities while supporting asymptotically valid inference under the conditions established in Section~\ref{sec:inference}.

\section{Discussion}
\label{sec:discussion}

In this work, we have established a general estimation framework that outputs one-step corrected and targeted minimum loss-based estimators through an automated algorithm for a wide range of graphical models with unmeasured/latent/hidden variables, satisfying the primal fixability criterion. Our framework is particularly well-suited for handling high-dimensional mediators of various types---binary, continuous, or mixed---without the need for mediator density estimation, which significantly enhances computational efficiency, especially in settings with a large number of mediators. Additionally, the proposed estimators seamlessly integrate statistical and machine learning models, offering flexibility for nuisance estimation and robustness against model misspecification. We have also examined the robustness properties and asymptotic behaviors of the estimators, characterizing conditions for achieving the desired asymptotic properties. Our framework supports sample splitting as an alternative to satisfying the Donsker conditions, which is particularly beneficial when using advanced models for nuisance estimations.  

We have developed the \href{https://github.com/annaguo-bios/flexCausal}{\texttt{flexCausal}} package in \textsf{R}, providing researchers and practitioners with a practical tool for implementing our advanced estimators. The package supports flexible modeling approaches and integrates machine learning techniques, thereby broadening the accessibility and application of sophisticated causal inference methods. 

The primal fixability criterion defines a rich class of causal models suitable in applied domains. This versatility makes it a particularly valuable tool for identifying causal relationships in settings with hidden variables. However, identifying causal effects outside the scope of primal fixability highlights opportunities to extend our methods beyond this class. Future work will explore causal effects that are identifiable even when the treatment is not primal fixable, as seen in the \textit{Napkin} graph \citep{pearl2018book, guo2026causal}, where traditional criteria do not apply. We also aim to extend our methods to more complex settings, including those with multiple treatments and dynamic treatment regimes, where the sequence and timing of interventions are crucial for determining outcomes.

The estimation framework proposed here relies on specifying the ADMG, yet determining the correct structure is itself a key challenge. Future work should contribute to the literature on causal discovery methods that blend expert knowledge with data-driven tools, and sensitivity analyses to assess robustness to misspecification and unmeasured confounding. Instrumental variables, already used to test assumptions in special cases like the front-door model \citep{bhattacharya2022testability}, could also help validate assumptions in the broader class of treatment primal fixable graphs. Together, advances in discovery, sensitivity analysis, and IV-based checks can strengthen the robustness and credibility of causal inference in complex settings.

Another important direction for future research is to extend our efficiency theory and estimation methods to incorporate generalized independence (Verma) constraints. Further investigation is needed to understand the impact of Verma constraints on the semiparametric efficiency bounds for causal estimators within ADMGs.




\bibliographystyle{plainnat}
\bibliography{references}

\newpage
\appendix  

\setcounter{figure}{0}
\renewcommand{\thefigure}{S\arabic{figure}}

\setcounter{table}{0}
\renewcommand{\thetable}{S\arabic{table}}

\setcounter{equation}{0}
\renewcommand{\theequation}{S\arabic{equation}} 

\noindent {\bf \LARGE Appendix} \vspace{0.25cm}

\noindent The appendix is structured as follows. 
\textbf{Appendix~\ref{app:notation}} offers a summary of the notations used throughout the manuscript for ease of  reference.
\textbf{Appendix~\ref{app:details}} provides additional details of the TMLE procedure for continuous and binary outcomes, discusses modifications for density estimation, and includes proofs for the validity of the loss functions and sub-models. It also outlines the TMLE implementation algorithm.  
\textbf{Appendix~\ref{app:proofs}} contains proofs related to identification, EIF derivation, and inference statements. 
\textbf{Appendix~\ref{app:examples}} illustrates the proposed estimation strategies and compares their nuisance representations with prior approaches.
\textbf{Appendix~\ref{app:sims}} describes the simulation data-generating processes and presents additional simulation details and results.

\begingroup
\let\newpage\relax
\let\clearpage\relax
\maketitle
\endgroup

\vspace{0.75cm}
\startcontents[supp]
\renewcommand{\baselinestretch}{1}\selectfont
\printcontents[supp]{}{1}{}

\newpage
\section{Glossary of terms and notations} 
\label{app:notation} 

A comprehensive list of notations used in the manuscript is provided in Table~\ref{tab:notations}. 

\begin{table}[!h]
\begin{center}
\caption{\centering Glossary of terms and notations}
\label{tab:notations}
\addtolength{\tabcolsep}{2pt}
\scalebox{0.8}{
\begin{tabular}{ll | ll}
    \hline  
    &&& \\
    \textbf{Symbol} & \textbf{Definition} & \textbf{Symbol} & \textbf{Definition}  
    \\ \hline 
    $A, \ Y$  & Treatment, outcome variables  & $P \in \pazocal{M}$ & Distribution $P$ in model $\pazocal{M}$      
    \\ 
    $O, \ U$ & Observed, unmeasured variables &  $\G(V)$ & DAG with vertices $V$ 
    \\
    $V$ & All variables $(O \cup U)$ &  $\G(O)$ & ADMG with vertices $O$
    \\
    $\pa_\G(V_i)$ & Parents of $V_i$ in $\G(V)$ & $Y(a)$ & Potential outcome $Y$ if $A=a$
    \\
    $a_0,\ a_1$ & Fixed treatment level, $1-a_0$ &  $\cal{D}(\G)$ & Set of districts in $\G(O)$
    \\
    $q_{D}(D \mid \pa_{\G}(D))$ & Kernel of district $D$ & $\tau$ &  A valid topological order over $O$
    \\
    $\mpg(O_i)$ & Markov pillow of $O_i$ & $\ch_\G(A)$ & Children of $A$
    \\
    $\X$ & Set of pre-treatment variables & $\bbL$ & $\{L_i\mid L_i\in\dis\{A\}, L_i\succeq A\}$
    \\
    $\M$ & $O\backslash\{\X,\bbL\}$ & $\mpgA{O_i}$ & $\mpg(O_i)\backslash A$
    \\
    $\Z$ & $\M\cup\bbL\backslash\{A,Y\}$ & $Z_1,\cdots,Z_K$ & $Z_k\in\Z$, ordered w.r.t. $\tau$
    \\
    $a_{Z_k}$ & 
    $a_0$ if $Z_k\in\M$ or $a_1$ if $Z_k\in\bbL$
    & $a_Y$ & $a_0$ if $Y\in\M$ or $a_1$ if $Y\in\bbL$
    \\
    $\mu(\mpg(Y))$ & $\E[Y \mid \mpg(Y)]$ & $\pi(a \mid \mpg(A))$ & $P(A=a \mid \mpg(A))$
    \\
    $f_{Z_k}(Z_k, \mpg(Z_k))$ & $P(Z_k \mid \mpg(Z_k))$ & $P_X$ & Distribution of $X$
    \\
    $Q$ & Collection of nuisances & $\mB{Z_k}$ & $\E\left[ \mB{Z_{k+1}} \mid \mpgA{Z_k}, a_{Z_k} \right]$ 
    \\
    $Z_{K+1}$ & $\equiv Y$ & $\mB{Z_{K+1}}$ & $\mu(\mpgA{Y}, a_Y)$ 
    \\
    $\psi^{\text{\tiny plug-in}}_{a_0}(\hat{Q})$ & Plug-in estimator of $\psi_{a_0}(Q)$ & $\epsilon_Y, \epsilon_{A},\epsilon_{Z_k}$ & Index of submodels for $\mu,\pi,\mB{Z_k}$
    \\
    $Pf$ & $\int f(o) \ dP(o)$ & $P_n f$ & $\frac{1}{n} \sum_{i = 1}^n f(O_i)$
    \\
    $\Phi_{a_0}$ & Influence function & $R_2(\hat{Q},Q)$ & Remainder terms
    \\
    $\M_{\prec Z_k}$ & Vertices in $\M$ that precede $Z_k$ & $\bbL_{\prec Z_k}$ & Vertices in $\bbL$ that precede $Z_k$
    \\
    &&& \\
    $\mathrm{mp}_{\G, \prec}(Z_k)$ & \makecell[l]{$\bigcup_{Z_i\in \M_{\prec Z_k}}\mpg(Z_i)$ if $Z_k\in\bbL$ \\ $\bigcup_{Z_i\in\bbL_{\prec Z_k}}\mpg(Z_i)$ if $Z_k\in\M$} & $\mathrm{mp}^{-A}_{\G, \prec}(Z_k)$ & $\mathrm{mp}_{\G, \prec}(Z_k)\backslash A$
    \\
    &&&
    \\
    $\fr{Z_k}$ & $\frac{\f{Z_k}(Z_k \mid \mpgA{Z_k}, a_{Z_k})}{\f{Z_k}(Z_k \mid \mpgA{Z_k}, 1- a_{Z_k})}$ & $\fr{A}(\mpg(A))$ & $\frac{\pi(a_1 \mid \mpg(A))}{\pi(a_0 \mid \mpg(A))}$
    \\
    &&& \\
    $\mR{Z_k}$ & \makecell[l]{$\prod_{M_i \in \M_{\prec Z_k}} \fr{M_i}$ if $Z_k\in\bbL$ \\ $\fr{A} \! \prod_{L_i \in \bbL_{\prec Z_k} \setminus A} \fr{L_i}$ if $Z_k\in\M$} & $\psi_{a_0}(P)$ & Parameter of interest: $\E[Y(a_0)]$
    \\
    &&&
    \\
    $\li{Z_k}$ & $P(a_{Z_k} \mid Z_k,\mpgA{Z_k})$ & $\lex{Z_k}(\mpgA{Z_k})$ & $P(a_{Z_k}\mid \mpgA{Z_k})$
    \\
    $\psi^{+}_{a_0}(\hat{Q})$ & One-step estimator of $\psi_{a_0}(Q)$ & $\hat{Q}^*$ & Estimate of $Q$ in TMLE
    \\
    $\psi_{a_0}(\hat{Q}^*)$ & TMLE of $\psi_{a_0}(Q)$ & $\Phi_{a_0,Z_k}(Q)$ & \makecell[l]{Projection of $\Phi_{a_0}(Q)$ \\ into the tangent space of $Z_k$} 
    \\
     $\pazocal{M}_{Q_i}$ & Functional space of $Q_i\in Q$ & $\hat{Q}^{(t)}$ & Estimate of $Q$ at iteration $t$
    \\
    $\psi_{a_0}(\hat{Q})$ & Either $\psi^{+}_{a_0}(\hat{Q})$ or $\psi_{a_0}(\hat{Q}^*)$ & $\pazocal{Q}$ & Functional space of $Q$
    \\
    $\{f-g\}(O)$ & $f(O)-g(O)$ & $|| f ||$ & $(Pf^2)^{1/2}$
    \\
    $\psi_{a_0, 1}(\hat{Q}^*)$ & \makecell[l]{TMLE with density ratio estimate} & $\Z_{\prec Z_k}$ & $\{Z_i: Z_i\in\Z, Z_i\prec Z_k\}$
    \\
    $b_Y$ & $|| \hat{\mu} - \mu || =\smallO(n^{-{1}/{b_Y}})$ & $b_{Z_k}$ & $|| \hmB{Z_k} - \mB{Z_k} || = o_P(n^{-{1}/{b_{Z_k}}})$
    \\
    $r_{Z_k}$ & $|| \hfr{Z_k} - \fr{Z_k} || = o_P(n^{-{1}/{r_{Z_k}}})$ & $r_A$ & $|| \hat{f}^r_{A} - \fr{A} || = o_P(n^{-{1}/{r_{A}}})$
    \\
    $t_{Z_k}$ & $|| \hli{Z_k} - \li{Z_k} || = o_P(n^{-{1}/t_{Z_k}})$ & $d_{Z_k}$ & $|| \hlex{Z_k} - \lex{Z_k} || = o_P(n^{-{1}/d_{Z_k}})$
    \\
    $\psi_{a_0, 2}(\hat{Q}^*)$ & \makecell[l]{TMLE with Bayes' rule} & $\psi_{a_0, 3}(\hat{Q}^*)$ & TMLE with density estimate
    \\
    $l_A$ & $|| \hat{\pi} - \pi || =\smallO(n^{-{1}/{l_A}})$ & $l_{Z_k}$ & $|| \hat{f}_{Z_k} - {f}_{Z_k} || = o_P(n^{-{1}/{{l_{Z_k}}}})$
    \\
    $S_k$ & Sub-sample $k$ in cross-fitting & $S^{(-k)}$ & nuisance models trained on $O\backslash S_k$
    \\
    $\hat{Q}^{-k}$ & Estimate from $S^{(-k)}$ and $S_k$ & $\hat{Q}^{\text{cf}}(O)$ & Cross-fitted estimate of $Q$
    \\
    $ \Phi^\text{eff}_{a_0,Z_k}$ & EIF piece for $Z_k$ in semipar models & $\{\prec Z_k\}^{-A}$ & Variables precede $Z_k$, excluding $A$ \\ 
    $\mBarg{Z_k}$ & $\{ \cup_{V_i\succeq Z_k}\mpgA{V_i}\} \cap\{\prec Z_k\}$ & &\\
    &&& \\
\end{tabular}
}
\end{center}
\end{table}

\newpage
\section{Detailed TMLE procedures and validity proofs}
\label{app:details} 

\subsection{Details of TMLE procedure} 
\label{app:tmle_details}

TMLE procedure has been detailed in various textbooks and tutorials. Here, we provide a brief overview from \citep{guo2026flexible}, which offers a general description of the TMLE approach in a study focused on the front-door model. 

The TMLE procedure comprises two main steps: the \textit{initialization} step, where the initial estimate $\hat{Q}$ is obtained, and the subsequent \textit{targeting} step, where $\hat{Q}$ is adjusted to a new estimate $\hat{Q}^*$. In the \textit{initialization} step, we obtain an initial estimate of $Q$ based on a collection of estimates for each nuisance parameter individually, $\hat{Q} = (\hat{Q}_1, \ldots, \hat{Q}_J)$. In the \textit{targeting} step, we require (i) a \emph{submodel} and (ii) a \emph{loss function} for each component $Q_j$ of $Q$. For requirement (i), with an estimate $\hat{Q}$ of $Q$, we define a submodel $\{ \hat{Q}_{j}(\varepsilon_j; \hat{Q}_{-j}), \ \varepsilon_j \in \R \}$ within $\pazocal{M}_{Q_j}$. This submodel is indexed by a univariate real-valued parameter $\varepsilon_j$ and may also depend on $\hat{Q}_{-j}$ (the components of $\hat{Q}$ excluding component $j$) or a subset of $\hat{Q}_{-j}$ (including the possibility of an empty subset). For requirement (ii), with a given $\tilde{Q}$, we denote a loss function for $\tilde{Q}_j$ by $L(\tilde{Q}_j; \tilde{Q}_{-j})$. 
Note that the loss function for $\tilde{Q}_j$ can also be indexed using $\tilde{Q}_{-j}$, or possibly by a subset of $\tilde{Q}_{-j}$, which may sometimes be an empty set. The submodel and loss function must be chosen to satisfy three conditions:
\begin{itemize}
    \item[]\textbf{(C1)} \ $\hat{Q}_j(0; \hat{Q}_{-j}) = \hat{Q}_j$ \ , 
    \item[] \textbf{(C2)} \ $Q_j = \argmin_{\tilde{Q}_j \in \pazocal{M}_{Q_j}} \displaystyle \int L(\tilde{Q}_j; Q_{-j})(o) \ p(o) \ do$ \ , 
    \item[] \textbf{(C3)} \ $\frac{\partial
    }{\partial\varepsilon_j} L\left(\hat{Q}_j(\varepsilon_j; \hat{Q}_{-j}); \hat{Q}_{-j}\right)\Big|_{\varepsilon_j=0} = \Phi_j(\hat{Q})$ \ . 
\end{itemize}
Condition (C1) ensures that the submodel aligns with the given estimate $\hat{Q}_j$ at $\varepsilon_j=0$. Condition (C2) ensures that the true nuisance parameter minimizes the expected loss. Finally, condition (C3) implies that the derivative of the loss function with respect to $\varepsilon_j$, evaluated at $\varepsilon_j=0$ is equivalent to the mapping of EIF noto the tangent space of $\hat{Q}$. 

In the following, we demonstrate the validity of loss functions and submodels in Section~\ref{subsec:tmle} by establishing their alignment with conditions (C1)-(C3). This proof can be readily extended to establish the validity of loss functions and submodels in the scenario of binary outcome discussed in Appendix Section~\ref{app:tmle_binary_y}.

The proof establishing the validity of the loss function and submodel for the sequential regressions $\mathscr{B}_{Z_i},\ i=1,\cdots,K$ closely resembles that for the outcome regression $\mu$ . Therefore, we will focus on detailing the proof for $\mu$ and $\pi$.

\noindent \underline{Loss function and submodel combination used for updating $\hat{\pi}(a_1\mid \mpg{(A)})$}:
\begin{align*}
    &L_{A}(\tilde{\pi}) = - \I(A=a_1)\log \tilde{\pi} + (1-\I(A=a_1))\log (1-\tilde{\pi})  \ , 
    \\
    &\tilde{\pi}(\varepsilon_A; \tilde{\mathscr{B}}_{Z_1}) = \operatorname{expit} \Big[\operatorname{logit} \ \tilde{\pi} + \varepsilon_A\ \tilde{\mathscr{B}}_{Z_1} \Big] \ . 
\end{align*}

\noindent \textit{Proof of (C1):} \ 
    $\tilde{\pi}(\varepsilon_A=0; \tilde{\mathscr{B}}_{Z_1}) = \operatorname{expit} \Big[\operatorname{logit} \ \tilde{\pi}\Big] = \tilde{\pi}.$

\noindent \textit{Proof of (C2):} \ 
    $\E[L_{A}(\tilde{\pi})]=\E[- \log \tilde{\pi}]
    =\int \big[- \sum_a \pi(a\mid x )\log \tilde{\pi}(a\mid x)\big] d P(x). $

The above is minimized if $- \sum_{a} \pi(a\mid x) \log \tilde{\pi}(a \mid x)$ is minimized for any $x \in \mathcal{X}$. According to the following relation 
\begin{align*}
    - \sum_{a} \pi(a\mid x) \log \tilde{\pi}(a \mid x)
    &= - \sum_{a} \pi(a\mid x) \log \left( \frac{\tilde{\pi}(a \mid x)}{\pi(a\mid x)} \times \pi(a\mid x) \right) \\
    &= - \sum_{a} \pi(a\mid x) \log  \frac{\tilde{\pi}(a \mid x)}{\pi(a\mid x)} - \sum_{a} \pi(a\mid x) \log \pi(a\mid x) \ ,  
\end{align*}
we only need to focus on the minimization of $- \sum_{a} \pi(a\mid x) \log  \frac{\tilde{\pi}(a \mid x)}{\pi(a\mid x)}$, which corresponds to the Kullback-Leibler (KL) divergence  from $\pi(a\mid x)$ to $\tilde{\pi}(a \mid x)$, denoted by $D_\text{KL}(\pi \ || \  \tilde{\pi})$. This KL-divergence is minimized if $\tilde{\pi}(A \mid X=x) = \pi(A \mid X=x)$, for all $x \in \mathcal{X}.$ \\

\noindent \textit{Proof of (C3):}
\begin{align*}
    \frac{\partial
    }{\partial\varepsilon_A} L_A\left(\tilde{\pi}(\varepsilon_A; \tilde{\mathscr{B}}_{Z_1})\right)\Big|_{\varepsilon_A=0} &=  -\frac{\partial
    }{\partial\varepsilon_A}\Big\{\I(A=a_1)\log \tilde{\pi}+\left(1-\I(A=a_1)\right)\log(1-\tilde{\pi})\Big\}\Big|_{\varepsilon_A=0}
    \\
    &=-\I(A=a_1)\tilde{\mathscr{B}}_{Z_1}\frac{\tilde{\pi}\left(1-\tilde{\pi}\right)}{\tilde{\pi}} + \left(1 -\I(A=a_1)\right)\tilde{\mathscr{B}}_{Z_1}\frac{\tilde{\pi}\left(1-\tilde{\pi}\right)}{1-\tilde{\pi}}
    \\
    &=\left(\tilde{\pi}-\I(A=a_1)\right)\tilde{\mathscr{B}}_{Z_1} \propto \Phi_A(\tilde{Q})
\end{align*}

\noindent \underline{Loss function and submodel combination used for updating $\hat{\mu}(\mpg{(Y)}$}: 
\begin{align*}
    &L_Y(\tilde{{\mu}}; \tilde{\mathscr{R}}_Y) = \tilde{\mathscr{R}}_Y \{ Y - \tilde{\mu} \}^2 \ ,   \qquad
    \tilde{\mu}(\varepsilon_Y) = \tilde{\mu} + \varepsilon_Y \ .  
\end{align*}

\noindent \textit{Proof of (C1):} \ 
    $\tilde{\mu}(\varepsilon_Y=0) = \tilde{\mu}.$

\noindent \textit{Proof of (C2):} \ 
   $\E\big[L_Y(\tilde{{\mu}}; \tilde{\mathscr{R}}_Y)\big] = \E\left[\tilde{\mathscr{R}}_Y \{ Y - \tilde{\mu} \}^2\right]
    =\E\left[\tilde{\mathscr{R}}_Y  \{ \{ Y - \mu \}^2 + \{ \mu - \tilde{\mu} \}^2 \} \right],$ 
minimized at $\tilde{\mu}=\mu$.

\noindent \textit{Proof of (C3):} \  
    $\frac{\partial}{\partial\varepsilon_Y}L_Y(\tilde{{\mu}}; \tilde{\mathscr{R}}_Y)\Big|_{\varepsilon_Y=0} = 2\tilde{\mathscr{R}}_Y \{ Y - \tilde{\mu} \} \propto \Phi_Y(\tilde{Q}). $

\subsection{TMLE modifications with binary outcome} 
\label{app:tmle_binary_y}

In the case of a binary outcome, the TMLE procedure for the propensity score $\pi$ remains unchanged, while the choice of loss functions and submodels for $\mu$ and $\mB{Z_k},\ \forall Z_k\in\Z$ must account for the fact that these nuisance parameters now have values within the range $[0,1]$. We propose using the following pairs of loss functions and submodels: 
\begin{align*}
    &\begin{cases}
        L_Y(\tilde{\mu},\tilde{R}_Y) = -\log \left[Y\tilde{\mu} + (1-Y)\left(1-\tilde{\mu}\right)\right] 
        \\
        \tilde{\mu}\left(\varepsilon_Y;\tmB{Z_k}\right) = \expit\left[\operatorname{logit} \tilde{\mu} + \varepsilon_Y \tilde{\mathscr{B}}_{Z_1}\right] 
    \end{cases}
    \\
    & 
    \\
    &\begin{cases}
        L_{Z_k}(\tmB{Z_k}; \tmR{Z_k}, \tilde{\mathscr{B}}_{Z_{k+1}}) = - \log \left[\tilde{\mathscr{B}}_{Z_{k+1}}\tmB{Z_k}+ (1-\tilde{\mathscr{B}}_{Z_{k+1}})(1-\tilde{\mathscr{B}}_{Z_{k}})\right]  \\
        \tmB{Z_k}(\varepsilon_{Z_k};\tilde{R}_{Z_k}) = \expit \left[\operatorname{logit} \tmB{Z_k} + \varepsilon_{Z_k}\tmR{Z_k} \right]
    \end{cases}  \quad \forall Z_k \in \Z \ . 
\end{align*}
The justification for the loss functions and submodels provided above is directly derived from the proof for the propensity score $\pi$ detailed in Appendix Section~\ref{app:tmle_details}. Apart from these adjustments, all remaining steps in the TMLE procedure remain unchanged.

\subsection{TMLE algorithm} 
\label{app:tmle_alg}
\begin{algorithm}[t]
	\caption{\textproc{TMLE of $\psi_{a_0}(Q)$ with continuous outcome $Y$}}  
    \label{appalg:continuous}
    
    \begin{algorithmic}[1] 
		
		\State \textbf{Obtain initial nuisance estimates}: $\hat{Q}^{(0)}=\{\hat{\pi}^{(0)},\ \hat{\mu}^{(0)},\ \hmR{Y}{(0)},\ \{\hmR{Z_k}{(0)},\ \hmB{Z_k}{(0)},\ \forall Z_k\in\Z\}\}$.

        {\small (we denote nuisance estimate of $Q_j$ at $t^\text{th}$ iteration by $\hat{Q}_j^{(t)}$.) } 

		\vspace{0.1cm}
        \State \textbf{Define loss functions \& submodels}.
        {\small Given estimates $\tilde{Q} = \{\tilde{\pi},\ \tilde{\mu},\ \{\tmR{Z_k},\ \tmB{Z_k},\ \forall Z_k\in\Z\}\}$, within their respectively model space, define the pairs of loss function and submodel as follows: }
        {\small 
            \begin{align*}
                &\begin{cases}
                    L_Y(\tilde{{\mu}}; \tilde{\mathscr{R}}_Y) = \tilde{\mathscr{R}}_Y \{ Y - \tilde{\mu} \}^2  \\
                    \tilde{\mu}(\varepsilon_Y) = \tilde{\mu} + \varepsilon_Y 
                \end{cases}
                \ , \qquad   
                \begin{cases}
                    L_{A}(\tilde{\pi}) = - \log \tilde{\pi}   \\
                    \tilde{\pi}(\varepsilon_A; \tilde{\mathscr{B}}_{Z_1}) = \operatorname{expit} \Big[\operatorname{logit} \ \tilde{\pi} + \varepsilon_A\ \tilde{\mathscr{B}}_{Z_1} \Big]  
                \end{cases} \ , 
                \\ 
                &\hspace{2cm}
                \begin{cases}
                    L_{Z_k}(\tmB{Z_k}; \tmR{Z_k}, \tilde{\mathscr{B}}_{Z_{k+1}}) = \tmR{Z_k}\{\tilde{\mathscr{B}}_{Z_{k+1}}- \tmB{Z_k}\}^2  \\
                    \tmB{Z_k}(\varepsilon_{Z_k}) = \tmB{Z_k} + \varepsilon_{Z_k} 
                \end{cases}  \quad \forall Z_k \in \Z \ . 
            \end{align*}
        Starting with $\hat{Q}^{(0)}$, the nuisance estimates are updated in an \textit{iterative} manner. At iteration $t$, with the current nuisance estimates denoted as $\hat{Q}^{(t)}$, update towards $\hat{Q}^{(t+1)}$ is performed via the following targeting steps.
        }
		\vspace{0.1cm}
        \State \textbf{Update $\hat{\pi}^{(t)}$ and $\hat{\mu}^{(t)}$.}
        
        \vspace{0.15cm}
        {\small 
        \begin{itemize}
        \item Fit the following logistic regression without an intercept to update $\hat{\pi}^{(t)}$:
        \begin{align*}
            &\I(A=a_1)\sim\mathrm{offset}(\operatorname{logit}\ \hat{\pi}^{(t)}(\mpi(A)))+\hat{\mathscr{B}}^{(t)}_{Z_1} \ .  
        \end{align*} 
        The coefficient in front of $\hat{\mathscr{B}}^{(t)}_{Z_1}$ is the minimizer to $\hat{\varepsilon}_A \ = \ \argmin_{\varepsilon_A \in \mathbb{R}} \ P_n L_A\Big(\hat{\pi}\big(\varepsilon_A; \hmB{Z_1}{(t)}\big)\Big)$.    
        \item[] Update $\hat{\pi}^{(t)}$ to $ \hat{\pi}^{(t+1)} = \hat{\pi}^{(t)}(\hat{\varepsilon}_A; \hmB{Z_1}{(t)})$. 
        
        \item Update the density ratios $\hat{\mathscr{R}}^{(t)}_{Y}, \hat{\mathscr{R}}^{(t)}_{Z_k}$ accordingly as $\hat{\mathscr{R}}^{(t+1)}_{Y}, \hat{\mathscr{R}}^{(t+1)}_{Z_k},\ \forall Z_k\in \Z$.
        \item Fit the weighted regression to update $\hat{\mu}^{(t)}$: $Y \sim \mathrm{offset}(\hat{\mu}^{(t)})+1.$
        \item[] The coefficient of the intercept term is the minimizer to $\hat{\varepsilon}_Y \ = \ \argmin_{\varepsilon_Y \in \mathbb{R}} \ P_n L_Y\Big(\hat{\mu}\big(\varepsilon_Y; \hmR{Y}{(t+1)}\big)\Big)$.    
        \item[] Update $\hat{\mu}^{(t)}$ to $ \hat{\mu}^{(t+1)} = \hat{\mu}^{(t)}(\hat{\varepsilon}_Y)$. 

        \item Define $\hat{Q}^{(t, *)}$ as $\{\hat{\mu}^{(t+1)}, \hat{\pi}^{(t+1)}, \hmR{Y}{(t+1)}, \{ \hmR{Z_k}{(t+1)}, \hmB{Z_k}{(t)}, \  \forall Z_k \in \Z\} \}$.
        \end{itemize}
        }

		\vspace{0.3cm}
        \State \textbf{Update $\hmB{Z_k}{(t)}, \forall Z_k \in \Z$ sequentially, starting from $k=K$ down to $k=1$}. 

        \vspace{0.15cm}
        {\small
        \begin{itemize}
            \item Revise $\hmB{Z_k}{(t)}$ via fitting the regression: $\hmB{Z_{k+1}}{(t+1)} \sim \mpg{(Z_k)}$, evaluated at $a_{Z_k}$
            \item Fit the weighted regression to update $\hmB{Z_k}{(t)}$: $\hmB{Z_{k+1}}{(t+1)} \sim \mathrm{offset}(\hmB{Z_k}{(t)})+1.$
            \item[] The minimizer $\hat{\varepsilon}_{Z_k} \ = \ \argmin_{\varepsilon_{Z_k} \in \mathbb{R}} \ P_n L_{Z_k}\Big(\hmB{Z_k}{(t)}\big(\varepsilon_{Z_k}\big); \hmR{Z_k}{(t+1)}, \hmB{Z_{k+1}}{(t+1)}\Big)$ is the coefficient of the intercept term.
            \item[] Update $\hmB{Z_k}{(t)}$ to $\hmB{Z_k}{(t+1)} = \hmB{Z_k}{(t)}(\hat{\varepsilon}_{Z_k})$. 

            \item Update $\hat{Q}^{(t, *)}$ as $\{\hat{\mu}^{(t+1)}, \hat{\pi}^{(t+1)}, \hmR{Y}{(t+1)}, \{ \hmR{Z_k}{(t+1)}, \hmB{Z_k}{(t+1)}, \  \forall Z_k \in \Z\} \}$, and define $\hat{Q}^{(t+1)} = \hat{Q}^{(t, *)}$. 
        \end{itemize}}

		\vspace{0.3cm}
        \State \textbf{Obtain $\hat{Q}^*$}. If at iteration $t^*$, the magnitude of $|P_n \Phi_{a_0}(\hat{Q}^{(t^*)})|$ is less than a pre-specified threshold $C_n = o_P(n^{-1/2})$, then $\hat{Q}^*=\hat{Q}^{(t^*)}$. 
    
		\vspace{0.3cm}
		\State \textbf{Return} $\psi_{a_0}(\hat{Q}^*) = P_n\big[ \hat{\pi}^*( a_1 \mid \mpg(A)) \  \hat{\mathscr{B}}^*_{Z_1}(\mBarg{Z_1}, a_{Z_1})  + \I(A=a_0)Y \big]$ as the TMLE. 

	\end{algorithmic}
\end{algorithm}

\subsection{TMLE modifications under direct mediator densities estimation} 
\label{app:tmle_density}

In this TMLE procedure, the choice of loss functions and submodels for the propensity score $\pi$ and outcome regression $\mu$ remains consistent with previous implementations. However, new pairs of loss functions and submodels are needed to directly target the conditional densities for all $Z_k \in \Z$. For brevity, we will continue using the notation $\mB{Z_k}$ and $\mR{Z_k}$, and their estimated forms, $\forall Z_k \in \Z$. Note that for each step in the TMLE procedure, the estimates for these nuisance parameters should be derived from the most current estimates of $\pi,\ \mu$, and the conditional densities.

The loss function for this procedure is defined as the negative log-likelihood, as indicated in \eqref{appeq:f_loss}. There are two submodel choices, detailed in \eqref{appeq:f_submodel1} and \eqref{appeq:f_submodel2}. Risk minimization involving both submodels and the given loss function can no longer be solved through regression-based methods; instead, numerical optimization techniques are required. For each $Z_k \in \Z$, the submodel in \eqref{appeq:f_submodel1} necessitates finding a small real number $\delta_{Z_k}$ such that for all $\varepsilon_{Z_k} \in (-\delta_{Z_k}, \delta_{Z_k})$, the submodel $\tilde{f}_{Z_k}(\varepsilon_{Z_k})$ is positive. In contrast, the submodel in \eqref{appeq:f_submodel2} ensures a valid submodel for all $\varepsilon_{Z_k} \in \R$, although it may impose a higher computational burden during risk minimization. The choice between these two submodels depends on the specific practical requirements. The validity of the chosen loss function and submodels can be established by following proofs analogous to those outlined in Appendix Section~\ref{app:tmle_details}.
\begin{align}
    L_{Z_k}(\tilde{f}_{Z_k}) &= -\I(A=a_{Z_k})\ \log \tilde{f}_{Z_k} \ , \label{appeq:f_loss}\\
    \tilde{f}_{Z_k}(\varepsilon_{Z_k}; \tmR{Z_k},\tmB{Z_{k+1}}, \tmB{Z_k}) &= \tilde{f}_{Z_k}\times\left(1+\left(1 + \varepsilon_{Z_k} \tilde{\Phi}^{\text{eff}}_{a_0,Z_k}\right)\right),\  \forall \varepsilon_{Z_k}\in (-\delta_{Z_k},\delta_{Z_k}) \ ,\label{appeq:f_submodel1}
    \\
     \tilde{f}_{Z_k}(\varepsilon_{Z_k}; \tmR{Z_k},\tmB{Z_{k+1}}, \tmB{Z_k}) &= \frac{\tf{Z_k} \times  \exp{\left(\varepsilon_{Z_k} \tilde{\Phi}^{\text{eff}}_{a_0,Z_k}\right)}}{\int \exp{\big(\varepsilon_{Z_k} \tilde{\Phi}^{\text{eff}}_{a_0,Z_k}(z_k, \mpg(Z_k))\big)} \tf{Z_k}(z_k\mid \mpgA{Z_k},a_{Z_k}) dz_k} \ . \label{appeq:f_submodel2}
\end{align}

Estimates of $Q$ are then updated through an iterative risk minimization process. We begin with initial estimates, denoted by $\hat{Q}^{(0)}$. At each iteration $t$, with the current estimates denoted by $\hat{Q}^{(t)}$, we follow three targeting steps (T1-T3) to derive $\hat{Q}^{(t+1)}$. Steps T1 and T2, involving the updates for $\pi$ and $\mu$, align exactly with the description in Section~\ref{subsec:tmle}. To avoid redundancy, we omit the discussion of these steps here.

\noindent \emph{(T3): Sequential risk minimization for $f_{Z_k}, \ \forall Z_k\in\Z$.} 

The conditional densities are updated in reverse order, starting from $k=K$ down to $k=1$. Assuming that we have just completed updating the estimate for $f_{Z_{k+1}}$, resulting in the following estimates of the nuisances: $\hat{Q}^{(t,*)}=\{\hat{\pi}^{(t+1)},\hat{\mu}^{(t+1)},f^{(t)}_{1},\cdots,\hat{f}^{(t)}_{Z_{k}},\hat{f}^{(t+1)}_{Z_{k+1}},\cdots,\hat{f}^{(t+1)}_{Z_{K}}\}$. The following steps detail the process for updating the conditional density estimate at the $k$-th level:

\hspace{0.25cm} \emph{(T3-a):}  Perform numerical risk minimization, finding $\hat{\epsilon}_{Z_k}$ as the minimizer to the empirical risk defined as $\hat{\varepsilon}_{Z_k} \ = \ \argmin_{\varepsilon_{Z_k}} \ P_n L_{Z_k}(\hat{f}^{(t)}_{Z_k}(\varepsilon_{Z_k}) )$. Define $\hat{f}^{(t+1)}_{Z_k} = \hat{f}^{(t)}_{Z_k}(\hat{\epsilon}_{Z_k})$. Note that nuisances involved in the definition of submodels should align with current nuisance estimate $\hat{Q}^{(t,*)}$.

\hspace{0.25cm} \emph{(T3-b):}  Update $\hat{Q}^{(t,*)}=\{\hat{\pi}^{(t+1)},\hat{\mu}^{(t+1)},\hat{f}^{(t)}_{1},\cdots,\hat{f}^{(t)}_{Z_{k-1}},\hat{f}^{(t+1)}_{Z_{k}},\cdot,\hat{f}^{(t+1)}_{Z_{K}}\}$, and decrement $k$ by one. It is noted that $P_n\Phi_{Z_k}(\hat{Q}^{(t, *)}) = 0$.

Repeat steps (T3-a) and (T3-b) while $k > 1$. Once $k=1$, all conditional densities are updated. At the conclusion of the three targeting steps (T1-T3), we define $\hat{Q}^{(t+1)} = \hat{Q}^{(t, *)}$ which includes $\hat{\mu}^{(t+1)}, \hat{\pi}^{(t+1)}, \hat{f}^{(t+1)}_{Z_k}, \  \forall Z_k \in \Z$.

If at iteration $t^*$, the magnitude of $|P_n \Phi_{a_0}(\hat{Q}^{t^*})|$ is less than a pre-specified threshold $C_n = o_P(n^{-1/2})$, the TMLE procedure concludes, and $\hat{Q}^{*}=\hat{Q}^{(t^*)}$ becomes our final estimate of $Q$.

\newpage
\section{Proofs} 
\label{app:proofs} 

\subsection{Equivalent representation of the identifying functional}
\label{app:proofs_id}

\cite{tian02general} proved that the primal fixability is a necessary and sufficient condition for the identifiability of the causal effect of $A$ on all other variables $O \setminus A$. In an observed data distribution $P(O)$ that district factorizes according to an ADMG $\G(O)$ where $A$ is primal fixable, the resulting identifying functional for $\E[Y(a_0)]$, denoted by $\psi_{a_0}(P)$, is given by:
\begin{align*}
    \psi_{a_0}(P) =  \int y \times \prod_{M_i \in  \M} dP(m_i \mid \mpga{m_i}, a_0) \times \!\! \sum_{a \in \{a_0, a_1\}} \prod_{L_i \in \bbL} dP(\ell_i \mid \mpga{\ell_i}, a) \times  
    dP(x) \ , 
\end{align*}
where $dP(x) = \prod_{X_i \in \X}dP(x_i\mid \mpg(x_i))$; see Section~4 in \cite{bhattacharya2022semiparametric} for more details. Since $A$ is binary, we can write the above as: 
{\small 
\begin{align*}
    \psi_{a_0}(P) 
    &= \int y \times \!\! \prod_{M_i \in \M} dP(m_i \mid \mpga{m_i}, a_0) \times  \!\! \prod_{L_i \in \bbL \setminus A} \!\! dP(\ell_i \mid \mpga{\ell_i}, a_1) \times P(a_1 \mid \mpg(a)) \times dP(x)   \\
    &\hspace{0.25cm} + \int y \times \!\! \prod_{M_i \in  \M} dP(m_i \mid \mpga{m_i}, a_0) \times \!\! \prod_{L_i \in \bbL \setminus A} \!\! dP(\ell_i \mid \mpga{\ell_i}, a_0) \times  P(a_0 \mid \mpg(a)) \times dP(x)   \ . 
\end{align*}
}

Let $\Z = \M \cup \bbL \setminus \{A, Y\}$ denote the collection of post-treatment pre-outcome variables. We index variables in $\Z$ according to $\tau$, as $Z_1, \ldots, Z_K$.  For any $Z_k \in \Z$, let $a_{Z_k} = a_0$ if $Z_k \in \M$ and $a_{Z_k} = a_1$ if $Z_k \in \bbL$. For outcome $Y$, we define $a_Y$ similarly. With this notation, we can simplify $\psi_{a_0}(P)$ by combining the terms using the set $\Z$:

{\small 
\begin{equation*}
    \begin{aligned}
        \psi_{a_0}(P) 
        &= \int \ y \ dP(y \mid \mpga{y}, a_{Y}) \times \prod_{Z_k \in \Z} \ dP(z_k \mid \mpga{z_k}, a_{Z_k}) \times P(a_1 \mid \mpg(a)) \times dP(x)   \\
        &\hspace{0.5cm} +  \int y \ dP(y \mid \mpga{y}, a_0)\ \times \prod_{Z_k \in  \Z} \ dP(z_k \mid \mpga{z_k}, a_0) \times P(a_0 \mid \mpg(a))\times dP(x) 
        \\
        &= \int \ y \ dP(y \mid \mpga{y}, a_{Y}) \times \prod_{Z_k \in \Z} \ dP(z_k \mid \mpga{z_k}, a_{Z_k}) \times P(a_1 \mid \mpg(a)) \times dP(x)   \\
        &\hspace{0.5cm} +  \int \ \I(a=a_0)\ y \ \underbrace{dP(y \mid \mpg(y))\ \times \prod_{Z_k \in  \Z} \ dP(z_k \mid \mpg(z_k)) \ \times dP(a \mid \mpg(a))\times dP(x) }_{dP(o)} \\
        &= \int P(a_1 \mid \mpg(a)) \times \Big\{ \int \ y \ dP(y \mid \mpga{y}, a_{Y}) \times \prod_{Z_k \in \Z} \ dP(z_k \mid \mpga{z_k}, a_{Z_k}) \Big\} \times dP(x)   \\
        &\hspace{0.5cm} +  \underbrace{\int \I(a=a_0)\ y \ dP(o)}_{ = \ \int \I(a=a_0)\ y \ dP(a, y)} \\
        &=\E\bigg[ P(a_1 \mid \mpg(A))  \int  y \ dP(y \mid \mpga{y}, a_{Y})  \ \prod_{Z_k \in \Z} dP(z_k \mid \mpga{z_k}, a_{Z_k})  \bigg] + \E\left[\I(A=a_0)Y\right]  \ . 
    \end{aligned}
\end{equation*}%
}

To illustrate the reformulation idea, consider the ADMG in Figure~\ref{fig:graphs_primal}(d), where $\M=\{M,Y\}$, $\bbL=\{A,L\}$, and $\X=\{X\}$. The identification functional for $\E[Y(a_0)]$, denoted by $\psi_{a_0}(P)$, is:
\begin{align*}
    \psi_{a_0}(P) & = \int y\ dP(y\mid l,m,a_0,x)\ dP(l\mid m,a,x)\ dP(m\mid a_0,x)\ dP(a\mid x)\ dP(x).
\end{align*}
By separating out the summation over $A$, we have
\begin{align*}
    \psi_{a_0}(P) & = \int y\ dP(y\mid l,m,a_0,x)\ dP(l\mid m,a_1,x)\ dP(m\mid a_0,x)\ P(a_1\mid x)\ dP(x)
    \\
    &\hspace{0.5cm}+\int y\ dP(y\mid l,m,a_0,x)\ dP(l\mid m,a_0,x)\ dP(m\mid a_0,x)\ P(a_0\mid x)\ dP(x)
    \\
    &=\int y\ dP(y\mid l,m,a_Y,x)\ dP(l\mid m,a_L,x)\ dP(m\mid a_M,x)\ P(a_1\mid x)\ dP(x)
    \\
    &\hspace{0.5cm}+\int \I(a=a_0)\ y\ dP(y,l,m,a,x)
    \\
    &=\E\Big[P(a_1\mid \mpg(A))\int y\ dP(y\mid l,m,a_Y,x) \prod_{Z_k\in\Z}dP(z_k\mid\mpga{z_k},a_{Z_k})\Big]
    \\
    &\hspace{0.5cm}+\E[\I(A=a_0)\ Y] \ .
\end{align*}

\subsection{Equivalence of sequential regressions and conditional density integrations}
\label{app:proofs_seqreg}
In this section, we recursively prove the equivalence of the regression representation $\mB{Z_k}$ and the integral of $\mB{Z_{k+1}}$ with respective to the conditional density $f_{Z_k}(Z_k\mid \mpgA{Z_k})$. This result proves the consistency of the plug-in estimator in \eqref{eq:plug-in}, assuming consistent estimation of the nuisance parameters.
\begin{equation}\label{appeq:seqreg_equivalence}
    \begin{aligned}
    \mB{Z_k}(\mBarg{Z_k}, a_{Z_k})&=\E\left[ \mB{Z_{k+1}}(\mBarg{Z_{k+1}}, a_{Z_{k+1}}) \mid \mBarg{Z_k}, a_{Z_k} \right]
    \\
    &=\int \mB{Z_{k+1}}(\mBarg{Z_{k+1}}, a_{Z_{k+1}}) \ f_{Z_k}(z_k\mid \mpga{Z_k},a_{Z_k})\ d z_k \ .
\end{aligned}
\end{equation}
We begin the recursion at step $K+1$. By convention, we define $\mB{Z_{K+1}}(\mBarg{Z_{K+1}}, a_{Z_{K+1}}) = \mu(\mpgA{Y}, a_Y)$. Since no variable succeeds $Y$, the set $\mBarg{Z_{K+1}}=\mpgA{Y}$, given by the intersection of $\mpgA{Y}$ and the set of variables that precede $Y$, excluding $A$.

To complete the argument, we show that \eqref{appeq:seqreg_equivalence} holds for any $Z_k\in\Z$. From the definition of $\mBarg{Z_k}$, it follows that $\mBarg{Z_{k+1}}\subseteq \{\mBarg{Z_k},Z_k\}$. This implies that the regression $\mB{Z_k}$ can be expressed as the integral of $\mB{Z_{k+1}}$ with respect to the conditional density $p(Z_k \mid \mBarg{Z_k}, a_{Z_k})$:
\begin{equation}\label{appeq:seqreg_int}
    \begin{aligned}
    \mB{Z_k}(\mBarg{Z_k}, a_{Z_k}) &=\int \mB{Z_{k+1}}(\mBarg{Z_{k+1}}, a_{Z_{k+1}}) \ p(z_k\mid \mBarg{Z_k},a_{Z_k})\ d z_k \ .
\end{aligned}
\end{equation}

Furthermore, by construction, we have $\mpgA{Z_k} \subseteq \mBarg{Z_k}$. Hence, by the local Markov property, the following relationship holds:
\begin{equation}\label{appeq:seqreg_markov}
    \begin{aligned}
    p(Z_k\mid \mBarg{Z_k}, a_{Z_k})=f_{Z_k}(Z_k\mid \mpgA{Z_k},a_{Z_k}) \ . 
\end{aligned}
\end{equation}
Equations~\eqref{appeq:seqreg_int} and \eqref{appeq:seqreg_markov} together establish the validity of \eqref{appeq:seqreg_equivalence}.

\subsection{Sequential-regression representation of the nonparametric EIF}
\label{app:proofs_eif}

The EIF from \cite{bhattacharya2022semiparametric} is displayed in \eqref{eq:EIF_org}. In the following, we show the necessary steps for reformulating the EIF from \eqref{eq:EIF_org} to the form presented in Section~\ref{subsec:one-step}, Lemma~\ref{lem:eif}, which circumvent the need of density estimation.

The EIF presented in \cite{bhattacharya2022semiparametric} is outlined as follows, using slightly different notations. In their framework, $L_i$ and $M_i$ denote variables in $\bbL$ and $\M$, respectively. $\{\succ L_i\}$ and $\{\succ M_i\}$ denote variables in $O$ that succeed $L_i$ and $M_i$. Additionally, $\{\succeq L_i\}$ and $\{\succeq M_i\}$ denote the sets formed by including $L_i$ and $M_i$ on top of $\{\succ L_i\}$ and $\{\succ M_i\}$.
\begin{equation}
\label{eq:EIF_org}
    \begin{aligned}
   & \Phi_{a_0}(Q)(O) \\
   &\hspace{0.25cm} =\sum_{M_i \in \M}\Big\{\frac{\mathbb{I}(A=a_0)}{{\prod_{L_i \prec M_i} p(L_i \mid \mpg(L_i))}} \times\big(\sum_{A \cup \{\succ M_i\}} Y  {\prod_{\substack{O_i \in \bbL \cup \{\succ M_i\}}} p (O_i \mid \mpg(O_i))\big|_{A=a_0 \text { if } O_i \in \M}}
    \\
    &\hspace{4cm} -\sum_{A \cup \{\succeq M_i\}} Y \times{\prod_{\substack{O_i \in \bbL \cup \{\succeq M_i\}}} p(O_i \mid \mpg(O_i))\big|_{A=a_0 \text { if } O_i \in \M}} \big)\Big\}\qquad\text{\textbf{(A)}}
    \\
    &\hspace{0.5cm}+\sum_{L_i \in \bbL \backslash A}\Big\{\frac{\prod_{M_i\prec L_i} p(M_i \mid \mpg(M_i))\big|_{A=a_0}}{\prod_{M_i \prec L_i} p(M_i \mid \mpg(M_i))} \times (\sum_{\{\succ L_i\}} Y \times {\prod_{O_i \succ L_i} p(O_i \mid \mpg(O_i))\big|_{A=a_0 \text { if } O_i \in \M}}
    \\
    &\hspace{2cm}-\sum_{\{\succeq L_i\}} Y \times{\prod_{O_i \succeq L_i} p(O_i \mid \mpg(O_i))\big|_{A=a_0 \text { if } O_i \in \M}}\big)\Big\}\qquad\text{\textbf{(B)}}
    \\
    &\hspace{0.5cm}+\sum_{\{Y,\Z\}} Y \times{\prod_{M_i \in \M} p\left(M_i \mid \mpg\left(M_i\right)\right)\big|_{A=a_0} \times \prod_{L_i \in \bbL \backslash A} p\left(L_i \mid \mpg\left(L_i\right)\right)}\qquad\text{\textbf{(C)}}
    \\
    &\hspace{0.5cm}-\psi_{a_0}(Q) \ .
    \end{aligned}
\end{equation}

The core idea of the reformulation is to expand the equation by explicitly decomposing the summation across the two levels of $A$ as follows:
{\small\begin{align*}
    & \Phi_{a_0}(Q)(O)=\sum_{M_i \in \M}\Big\{\frac{\mathbb{I}(A=a_0)}{{\prod_{L_i \prec M_i} p(L_i \mid \mpga{L_i},a_0)}} \times\big(\sum_{\{\succ M_i\}} Y \times {\prod_{\substack{O_i \in \bbL \cup \{\succ M_i\}}} p (O_i \mid \mpga{O_i},a_0)}
    \\
    &\hspace{4cm} -\sum_{\{\succeq M_i\}} Y \times{\prod_{\substack{O_i \in \bbL \cup \{\succeq M_i\}}} p(O_i \mid \mpga{O_i},a_0)} \big)\Big\} \qquad \text{\textbf{(A1)}}
    \\
    &\hspace{2cm}+\sum_{M_i \in \M}\Big\{\frac{\mathbb{I}(A=a_0)}{{\prod_{L_i \prec M_i} p(L_i \mid \mpga{L_i},a_0)}} \times\big(\sum_{\{\succ M_i\}} Y \times {\prod_{\substack{O_i \in \bbL \cup \{\succ M_i\}}} p (O_i \mid \mpga{O_i},a_{O_i})}
    \\
    &\hspace{4cm} -\sum_{\{\succeq M_i\}} Y \times{\prod_{\substack{O_i \in \bbL \cup \{\succeq M_i\}}} p(O_i \mid \mpga{O_i},a_{O_i})} \big)\Big\}\qquad \text{\textbf{(A2)}}
    \\ 
    &\hspace{2cm}+\sum_{L_i \in \bbL \backslash A}\Big\{\I(A=a_0)\cancelto{1}{\frac{\prod_{M_i\prec L_i} p(M_i \mid \mpg(M_i))\big|_{A=a_0}}{\prod_{M_i \prec L_i} p(M_i \mid \mpg(M_i))\big|_{A=a_0}}} \times (\sum_{\{\succ L_i\}} Y \times {\prod_{O_i \succ L_i} p(O_i \mid \mpga{O_i},a_0)}
    \\
    &\hspace{4cm}-\sum_{\{\succeq L_i\}} Y \times{\prod_{O_i \succeq L_i} p(O_i \mid \mpga{O_i},a_0)}\big)\Big\}\qquad \text{\textbf{(B1)}}
    \\
    &\hspace{2cm}+\sum_{L_i \in \bbL \backslash A}\Big\{\I(A=a_1)\frac{\prod_{M_i\prec L_i} p(M_i \mid \mpga{M_i},a_0)}{\prod_{M_i \prec L_i} p(M_i \mid \mpga{M_i},a_1)} \times (\sum_{\{\succ L_i\}} Y \times {\prod_{O_i \succ L_i} p(O_i \mid \mpga{O_i},a_{O_i})}
    \\
    &\hspace{4cm}-\sum_{\{\succeq L_i\}} Y \times{\prod_{O_i \succeq L_i} p(O_i \mid \mpga{O_i},a_{O_i})}\big)\Big\}\qquad \text{\textbf{(B2)}}
    \\
    &\hspace{2cm}+\I(A=a_0)\sum_{\{Y,\Z\}} Y \times{\prod_{M_i \in \M} p\left(M_i \mid \mpga{M_i},a_0\right) \times \prod_{L_i \in \bbL \backslash A} p\left(L_i \mid \mpga{L_i},a_0\right)}\qquad \text{\textbf{(C1)}}
    \\
    &\hspace{2cm}+\I(A=a_1)\sum_{\{Y,\Z\}} Y \times{\prod_{M_i \in \M} p\left(M_i \mid \mpga{M_i},a_0\right) \times \prod_{L_i \in \bbL \backslash A} p\left(L_i \mid \mpga{L_i},a_1\right)}\qquad \text{\textbf{(C2)}}
    \\
    &\hspace{2cm}+\sum_{\{Y,\Z\}} Y \times \pi(a_1\mid \mpg(A)) \times{\prod_{O_i \in  \{Y,\Z\}} p\left(O_i \mid \mpga{O_i},a_{O_i}\right)}\qquad \text{\textbf{(D1)}}
    \\
    &\hspace{2cm}-\sum_{\{Y,\Z\}} Y \times \pi(a_1\mid \mpg(A)) \times{\prod_{O_i \in  \{Y,\Z\}} p\left(O_i \mid \mpga{O_i},a_{O_i}\right)}\qquad \text{\textbf{(D2)}}
    \\
    &\hspace{2cm}-\psi_{a_0}(Q)
    \\
    &\hspace{2cm}=\sum_{M_i \in \M}\Bigg\{\I(A=a_0)\frac{\prod_{L_i \prec M_i} p(L_i \mid \mpga{L_i},a_1)}{\prod_{L_i \prec M_i} p(L_i \mid \mpga{L_i},a_0)} \times\Big(\sum_{\{\succ M_i\}} Y \times \prod_{\substack{O_i \succ M_i}} p (O_i \mid \mpga{O_i},a_{O_i})
    \\
    &\hspace{4cm} -\sum_{\{\succeq M_i\}} Y \times {\prod_{\substack{O_i \in \{\succeq M_i\}}} p(O_i \mid \mpga{O_i}, a_{O_i})} \Big)\Bigg\}
    \\
    &\hspace{2cm}+\sum_{L_i \in \bbL \backslash A}\Bigg\{\I(A=a_1)\frac{\prod_{M_i \prec L_i} p(M_i \mid \mpga{M_i},a_{0})}{\prod_{M_i \prec L_i} p(M_i \mid \mpga{M_i},a_1)} \times \Big(\sum_{\{\succ L_i\}} Y \times {\prod_{O_i \succ L_i} p(O_i \mid \mpga{O_i},a_{O_i})}
    \\
    &\hspace{4cm}-\sum_{\{\succeq L_i\}} Y \times{\prod_{O_i \succeq L_i} p(O_i \mid \mpga{O_i},a_{O_i})}\Big)\Bigg\}
    \\
    &\hspace{2cm}+\left\{\I(A=a_1) - \pi(a_1\mid \mpg(A))\right\}\times\sum_{\{Y,\Z\}} Y \times{\prod_{O_i \in \{Y,\Z\}} p\left(O_i \mid \mpga{O_i},a_{O_i}\right)}
    \\
    &\hspace{2cm}+\sum_{\{Y,\Z\}} Y \times \pi(a_1\mid \mpg(A)) \times{\prod_{O_i \in  \{Y,\Z\}} p\left(O_i \mid \mpga{O_i},a_{O_i}\right)}-\psi_{a_0}(Q) + \I(A=a_0)Y.
\end{align*}}
We note that (A1) and (A2) are the respective decompositions of line (A) in \eqref{eq:EIF_org} evaluated at $A = a_0$ and $A = a_1$. Similarly, (B1) and (B2) decompose line (B), and (C1) and (C2) decompose line (C) at $A = a_0$ and $A = a_1$, respectively. We further introduce terms (D1) and (D2), which sum to zero, to isolate the component of the EIF that resides in the tangent space defined by the factorization component $p(A \mid \mpg(A))$.

To obtain the final form of the reformulated EIF, we further simplify lines (A1) and (A2) by decomposing the product of densities. Specifically, for line (A1), we perform the following expansion:
\begin{align*}
    \prod_{\substack{O_i \in \bbL \cup \{\succ M_i\}}} p (O_i \mid \mpga{O_i},a_0)&=\prod_{\substack{L_i \prec M_i}} p (L_i \mid \mpga{L_i},a_0)\times \prod_{\substack{O_i \succ M_i}} p (O_i \mid \mpga{O_i},a_0).
\end{align*}
An analogous expansion is applied to line (A2).

With the expansions, lines (A1) and (A2) can be simplified as follows:
\begin{align*}
    &\text{\textbf{(A1)}}=\sum_{M_i \in \M}\Big\{\mathbb{I}(A=a_0)\cancelto{1}{\frac{\prod_{L_i \prec M_i} p(L_i \mid \mpga{L_i},a_0)}{{\prod_{L_i \prec M_i} p(L_i \mid \mpga{L_i},a_0)}}}\times\big(\sum_{\{\succ M_i\}} Y \times {\prod_{\substack{O_i \succ M_i}} p (O_i \mid \mpga{O_i},a_0)}
    \\
    &\hspace{4cm} -\sum_{\{\succeq M_i\}} Y \times{\prod_{\substack{O_i \succeq M_i}} p(O_i \mid \mpga{O_i},a_0)} \big)\Big\}
    \\
    &\text{\textbf{(A2)}}=\sum_{M_i \in \M}\Big\{\mathbb{I}(A=a_0)\frac{\prod_{L_i \prec M_i} p(L_i \mid \mpga{L_i},a_1)}{{\prod_{L_i \prec M_i} p(L_i \mid \mpga{L_i},a_0)}} \times\big(\sum_{\{\succ M_i\}} Y \times {\prod_{\substack{O_i \succ M_i}} p (O_i \mid \mpga{O_i},a_{O_i})}
    \\
    &\hspace{4cm} -\sum_{\{\succeq M_i\}} Y \times{\prod_{\substack{O_i \succeq M_i}} p(O_i \mid \mpga{O_i},a_{O_i})} \big)\Big\}.
\end{align*}
One can then show that $\text{{(A1)}}+\text{{(B1)}}+\text{{(C1)}}=\I(A=a_0)\ Y$, which, when combined with lines (A2), (B2), (C2), (D1), (D2), and the constant term $-\psi_{a_0}(Q)$, yields the reformulated EIF as derived above. Re-writing this expression using the notation of $\mR{Z_k}$ and $\mB{Z_k}$ results in the formulation of EIF in Lemma~\ref{lem:eif}.

\subsection{Remainder analyses, robustness, and asymptotic linearity}
\label{app:proofs_r2}

To achieve asymptotic linearity of either one-step estimator or TMLE, denoted by $\psi_{a_0}(\hat{Q})$, we require the following conditions to be simultaneously met:
\begin{itemize}
    \item[(C1)] \textit{Donsker Condition:} $\Phi_{a_0}(\hat{Q})-\Phi(Q)$ falls in a P-Donsker class with probability 1, and 
    \item[(C2)] \textit{Consistent estimation of influence function:} $P\{\Phi_{a_0}(\hat{Q})-\Phi_{a_0}(Q)\}^2=o_{P}(1)$.
\end{itemize}
As an alternative of condition (C1), we adopted the cross-fitting procedure for nuisance estimation in Section~\ref{subsec:cross_fit}. Note that the TMLE $\psi_{a_0}(\hat{Q}^*)$ also requires the following:
\begin{itemize}
    \item[(C3)] \textit{Successful targeting of nuisances:} $P_{n} \Phi_{a_0}(\hat{Q}^*)=o_{P}(n^{-1 / 2})$. 
\end{itemize}

For all derivations related to $R_2$ in this section, we assume that the nuisance estimates converge at the rate specified below: 
\begin{equation}
\label{appeq:cvg_rate}
    \begin{aligned}
    &|| \hat{\mu} - \mu || =\smallO(n^{-{1}/{b_Y}}),\qquad &&|| \hat{f}^r_{A} - \fr{A} || = o_P(n^{-{1}/{r_{A}}}),\\
    &|| \hmB{Z_k} - \mB{Z_k} || = o_P(n^{-{1}/{b_{Z_k}}}),\qquad &&|| \hfr{Z_k} - \fr{Z_k} || = o_P(n^{-{1}/{r_{Z_k}}}),\\
    &|| \hli{Z_k} - \li{Z_k} || = o_P(n^{-{1}/{t_{Z_k}}}),\qquad &&|| \hlex{Z_k} - \lex{Z_k} || = o_P(n^{-{1}/{d_{Z_k}}}),\\
    &||\hat{\pi} - \pi || =\smallO(n^{-{1}/l_A}),\qquad &&|| \hat{f}_{Z_k} - {f}_{Z_k} || = o_P(n^{-{1}/{{l_{Z_k}}}}) ,
    \end{aligned}
\end{equation}
for all $Z_k \in \Z$.


\subsubsection{The second-order remainder term for $\psi_{a_0}(\hat{Q})$}

For $\psi_{a_0}(\hat{Q})$, the set of nuisances is $Q=\{\mu, \pi, \mR{Y}, \{\mR{Z_k},\mB{Z_k},\  \forall Z_k\in\Z\}\}$.

Based on the von Mises expansion, we have the following form of $R_2$. 
\begin{align*}
    &R_{2}(\hat{Q}, Q) 
    =\psi_{a_0}(\hat{Q})-\psi_{a_0}(Q)+\int \Phi_{a_0}(\hat{Q}) dP(o)
    \\ 
    &=\sum_{Z_k \in \Z} \int\I(A = a_{Z_k}) \ \hmR{Z_k}(\mathrm{mp}^{-a}_{\G, \prec}(z_k))\Big( \E\left[\hmB{Z_{k+1}}(\mBarg{Z_{k+1}}, a_{Z_{k+1}})\mid\mBarga{z_k},a_{Z_k} \right]  \\
    &\hspace{9cm}- \hmB{Z_{k}}(\mBarga{z_k}, a_{Z_k})  \Big)\ d P(o)
    \\ 
    &\hspace{0.25cm}+\int\I(A=a_{Y})\ \hmR{Y}(\mathrm{mp}^{-a}_{\G, \prec}(y)) \times\left(\mu(\mpga{y},a_Y)-\hat{\mu}(\mpga{y},a_Y)\right) d P(o)
    \\ 
    &\hspace{0.25cm} + \int\Big\{\pi(a_1\mid \mpg(a)) - \hat{\pi}(a_1\mid \mpg(a)) \Big\} \ \hmB{Z_1}(\mBarga{z_1}, a_{Z_1}) \ d P(o)
    \\
    &\hspace{0.25cm} + \int \hat{\pi}(a_1\mid \mpg(a)) \ \hmB{Z_1}(\mBarga{z_1}, a_{Z_1})\ dP(o) + \cancel{\E[\I(A=a_0)Y] }
    \\ 
    &\hspace{0.25cm}-  \int \pi(a_1\mid \mpg(a)) \ \mB{Z_1}(\mBarga{z_1}, a_{Z_1})\ dP(o) \cancel{- \E[\I(A=a_0)Y]} \ 
    \\ 
    &=\sum_{Z_k \in \Z}  \int\I(A = a_{Z_k}) \ \left\{\hmR{Z_k}(\mathrm{mp}^{-a}_{\G, \prec}(z_k))-\mR{Z_k}(\mathrm{mp}^{-a}_{\G, \prec}(z_k))\right\}
    \\
    &\hspace{0.5cm}\times\Big( \E\left[\hmB{Z_{k+1}}(\mBarg{Z_{k+1}}, a_{Z_{k+1}})\mid\mBarga{z_k},a_{Z_k} \right] - \hmB{Z_{k}}(\mBarga{z_k}, a_{Z_k})  \Big) d P(o)
    \\
    &\hspace{0.25cm} + \sum_{Z_k \in \Z} \int\I(A = a_{Z_k}) \ \mR{Z_k}(\mathrm{mp}^{-a}_{\G, \prec}(z_k)) 
    \\
    &\hspace{0.5cm}\times \Big( \E\left[\hmB{Z_{k+1}}(\mBarg{Z_{k+1}}, a_{Z_{k+1}})\mid\mBarga{z_k},a_{Z_k} \right] - \hmB{Z_{k}}(\mBarga{z_k}, a_{Z_k})  \Big) d P(o)\tag{1}
    \\ 
    &\hspace{0.25cm}+\int \I(A=a_{Y})\left\{\hmR{Y}(\mathrm{mp}^{-a}_{\G, \prec}(y))-\mR{Y}(\mathrm{mp}^{-a}_{\G, \prec}(y))\right\}\\
    &\hspace{3cm}\times\left(\mu(\mpga{y},a_Y)-\hat{\mu}(\mpga{y},a_Y)\right) d P(o)
    \\
    &\hspace{0.25cm}+\int\I(A=a_{Y})\ \mR{Y}(\mathrm{mp}^{-a}_{\G, \prec}(y)) \times\left(\mu(\mpga{y},a_Y)-\hat{\mu}(\mpga{y},a_Y)\right) d P(o)\tag{2}
    \\ 
    &\hspace{0.25cm} + \int\Big\{\pi(a_1\mid \mpg(a)) - \hat{\pi}(a_1\mid \mpg(a))\Big\}
    \\
    &\hspace{3cm}\times \left\{\hmB{Z_1}(\mBarga{z_1}, a_{Z_1})-\mB{Z_1}(\mBarga{z_1}, a_{Z_1})\right\} \ dP(o)\tag{3}
    \\
    &\hspace{0.25cm}+\int\Big\{\pi(a_1\mid \mpg(a)) - \hat{\pi}(a_1\mid \mpg(a))\Big\} \ \mB{Z_1}(\mBarga{z_1}, a_{Z_1}) \ dP(o)\tag{4}
    \\
    &\hspace{0.25cm} +\int \hat{\pi}(a_1\mid \mpg(a)) \ \hmB{Z_1}(\mBarga{z_1}, a_{Z_1})\ dP(o) \tag{5}
    \\
    &\hspace{0.25cm} -  \int \pi(a_1\mid \mpg(a)) \ \mB{Z_1}(\mBarga{z_1}, a_{Z_1})\ dP(o)  \tag{6}
    \\ 
    &=\sum_{Z_k \in \Z} \int \{\hmR{Z_k} -\mR{Z_k}\}(\mathrm{mp}^{-a}_{\G, \prec}(z_k)) \times \{\mB{Z_k} -\hmB{Z_k}\}(\mBarga{z_k}, a_{Z_k})\ dP(\{\prec z_k\}^{-a},a_{Z_k})
        \\
        &\hspace{0.25cm}+\int \{\hmR{Y} -\mR{Y} \}(\mathrm{mp}^{-a}_{\G, \prec}(y)) \times \{\mu -\hat{\mu} \}(\mpga{y}, a_{Y})\ dP(\{\prec y\}^{-a},a_Y). 
\end{align*}
The terms labeled (1), (2) and (4) cancel out with the terms labeled (3), (5) and (6), yielding the formulation of $R_2$ in Lemma~\ref{lem:r2_compact}.

\subsubsection{The second-order remainder term for $\psi_{a_0,1}(\hat{Q})$}
\label{appsubsec:r2_ratio}

For $\psi_{a_0,1}(\hat{Q})$, the set of nuisances is $Q=\{\mu, \fr{A}, \{\fr{Z_k},\mB{Z_k}  \forall Z_k\in\Z\}\}$. Given the von Mises expansion, we can write:
\begin{align*}
    R_{2}\left(\hat{Q}, Q\right)&=\psi_{a_0}(\hat{Q})-\psi_{a_0}(Q)+\int \Phi_{a_0}(\hat{Q}) d P(o)
    \\ 
    &=\sum_{Z_k\in\Z} \Bigg[\int \bigg\{\prod_{\substack{Z_i \in \{\Z_{\prec Z_k},A\} \\ \text{s.t. } a_{Z_i} \neq a_{Z_k}}}\hfr{Z_i}(z_i, \mpga{z_i}) - \prod_{\substack{Z_i \in \{\Z_{\prec Z_k},A\} \\ \text{s.t. } a_{Z_i} \neq a_{Z_k}}} \fr{Z_i}(z_i, \mpga{z_i})\bigg\}\\
    &\hspace{2cm}\times\left( \mB{Z_k}(\mBarga{z_k}, a_{Z_k}) - \hmB{Z_{k}}(\mBarga{z_k}, a_{Z_k}) \right)\ dP(o) \Bigg]
    \\
    &\hspace{0.25cm}+\int \big\{\prod_{\substack{Z_i \in \{\Z,A\} \\ \text{s.t. } a_{Z_i} \neq a_{Y}}} \hfr{Z_i}(z_i, \mpga{z_i}) - \prod_{\substack{Z_i \in \{\Z,A\} \\ \text{s.t. } a_{Z_i} \neq a_{Y}}} \fr{Z_i}(z_i, \mpga{z_i})\big\}\\
    &\hspace{4cm}\times\left(\mu(\mpga{y},a_Y)-\hat{\mu}(\mpga{y},a_Y)\right) \ d P(o)
    \\
    &{\text{using the fact that }\prod_{Z_i}\hfr{Z_i}-\prod_{Z_i}\fr{Z_i}=\sum_{Z_i}\Big\{\left(\hfr{Z_i}-\fr{Z_i}\right) \prod_{j<i} \fr{Z_j} \prod_{h>i} \hfr{Z_h}\Big\}}
    \\
    &=\sum_{Z_k\in\Z} \sum_{\substack{Z_i \in \{\Z_{\prec Z_k},A\} \\ \text{s.t. } a_{Z_i} \neq a_{Z_k}}}\Bigg[\int \Big\{\left(\hfr{Z_i}(z_i, \mpga{z_i})-\fr{Z_i}(z_i, \mpga{z_i})\right)
    \\
    &\hspace{2.5cm}\times\prod_{\{\substack{j<i:Z_j \in \{\Z_{\prec Z_k},A\} \\ \text{s.t. } a_{Z_j} \neq a_{Z_k}}\}} \fr{Z_j}(z_j, \mpga{z_j})\prod_{\{\substack{h>i:Z_h \in \Z_{\prec Z_k} \\ \text{s.t. } a_{Z_h} \neq a_{Z_k}}\}} \hfr{Z_h}(z_h, \mpga{z_h})\Big\}
    \\
    &\hspace{2.5cm}\times \left( \mB{Z_k}(\mBarga{z_k}, a_{Z_k}) - \hmB{Z_{k}}(\mBarga{z_k}, a_{Z_k})  \right)\ dP(o) \Bigg]
    \\
    &\hspace{0.25cm}+\int \sum_{\substack{Z_i \in \{\Z,A\} \\ \text{s.t. } a_{Z_i} \neq a_{Y}}}\Big\{\left(\hfr{Z_i}(z_i, \mpga{z_i})-\fr{Z_i}(z_i, \mpga{z_i})\right)
    \\
    &\hspace{2cm}\times\prod_{\{\substack{j<i:Z_j \in \{\Z,A\} \\ \text{s.t. } a_{Z_j} \neq a_{Y}}\}} \fr{Z_j}(z_j, \mpga{z_j}) \prod_{\{\substack{h>i:Z_h \in \Z \\ \text{s.t. } a_{Z_h} \neq a_{Y}}\}} \hfr{Z_h}(z_h, \mpga{z_h})\Big\}
    \\
    &\hspace{2cm}\times\left(\mu(\mpga{y},a_Y)-\hat{\mu}(\mpga{y},a_Y)\right) \ d P(o) \ .
\end{align*}

\underline{[regularity conditions]} Assume $\exists\ C\in\R$ such that the following holds
\begin{equation}\label{appeq:r2_ratio_condition}
    \begin{aligned}
        &\Big|\prod_{\{\substack{j<i:Z_j \in \{\Z_{\prec Z_k},A\} \\ \text{s.t. } a_{Z_j} \neq a_{Z_k}}\}} \fr{Z_j}\prod_{\{\substack{h>i:Z_h \in \Z_{\prec Z_k} \\ \text{s.t. } a_{Z_h} \neq a_{Z_k}}\}} \hfr{Z_h} \Big| < C, \ \forall Z_k\in\{\Z,Y\} \ .
    \end{aligned}
\end{equation}

Under the boundedness conditions of \eqref{appeq:r2_ratio_condition}, we apply the Cauchy–Schwarz inequality to each term in $R_2$,
leading to the following inequality
\begin{align*}
    R_2(\hat{Q}, Q) & \leq C\bigg[ \sum_{Z_k\in\Z}\sum_{\substack{Z_i \in \Z_{\prec Z_k} \\ \text{s.t. } a_{Z_i} \neq a_{Z_k}}} ||\hfr{Z_i}-\fr{Z_i} || \times ||\mB{Z_k}-\hmB{Z_k}||
    \\
    &\hspace{0.5cm} + \sum_{\substack{Z_i \in\Z \\ \text{s.t. } a_{Z_i} \neq a_{Y}}}||\hfr{Z_i}-\fr{Z_i}|| \times ||\mu-\hat{\mu}||
    \\
    &\hspace{0.5cm} + \sum_{Z_k\in\M} ||\hfr{A} -\fr{A} || \times ||\mB{Z_k}-\hmB{Z_k}||\bigg] \ .
\end{align*}

Given the nuisance convergence rates in \eqref{appeq:cvg_rate}, let $r^*_1$ be the minimal value in the joint of the following sets:
\begin{align*}
    &\left\{\frac{1}{b_{Z_k}}+\frac{1}{r_{Z_i}},\forall Z_k\in\Z\text{ and }Z_i\in\Z_{\prec Z_k}\text{ s.t. }a_{Z_i}\neq a_{Z_k}\right\},\quad \left\{\frac{1}{b_Y}+\frac{1}{r_{Z_i}},\forall Z_i\in\Z\text{ s.t. }a_{Z_i}\neq a_{Y}\right\},\\
    &\left\{\frac{1}{r_A}+\frac{1}{b_{Z_i}},\forall Z_i\in\M\right\}.
\end{align*}
Given above, we get $R_2\left(\hat{Q}^{\star}, Q\right) \leq o_P\left(n^{-r^*_1}\right).$

\subsubsection{The second-order remainder term for $\psi_{a_0,2}(\hat{Q})$}
\label{appsubsec:r2_bayes}
For $\psi_{a_0,2}(\hat{Q})$, the set of nuisances is $\{\mu, \{\li{Z_k},\lex{Z_k},\mB{Z_k},\  \forall Z_k\in\Z\}\}$. The formulation of $R_2$ for $\psi_{a_0,2}(\hat{Q})$ closely resembles that for $\psi_{a_0,1}(\hat{Q})$, with some minor modifications. Specifically,$\hfr{A}$ and $\hfr{Z_k},\ \forall Z_k\in\Z$ are replaced by functions of $\hli{Z_k}$ and $\hlex{Z_k}$. These replacements are defined as follows:
\begin{align*}
    \hfr{Z_k} = \frac{\hli{Z_k}}{1-\hli{Z_k}}\frac{1-\hlex{Z_k}}{\hlex{Z_k}} \ , \ \forall Z_k\in\Z,\qquad \hfr{A}=\frac{1-\hlex{Z_1}}{\hlex{Z_1}} \ .
\end{align*}

\underline{[regularity conditions]} Assume the following quantities are bounded by some finite number in $\R$ and the positivity assumption holds for $\hli{Z_k},\hlex{Z_k}, \li{Z_k}$, and $\lex{Z_k}$ for all $Z_k\in\{\Z,A\}$.
\begin{equation}\label{appeq:r2_bayes_condition}
    \begin{aligned}
        &\Big|\prod_{\{\substack{j<i:Z_j \in \{\Z_{\prec Z_k},A\} \\ \text{s.t. } a_{Z_j} \neq a_{Z_k}}\}} \fr{M_j}\prod_{\{\substack{h>i:Z_h \in \{\Z_{\prec Z_k},A\} \\ \text{s.t. } a_{Z_h} \neq a_{Z_k}}\}} \hfr{M_h} \Big| , \ \forall Z_k\in\{\Z,Y\} \ . 
    \end{aligned}
\end{equation}

Under the boundedness conditions of \eqref{appeq:r2_bayes_condition}, we apply the Cauchy–Schwarz inequality to each term in $R_2$. There exists finite number $C\in\R$ such that the following inequality holds
\begin{align*}
    R_2(\hat{Q}^\star, Q) & \leq C\bigg[ \sum_{Z_k\in\Z}\sum_{\substack{Z_i \in \Z_{\prec Z_k} \\ \text{s.t. } a_{Z_i} \neq a_{Z_k}}} ||\hli{M_i}-\li{M_i} || \times ||\mB{Z_k}-\hmB{Z_k}||
    \\
    &\hspace{0.5cm}+\sum_{Z_k\in\Z}\sum_{\substack{Z_i \in \Z_{\prec Z_k} \\ \text{s.t. } a_{Z_i} \neq a_{Z_k}}} ||\hlex{M_i}-\lex{M_i} || \times ||\mB{Z_k}-\hmB{Z_k}||
    \\
    &\hspace{0.5cm}+\sum_{\substack{Z_i \in \Z, \\ \text{s.t. } a_{Z_i} \neq a_{Y}}}||\hli{Z_i}-\li{Z_i} || \times ||\mu-\hat{\mu}|| 
    \\
    &\hspace{0.5cm}+\sum_{\substack{Z_i \in \Z, \\ \text{s.t. } a_{Z_i} \neq a_{Y}}}||\hlex{Z_i}-\lex{Z_i} || \times ||\mu-\hat{\mu}|| 
    \\
    &\hspace{0.5cm}+\sum_{Z_k\in\M} ||\hlex{Z_1} - \lex{Z_1} || \times ||\mB{Z_k}-\hmB{Z_k}||\bigg] \ .
\end{align*}

Given the nuisance convergence rates in \eqref{appeq:cvg_rate}, let $r^*_2$ be the minimal value in the joint of the following sets:
\begin{align*}
    &\left\{\frac{1}{b_{Z_k}}+\frac{1}{d_{Z_i}},\forall Z_k\in\Z\text{ and }Z_i\in\Z_{\prec Z_k}\text{ s.t. }a_{Z_i}\neq a_{Z_k}\right\},\quad \left\{\frac{1}{b_Y}+\frac{1}{d_{Z_i}},\forall Z_i\in\Z\text{ s.t. }a_{Z_i}\neq a_{Y}\right\},\\
    &\left\{\frac{1}{b_{Z_k}}+\frac{1}{t_{Z_i}},\forall Z_k\in\Z\text{ and }Z_i\in\Z_{\prec Z_k}\text{ s.t. }a_{Z_i}\neq a_{Z_k}\right\},\quad \left\{\frac{1}{b_Y}+\frac{1}{t_{Z_i}},\forall Z_i\in\Z\text{ s.t. }a_{Z_i}\neq a_{Y}\right\},\\
    &\left\{\frac{1}{d_{Z_1}}+\frac{1}{b_{Z_i}},\forall Z_i\in\M\right\}.
\end{align*}
Given above, we get: 
$R_2\left(\hat{Q}^{\star}, Q\right) \leq o_P\left(n^{-r^*_2}\right).$

\subsubsection{The second-order remainder term of $\psi_{a_0,3}(\hat{Q})$}
\label{appsubsec:r2_density}

For $\psi_{a_0,3}(\hat{Q})$, the set of nuisances is $Q=\{\pi,\mu,f_{Z_k}\ \forall Z_k\in\Z\}$. In the following derivation, for simplicity of notation, we interchangeably use $\pi(a\mid \mpg(A))$ with $f_{A}(a\mid \mpgA{A},a_{A})$ for all $a\in\{a_0,a_1\}$, as well as their estimates.
Given the von Mises expansion, we can write:
{\begin{align*}
    R_{2}(\hat{Q}, Q)&=\psi_{a_0}(\hat{Q})-\psi_{a_0}(Q)+\int \Phi_{a_0}(\hat{Q}) d P(o)
    \\
    &=\sum_{Z_k\in\Z} \sum_{\substack{Z_i \in \{\Z_{\prec Z_k},A\} \\ \text{s.t. } a_{Z_i} \neq a_{Z_k}}}\Bigg[\int \Big\{\left(\frac{\hf{Z_i}(z_i\mid \mpga{z_i},a_{Z_i})}{\hf{Z_i}(z_i\mid \mpga{z_i},1-a_{Z_i})}-\fr{Z_i}(z_i,\mpga{z_i})\right)
    \\
    &\hspace{1cm}\times\prod_{\{\substack{j<i:Z_j \in \{\Z_{\prec Z_k},A\} \\ \text{s.t. } a_{Z_j} \neq a_{Z_k}}\}} \fr{Z_j}(z_j, \mpga{z_j}) \prod_{\{\substack{h>i:Z_h \in \Z_{\prec Z_k} \\ \text{s.t. } a_{Z_h} \neq a_{Z_k}}\}} \frac{\hf{Z_h}(z_h\mid \mpga{z_h},a_{Z_h})}{\hf{Z_h}(z_h\mid \mpga{z_h},1-a_{Z_h})}\Big\}
    \\
    &\hspace{1cm}\times \hat{\mu}(\mpga{y},a_Y)\left(f_{Z_k}(z_k\mid \mpga{z_k},a_{Z_k})-\hf{Z_k}(z_k\mid \mpga{z_k},a_{Z_k})\right)
    \\
    &\hspace{1cm}\times\prod_{k<l\leq K} \left[\hf{Z_l}(z_l\mid \mpga{z_l},a_{Z_l})\ d z_{l}\right]\ dP(o) \Bigg]
    \\
    &\hspace{0.25cm}+\int \sum_{\substack{Z_i \in \{\Z,A\} \\ \text{s.t. } a_{Z_i} \neq a_{Y}}}\Bigg\{\left(\frac{\hf{Z_i}(z_i\mid \mpga{z_i},a_{Z_i})}{\hf{Z_i}(z_i\mid \mpga{z_i},1-a_{Z_i})}-\fr{Z_i}(z_i, \mpga{z_i})\right) 
    \\
    &\hspace{1cm}\times\prod_{\{\substack{j<i:Z_j \in \{\Z,A\} \\ \text{s.t. } a_{Z_j} \neq a_{Y}}\}} \fr{Z_j}(z_j, \mpga{z_j})
    \\
    &\hspace{1cm}\times\prod_{\{\substack{h>i:Z_h \in \Z \\ \text{s.t. } a_{Z_h} \neq a_{Y}}\}} \frac{\hf{Z_h}(z_h\mid \mpga{z_h},a_{Z_h})}{\hf{Z_h}(z_h\mid \mpga{z_h},1-a_{Z_h})}\Bigg\}
    \\
    &\hspace{1cm}\times\left(\mu(\mpga{y},a_Y)-\hat{\mu}(\mpga{y},a_Y)\right) \ d P(o) \ . 
\end{align*}}

\underline{[regularity conditions]} Assume $\forall Z_k\in\{\Z,Y\}$ and $a\in\{a_0,a_1\}$, $\exists\ C\in\mathbb{R}$ such that:
\begin{equation}\label{appeq:r2_density_condition}
    \begin{aligned}
        &\Big|\prod_{\{\substack{j<i:Z_j \in \{\Z_{\prec Z_k},A\} \\ \text{s.t. } a_{Z_j} \neq a_{Z_k}}\}} \fr{Z_j}(Z_j\mid \mpgA{Z_j},a_{Z_j}) \prod_{\{\substack{h>i:Z_h \in \Z_{\prec Z_k} \\ \text{s.t. } a_{Z_h} \neq a_{Z_k}}\}} \frac{\hf{Z_h}(Z_h\mid \mpga{Z_h},a_{Z_h})}{\hf{Z_h}(Z_h\mid \mpga{Z_h},1-a_{Z_h})} \Big| < C,
        \\
        &\hspace{9cm} \forall Z_i \in \{\Z_{\prec Z_k},A\} \ \text{s.t. } a_{Z_i} \neq a_{Z_k},
        \\
        &\big|f_{Z_k}(Z_k\mid \mpga{Z_k},a)\big| < C, \big|\hf{Z_k}(Z_k\mid \mpga{Z_k},a)\big| < C ,\ \forall Z_k\in\{\Z,A\},
        \\
        &\big|\hat{\mu}(\mpgA{Y},a_Y)\big| < C.
    \end{aligned}
\end{equation}

Under the boundedness conditions of \eqref{appeq:r2_density_condition}, we apply the Cauchy–Schwarz inequality to each term in $R_2$,
leading to the following inequality
{\small\begin{align*}
    R_2(\hat{Q}, Q) & \leq C\bigg[ \sum_{Z_k\in\Z}\sum_{\substack{Z_i \in \Z_{\prec Z_k} \\ \text{s.t. } a_{Z_i} \neq a_{Z_k}}} ||\hf{Z_i}(Z_i\mid \mpgA{Z_i},a_{Z_i})-f_{Z_i}(Z_i\mid \mpgA{Z_i},a_{Z_i}) || 
    \\
    &\hspace{3cm}\times ||f_{Z_k}(Z_k\mid \mpgA{Z_k},a_{Z_k})-\hf{Z_k}(Z_k\mid \mpgA{Z_k},a_{Z_k})|| 
    \\
    &\hspace{0.5cm}+\sum_{\substack{Z_i \in \Z \\ \text{s.t. } a_{Z_i} \neq a_{Z_k}}}||\hf{Z_i}(Z_i\mid \mpgA{Z_i},a_{Z_i})-f_{Z_i}(Z_i\mid \mpgA{Z_i},a_{Z_i})||
    \\
    &\hspace{3cm}\times ||\mu(\mpgA{Y},a_Y)-\hat{\mu}(\mpgA{Y},a_Y)||,
    \\
    &\hspace{0.5cm}+\sum_{Z_k\in\M} ||\hat{\pi}(a_1\mid\mpg(A)) - \pi(a_1\mid\mpg(A))|| \\ 
    &\hspace{3cm}\times ||f_{Z_k}(Z_k\mid \mpgA{Z_k},a_{Z_k})-\hf{Z_k}(Z_k\mid \mpgA{Z_k},a_{Z_k})|| \bigg] \ .
\end{align*}}

Given the convergence rates in \eqref{appeq:cvg_rate}, let $r^*_3$ be the minimal value in the joint of the following sets:
\begin{align*}
    &\left\{\frac{1}{l_{Z_i}} + \frac{1}{l_{Z_k}},\forall Z_k\in\Z\text{ and }Z_i\in\Z_{\prec Z_k}\text{ s.t. }a_{Z_i}\neq a_{Z_k}\right\},\quad \left\{\frac{1}{l_{Z_i}} + \frac{1}{b_{Y}},\forall Z_i\in\Z\text{ s.t. }a_{Z_i}\neq a_{Y}\right\},\\
    &\left\{\frac{1}{l_{A}}+\frac{1}{l_{Z_k}},\forall Z_i\in\M\right\}.
\end{align*}
Given above, we get $R_2\left(\hat{Q}^{\star}, Q\right) \leq o_P\left(n^{-r^*_3}\right).$

\subsection{Semiparametric EIF under ordinary conditional independence restrictions}
\label{app:proofs_eif_semi}
Focusing on \textit{mb-shielded} ADMGs that imply ordinary independence constraints related to variables in $\{\Z,Y\}$, stating that $Z_k$ is independent of $Z_j$ given $\mpg(Z_k)$, where $Z_j \prec Z_k$ and $Z_j \notin \mpg(Z_k)$, the form of semiparametric EIF guiding estimation is shown in \eqref{eq:EIF_semi}. Align with the discussion in Section~\ref{sec:semiparam}, the following derivation excludes $\E[\I(A=a_0)Y]$ from the projection into the reduced tangent space to achieve substantial improvement in computational efficiency.

\begin{equation}
\label{eq:EIF_semi}
    \begin{aligned}
    \Phi(P)(O) 
    & = \E\bigg[ \I(A=a_{Y})\mR{Y}\big(\mathrm{mp}^{-A}_{\G, \prec}(Y)\big) \mid \mpi(Y)\bigg]\times\left(Y-\mu(\mpgA{Y},a_Y)\right) 
    \\
    &\hspace{0.25cm}+  \sum_{Z_k \in \Z} \E\bigg[ \I(A = a_{Z_k}) \ \mR{Z_k}\big(\mathrm{mp}^{-A}_{\G, \prec}(Z_k)\big)\times \bigg( \mB{Z_{k+1}} - \mB{Z_{k}} \bigg) \ \bigg| \ Z_k, \mpg(Z_k) \bigg]
    \\
    &\hspace{0.25cm} + \left\{ \I(A = a_1) - \pi(a_1 \mid \mpg(A)) \right\} \ \mB{Z_1}
    \\
    &\hspace{0.25cm} + \pi(a_1 \mid \mpg(A)) \ \mB{Z_1}
    \\
    &\hspace{0.25cm} + \E[\I(A=a_0)Y] - \psi_{a_0}(Q) \ .
\end{aligned}
\end{equation}

For brevity, we omit the argument of $\mB{Z_k}$ in the EIF above. Without this omission, $\mB{Z_k}$ is defined on $\mBarg{Z_k}$.

\newpage
\section{A worked example}
\label{app:examples}

Both \citet{jung2024unified} and our work propose estimators that avoid direct density estimation, but they take distinct approaches. \citet{jung2024unified} employ the empirical bifurcation method \citep{chernozhukov2023simple, xu2022neural}, introducing an independent copy of the treatment variable $A' \sim P(A)$ to circumvent density estimation. In contrast, we decompose the target functional by treatment levels and apply sequential regression reparameterizations. While both strategies yield one-step estimators with desirable properties, such as double robustness and asymptotic linearity under appropriate rate conditions, they lead to different classes of estimators, including plug-in and one-step versions. Below, we discuss the differences in terms of compatibility in parametric models, nuisance characterizations, and theoretical conditions. Moreover, beyond various one-step estimators, we also propose TMLEs, whose construction relies on expressing the EIF as a sum of components that each lie within the tangent space of a corresponding nuisance parameter. Incorporating both $A$ and its independent copy $A'$ complicates the tangent space formulation, the TMLE construction, and the discussion of semiparametric efficiency. 

Using the ADMG in Figure~\ref{fig:graphs_primal}(d), we compare their approach with ours by presenting the identification functional reformulation in terms of sequential regression, plug-in estimators, the one-step estimator, nuisance rate conditions, and robustness results. 
    
In Figure~\ref{fig:graphs_primal}(d), we have $\M=\{M,Y\}$, $\bbL=\{A,L\}$, and $\Z=\{M,L\}$. The identification functional (given in Lemma~\ref{lem:reID}) takes the following form:
{\small
\begin{align*}
    \psi_{a_0}(P) 
    &=  \E\Big[ P(a_1 \mid X)  \int  y \ dP(y \mid l,m, a_{0},X)  \ dP(l \mid m, a_{1},X) \ dP(m\mid a_{0},X)  \Big] 
    + \E\left[\I(A=a_0)Y\right] \ .
\end{align*}
}

We can rewrite the above as follows: 
{\small
\begin{align*}
    \psi_{a_0}(P) 
    &= \E\Big[ \pi(a_1 \mid X) \mB{Z_1}(X, a_{0}) \Big] + \E\left[\I(A=a_0)Y\right] \ .
\end{align*}
}
where $Z_1$ represents $M$, $\mB{Z_1}(X, a_{0}) \coloneqq \E[ \mB{Z_2}(M, X, a_{1}) \mid X, a_0]$, $Z_2$ represents $L$, $\mB{Z_2}(M, X, a_{1}) \coloneqq \E[ \mu_Y(L, M, X, a_0) \mid M, X, a_1]$, and $\mu_Y(L, M, X, a_0) \coloneqq \E[Y \mid L, M, X, a_1]$. 

Therefore, our plug-in estimator (given by \eqref{eq:plug-in}) requires fitting $\mu_Y$, $\mB{Z_2}$, and $\mB{Z_1}$, and it will be consistent provided all these regressions are correctly specified. Note that estimation of $\mB{Z_1}$ relies on treating $\mB{Z_2}$ as a pseudo-outcome and regressing it on $A$ and $X$, and evaluating the regression at $A=a_0$ (or alternatively fitting the regression using only rows where $A=a_0$.) 

This approach differs from \citet{jung2024unified}, which employs an independent copy of $A$, denoted $A' \sim P(A)$. Let $\mu_Y(L, M, a_0, X) \coloneqq \E[Y |  L, M, a_0, X]$, $\mu_L(M, A, X) \coloneqq \E[\mu_Y(L, M, a_0, X) \, | \, M, A, X]$, and $\mu_M(A^{\prime}, X) \coloneqq \E[\mu_L(M, A^{\prime}, X) \, | \, a_0, A^{\prime}, X]$. Their plug-in estimator is obtained as follows. First, estimate $\mu_Y$ by regressing $Y$ on $(L,M,A,X)$ and evaluating at $A=a_0$, yielding $\hat{\mu}_Y(L,M,a_0,X)$. Next, regress this pseudo-outcome on $(M,A,X)$ to obtain $\hat{\mu}_L$. To estimate $\mu_M$, evaluate $\hat{\mu}_L$ at $A'$, giving $\hat{\mu}_L(M,A',X)$, which is then treated as a pseudo-outcome and regressed on $(A,A',X)$. The fitted regression is evaluated at $A=a_0$ to produce $\hat{\mu}_M(A',X)$. Finally, their plug-in estimator of $\psi_{a_0}(P)$ is the empirical mean $\frac{1}{n}\sum{j=1}^n \hat{\mu}_M(A'_j,X_j)$. 

In our plug-in estimator, because each regression corresponds directly to a conditional density in the joint factorization, compatibility is not a concern in our setting. By contrast, a plug-in estimator based on the approach in \cite{jung2024unified} faces such issues, since consistent estimation of $\mu_M(A', X)$ depends on the specification of $P(M \mid A, A', X)$, which must also be compatible with $P(M \mid A, X)$. In our case, however, consistent estimation of $\mB{Z_1}(a_0, X)$ depends only on the form of $P(M \mid A, X)$, thereby avoiding this additional compatibility requirement. It is important to emphasize that in the context of parametric models in statistics, compatibility is especially critical: if the specified regressions are not mutually compatible, the estimator may be inconsistent, with one part of the model implying a form for $P(M \mid A, X)$ that conflicts with the form used to identify or estimate $\mB{Z_1}$, leading to bias and/or invalid inference.  

The issue of incompatibility is less critical  for influence-function-based estimators in semi/non-parametric models. However, the one-step estimators in \cite{jung2024unified} and those in our current work differ in important ways. In our work, we focus on studying how the influence function can be reparametrized, and on analyzing the consequences of such reparametrization for the nuisance functions: which functions must be estimated, how accurately each must be learned relative to the others, and how these rates ensure that the remainder term vanishes at a desirable rate and yields an asymptotically linear estimator, which we summarize below. 

In the above example, the one-step estimator (given in Lemma~\ref{lem:eif}) takes the following form:
{\scriptsize \begin{align*}
    \psi^{+}_{a_0}(\hat{Q}) 
    &= \frac{1}{n}\sum_{j = 1}^n \bigg[ \I(A_j = a_0) \  \underbrace{\frac{\hat{\pi}(a_1\mid X_j)}{\hat{\pi}(a_0\mid X_j)}\times \frac{\hat{f}_L(L_j\mid M_j,a_1,X_j)}{\hat{f}_L(L_j\mid M_j,a_0,X_j)}}_{\hmR{Y}\big(\mathrm{mp}^{-A}_{\G, \prec}(Y_j)\big)}  \left\{ Y_j - \hat{\mu}(L_j, M_j, a_0,X_j) \right\} 
    \notag \\
    &\hspace{-0.7cm} + \I(A_j = a_{1}) \underbrace{\frac{\hat{f}_M(M_j\mid a_0,X_j)}{\hat{f}_M(M_j\mid a_1,X_j)}}_{\hmR{L}\big(\mathrm{mp}^{-A}_{\G, \prec}(L_{j})\big)}  \Big\{  \underbrace{\hat{\mu}(L_j, M_j, a_0,X_j)}_{\hmB{Y}(\mBarg{Y_{j}}, a_{Y})} - \underbrace{\hat{\E}\big[\hat{\mu}(L, M, a_0,X)\mid M_j,a_1,X_j\big]}_{\hmB{L}(\mBarg{L_{j}}, a_{L})} \Big\}
    \\
    &\hspace{-0.7cm} + \I(A_j = a_{0}) \underbrace{\frac{\hat{\pi}(a_1\mid X_j)}{\hat{\pi}(a_0\mid X_j)}}_{\hmR{M}\big(\mathrm{mp}^{-A}_{\G, \prec}(M_{j})\big)} \Big\{  \underbrace{\hat{\E}\big[\hat{\mu}(L, M, a_0,X)\mid M_j,a_1,X_j\big]}_{\hmB{L}(\mBarg{L_{j}}, a_{L})} - \underbrace{\hat{\E}\Big[\hat{\E}\big[\hat{\mu}(L, M, a_0,X)\mid M,a_1,X\big]\mid a_0,X_j\Big]}_{\hmB{M}(\mBarg{M_{j}}, a_{M})} \Big\}
    \\
    &\hspace{-0.7cm} + \left\{ \I(A_j = a_1) - \hat{\pi}(a_1 \mid X_j) \right\} \times \underbrace{\hat{\E}\Big[\hat{\E}\big[\hat{\mu}(L, M, a_0,X)\mid M,a_1,X\big]\mid a_0,X_j\Big]}_{\hmB{M}(\mBarg{M_{j}}, a_0)}
    \\
    &\hspace{-0.7cm} + \hat{\pi}(a_1 \mid \mpg(A_j)) \ \underbrace{\hat{\E}\Big[\hat{\E}\big[\hat{\mu}(L, M, a_0,X)\mid M,a_1,X\big]\mid a_0,X_j\Big]}_{\hmB{M}(\mBarg{M_{j}}, a_0)} + \I(A_j = a_0) Y_j \bigg]  \ .
\end{align*}}

The above one-step estimator relies on key nuisance parameters involving the treatment propensity score,  conditional densities of mediators, and outcome regression. We can characterize these nuisances in three different ways, discussed in Section~4, and summarized below. 

\vspace{0.3cm}\textbf{(i) Direct density estimation:}\vspace{0.2cm} 
Assume the convergence rates of nuisance estimates are defined as $|| \hat{\mu} - \mu || =\smallO(n^{-{1}/{b_Y}})$, $|| \hat{\pi} - \pi || =\smallO(n^{-{1}/l_A})$, $|| \hat{f}_{L} - {f}_{L} || = o_P(n^{-{1}/{l_{L}}})$, and $|| \hat{f}_{M} - {f}_{M} || = o_P(n^{-{1}/{l_{M}}})$. 
According to Theorem~10, this one-step estimator and the corresponding TMLE is asymptotically linear if the following conditions are simultaneously true:

1. $\frac{1}{l_L}+\frac{1}{l_M}\geq \frac{1}{2}$, 
\quad 
2. $\frac{1}{b_Y}+\frac{1}{l_L}\geq \frac{1}{2}$, 
\quad 
3. $\frac{1}{l_A}+\frac{1}{l_M}\geq \frac{1}{2}$, 
\quad 
4. $\frac{1}{b_Y}+\frac{1}{l_A}\geq \frac{1}{2}$. 

According to Corollary 11, these estimators are \textit{consistent} if the following conditions are simultaneously true:
\begin{enumerate}[label=\arabic*.]
    \item $|| \hf{L} - f_{L} || = o_P(1)$ or $||\hf{M} - f_{M} || = o_P(1)$,
    \item $|| \hat{\mu} - \mu || = o_P(1)$ or $|| \hf{L} - f_{L} || = o_P(1)$, 
    \item $|| \hat{\pi} - \pi || = o_P(1)$ or $|| \hf{M} - f_{M} || = o_P(1)$, 
    \item $|| \hat{\mu} - \mu || = o_P(1)$ or $|| \hat{\pi} - \pi || = o_P(1)$.
\end{enumerate}

Thus, according to Theorem~10 and Corollary~11, these estimators enjoy the double robustness property: consistency is guaranteed if either $\{f_L, \pi\}$ or $\{f_M, \mu\}$ is consistently estimated. Moreover, asymptotic linearity can still be achieved under weaker rates of convergence; for example, if $|| \hat{\mu} - \mu || =\smallO(n^{-{1}/{8}})$, $|| \hat{\pi} - \pi || =\smallO(n^{-{3}/8})$, $|| \hat{f}_{L} - {f}_{L} || = o_P(n^{-{3}/{8}})$, and $|| \hat{f}_{M} - {f}_{M} || = o_P(n^{-{1}/{8}})$. That is, the slower rates do not necessarily have to be $\smallO(n^{-{1}/4})$.

\vspace{0.3cm}\textbf{(ii) Sequential regression and direct density ratio estimation:}     
For the one-step estimator and TMLE constructed via sequential regressions and direct density ratio estimations, Theorem 6 informs that the estimators are asymptotically linear if the following conditions are simultaneously true:

1. $\frac{1}{b_{L}}+\frac{1}{r_{M}}\geq \frac{1}{2}$, 
\quad 
2. $\frac{1}{b_Y}+\frac{1}{r_{L}}\geq \frac{1}{2}$, 
\quad 
3. $\frac{1}{r_{A}}+\frac{1}{b_{M}}\geq \frac{1}{2}$,
\quad 
4. $\frac{1}{b_Y}+\frac{1}{r_A}\geq \frac{1}{2}$, 

where $b_L$ and $b_M$ are convergence rates for sequential regressions associated with $L$ and $M$, defined as $|| \hmB{L} - \mB{L} || = o_P(n^{-{1}/{b_{L}}})$, and $|| \hmB{M} - \mB{M} || = o_P(n^{-{1}/{b_{M}}})$, respectively. $r_L$ and $r_M$ are convergence rates for density ratios associated with $L$ and $M$, defined as $|| \hfr{L} - \fr{L} || ,= o_P(n^{-{1}/{r_{L}}})$, and $|| \hfr{M} - \fr{M} || ,= o_P(n^{-{1}/{r_{M}}})$, respectively.

According to Corollary 7, these estimators are consistent if the following conditions are simultaneously true:
\begin{enumerate}[label=\arabic*.]
    \item $|| \hat{\mathscr{B}}_{L} - \mB{L} || = o_P(1)$ or $||\hat{f}^r_{M} - f^r_{M} || = o_P(1)$\ ,
    \item $|| \hat{\mu} - \mu || = o_P(1)$ or $|| \hat{f}^r_{L} - f^r_{L} || = o_P(1)$\ ,
    \item $|| \hat{f}^r_{A} - f^r_{A} || = o_P(1)$ or $|| \hat{\mathscr{B}}_{M} - \mB{M} || = o_P(1)$\ , 
    \item $|| \hat{\mu} - \mu || = o_P(1)$ or $|| \hat{f}^r_{A} - f^r_{A} || = o_P(1)$.
\end{enumerate}

These results imply multiple ways for achieving consistency. In particular, these estimators are consistent if one of the four sets of nuisances are consistently estimated: $\{\fr{M},\mu, \mB{M}\}$, $\{\mB{L},\fr{L},\fr{A}\}$, $\{\mB{L},\mB{M},\mu\}$, or $\{\fr{L},\fr{M},\fr{A}\}$ (additional combinations beyond these four sets can also be listed.) Furthermore, Theorem 6 implies that asymptotic linearity can likewise be attained in multiple ways. One example is when the nuisance estimators satisfy the following convergence rates: $|| \hat{\mu} - \mu || =\smallO(n^{-{1}/{8}})$, $|| \hat{\pi} - \pi || =\smallO(n^{-{3}/8})$, $|| \hmB{L} - \mB{L} || = o_P(n^{-{1}/{8}})$, $|| \hmB{M} - \mB{M} || = o_P(n^{-{1}/{8}})$, $|| \hfr{L} - \fr{L} || = o_P(n^{-{3}/{8}})$, and $|| \hfr{M} - \fr{M} || = o_P(n^{-{3}/{8}})$.

\vspace{0.3cm}\textbf{(iii) Sequential regressions and Bayesian reparameterization of density ratios:} 
According to Theorem 8, the one-step estimator and TMLE based on sequential regressions and reparameterization using Bayes' rule are asymptotically linear if the following conditions are simultaneously true:
\begin{enumerate}[label=\arabic*.]
\item $\frac{1}{b_{L}}+\frac{1}{t_{M}}\geq \frac{1}{2}$, and $\frac{1}{b_{L}}+\frac{1}{d_{M}}\geq \frac{1}{2}$ \ ,  
\item $\frac{1}{b_Y}+\frac{1}{t_{L}}\geq \frac{1}{2}$, and $\frac{1}{b_Y}+\frac{1}{d_{L}}\geq \frac{1}{2}$ \ ,
\item $\frac{1}{d_{M}}+\frac{1}{b_{M}}\geq \frac{1}{2}$\ ,
\item $\frac{1}{b_Y}+\frac{1}{d_{M}}\geq \frac{1}{2} \ $\ ,
\end{enumerate}
where $d_L$, $d_M$, $t_L$, and $t_M$ are convergence rates defined as $|| \hlex{L} - \lex{L} || = o_P(n^{-{1}/d_{L}})$, $|| \hlex{M} - \lex{M} || = o_P(n^{-{1}/d_{M}})$, $|| \hli{L} - \li{L} || = o_P(n^{-{1}/t_{L}})$, and $|| \hli{M} - \li{M} || = o_P(n^{-{1}/t_{M}})$, where $\lex{L}=P(a_1\mid M,X)$, $\lex{M}=P(a_0\mid X)$, $\li{L}=P(a_1\mid L,M,X)$, and $\li{M}=P(a_0\mid M,X)$.

According to Corollary 9, the one-step estimator and TMLE based on sequential regressions and reparameterization using Bayes' rule are consistent if the following conditions are simultaneously true:
\begin{enumerate}[label=\arabic*.]
    \item $|| \hat{\mathscr{B}}_{L} - \mB{L} || = o_P(1)$ or both $||\hat{h}_{M} - h_{M} || = o_P(1)$ and $||\hat{g}_{M} - g_{M} || = o_P(1)$ \ , 
    \item $|| \hat{\mu} - \mu || = o_P(1)$ or both $||\hat{h}_{L} - h_{L} || = o_P(1)$ and $||\hat{g}_{L} - g_{L} || = o_P(1)$ \ ,
    \item $||\hat{g}_{M} - g_{M} || = o_P(1)$ or $|| \hat{\mathscr{B}}_{M} - \mB{M} || = o_P(1)$ \ ,
    \item $|| \hat{\mu} - \mu || = o_P(1)$ or $||\hat{g}_{M} - g_{M} || = o_P(1)$.
\end{enumerate}

These results imply multiple ways for achieving consistency and asymptotic linearity, particularly under the observation that in this setting $g_L = 1 - h_M$. Specifically, the one-step estimator and TMLE are consistent if any following sets of nuisance parameters is consistently estimated: $\{\mB{L},\mB{M},\mu\}$, $\{h_M, h_L, g_M\}$, $\{g_M,\mu,\mB{L}\}$. Moreover, consistency can also be obtained through additional combinations beyond those explicitly listed here. One example to achieve asymptotic linearity of these estimators is to have nuisance estimates converge to the respective truth as rates: $||\hat{\mu}-\mu||= o_P(n^{-{1}/{8}})$, $|| \hmB{L} - \mB{L} || = o_P(n^{-{1}/{8}})$, $|| \hmB{M} - \mB{M} || = o_P(n^{-{1}/{8}})$, $||\hat{h}_M - h_M ||= o_P(n^{-{3}/{8}})$, $||\hat{g}_M - g_M ||= o_P(n^{-{3}/{8}})$, and $||\hat{h}_L - h_L ||= o_P(n^{-{3}/{8}})$.


\newpage
\section{Additional details on the simulation studies}
\label{app:sims}

An ADMG can be the latent projection of infinitely many latent-variable DAGs \citep{verma90equiv}. For the simulations in this appendix, we choose latent-variable DAGs and specify their joint laws so that the relevant observed-data nuisance functions are tractable. This lets us specify the nuisance models correctly and examine robustness and asymptotic behavior within a common simulation framework. The order in which we sample variables below is a chain-rule factorization of the joint law; the corresponding latent-variable DAG specifies their causal relationships.

When we refer to generating data according to Figure~\ref{fig:graphs_primal}(b), we use the latent-variable DAG in Figure~\ref{fig:app:graphs_primal}(a). Its latent projection, shown in Figure~\ref{fig:app:graphs_primal}(b), coincides with Figure~\ref{fig:graphs_primal}(b). In the latent-variable DAG of Figure~\ref{fig:app:graphs_primal}(a), $X$, $U$, and $A$ form a complete DAG, so their joint distribution imposes no conditional independence restrictions. We therefore specify and sample this joint distribution using the convenient factorization $p(x)p(a\mid x)p(u\mid a,x)$, as in the DGPs below. 

Likewise, references to generating data according to Figure~\ref{fig:graphs_primal}(d) use the latent-variable DAG in Figure~\ref{fig:app:graphs_primal}(c), whose latent projection is Figure~\ref{fig:app:graphs_primal}(d) and coincides with Figure~\ref{fig:graphs_primal}(d). The variables $X$, $U_1$, and $A$ form a complete DAG, so we specify their joint law in the order $X$, $A\mid X$, and $U_1\mid A,X$. Likewise, $X$, $A$, $M$, and $U_2$ form a complete DAG. Given $X,A$, we generate $M\mid A,X$ followed by $U_2\mid A,M,X$. 

\begin{figure}[h] 
	\begin{center}
    \scalebox{0.65}{
    \begin{tikzpicture}[>=stealth, node distance=1.3cm]
        \tikzstyle{format} = [thick, circle, minimum size=1.0mm, inner sep=2pt]
        \tikzstyle{square} = [draw, thick, minimum size=4.5mm, inner sep=2pt]

    \begin{scope}[xshift=0.cm, yshift=0cm]
		\path[->, thick]
		
		node[] (a) {$A$}
		node[right of=a, xshift=0.5cm] (m) {$M$}
        node[right of=m, xshift=0.5cm] (l) {$L$}
		node[above right of=m, yshift=0.2cm] (x) {$X$}
        node[above left of=m, yshift=0.2cm] (u) {\red{$U$}}
		node[right of=l, xshift=0.5cm] (y) {$Y$}
		
		(x) edge[blue, bend left=0] (a) 
		(x) edge[blue] (m) 
        (x) edge[blue] (l) 
		(x) edge[blue, bend right=0] (y) 
		(a) edge[blue] (m) 
		(m) edge[blue] (l) 
        (l) edge[blue] (y)    
        (a) edge[blue, bend right=25] (l) 
        (m) edge[blue, bend right=25] (y)

        (x) edge[red] (u) 
        (u) edge[red] (a) 
        (u) edge[red] (y) 

        node[below right of=m, xshift=0.cm, yshift=-0.2cm] (t1) {(a)} ;
		
	\end{scope}
    \begin{scope}[xshift=6.5cm, yshift=0cm]
		\path[->, thick]
		
		node[] (a) {$A$}
		node[right of=a, xshift=0.5cm] (m) {$M$}
        node[right of=m, xshift=0.5cm] (l) {$L$}
		node[above right of=m, yshift=0.2cm] (x) {$X$}
		node[right of=l, xshift=0.5cm] (y) {$Y$}
		
		(x) edge[blue, bend left=0] (a) 
		(x) edge[blue] (m) 
        (x) edge[blue] (l) 
		(x) edge[blue, bend right=0] (y) 
		(a) edge[blue] (m) 
		(m) edge[blue] (l) 
        (l) edge[blue] (y)    
        (a) edge[blue, bend right=25] (l) 
        (m) edge[blue, bend right=25] (y)

        (a) edge[red, <->, looseness=0.75, out=90, in=70] (y) 

        node[below right of=m, xshift=0.cm, yshift=-0.2cm] (t1) {(b)} ;
		
	\end{scope}
    \begin{scope}[xshift=13.cm, yshift=0cm]
		\path[->, thick]
		
		node[] (a) {$A$}
		node[right of=a, xshift=0.5cm] (m) {$M$}
        node[right of=m, xshift=0.5cm] (l) {$L$}
		node[above right of=m, yshift=0.2cm] (x) {$X$}
		node[above left of=m, yshift=0.2cm] (u1) {\red{$U_1$}}
        node[above right of=l, yshift=0.2cm] (u2) {\red{$U_2$}}
		node[right of=l, xshift=0.5cm] (y) {$Y$}
		
		(x) edge[blue, bend left=0] (a) 
		(x) edge[blue] (m) 
        (x) edge[blue] (l) 
		(x) edge[blue, bend right=0] (y) 
		(a) edge[blue] (m) 
		(m) edge[blue] (l) 
        (l) edge[blue] (y)    
        (a) edge[blue, bend right=15] (y) 

        (x) edge[red] (u1) 
        (x) edge[red] (u2)
        (a) edge[red] (u2) 
        (u1) edge[red] (a)
        (u1) edge[red] (l) 
        (u2) edge[red] (m) 
        (u2) edge[red] (y) 

        node[below right of=m, xshift=0.cm, yshift=-0.2cm] (t2) {(c)} ;
		
	\end{scope}
    \begin{scope}[xshift=19.5cm, yshift=0cm]
		\path[->, thick]
		
		node[] (a) {$A$}
		node[right of=a, xshift=0.5cm] (m) {$M$}
        node[right of=m, xshift=0.5cm] (l) {$L$}
		node[above right of=m, yshift=0.2cm] (x) {$X$}
		node[above left of=m, yshift=0.cm] (u) {}
		node[right of=l, xshift=0.5cm] (y) {$Y$}
		
		(x) edge[blue, bend left=0] (a) 
		(x) edge[blue] (m) 
        (x) edge[blue] (l) 
		(x) edge[blue, bend right=0] (y) 
		(a) edge[blue] (m) 
		(m) edge[blue] (l) 
        (l) edge[blue] (y)    
        (a) edge[blue, bend right=15] (y) 
        (a) edge[red, <->, out=90, in=155] (l) 
        (m) edge[red, <->, out=25, in=90] (y) 

        node[below right of=m, xshift=0.cm, yshift=-0.2cm] (t2) {(d)} ;
		
	\end{scope}
	\end{tikzpicture}
	}
    \vspace{-0.2cm}
	\caption{Latent-variable DAGs used in the simulation studies and their latent projections. Panels (a) and (c) show the DAGs underlying the data-generating distributions for Figures~\ref{fig:graphs_primal}(b) and (d), respectively. Marginalizing the latent variables gives the ADMGs in panels (b) and (d), which coincide with the corresponding graphs in Figure~\ref{fig:graphs_primal}.}
    \label{fig:app:graphs_primal}
	\end{center}
\end{figure}

\subsection{Simulation 1: Asymptotic behavior and computational runtime}
\label{app:sims:consistency}

Detailed descriptions of the DGPs used in Simulation 1 are provided below.
For the scenario where $Y$ is in the district of $A$, as illustrated by Figure~\ref{fig:graphs_primal}(b), the DGP is shown in \eqref{sim1:dgp:YinL}.

\begin{equation}\label{sim1:dgp:YinL}
    \begin{aligned}
    &X\sim \uniform(0,1),\ A\sim \bin(\expit(1+X)), \ U\sim \N(1+A+X,1)
    \\
    &M=\left[
    \begin{matrix}
    M_1 \\
    M_2
    \end{matrix}\right]
     \stackrel{\operatorname{dim}=2}{\sim} \mathcal{N}\left(\left[\begin{array}{c}
    1+A+X \\
    -1-0.5 A+2 X
    \end{array}\right],\left[\begin{array}{ll}
    2 & 1 \\
    1 & 3
    \end{array}\right]\right),
    \\
    & L\sim\N(1+A+M_1+M_2+X,1), \ Y\sim\N(1+L+M_1+M_2+X+U,1),
\end{aligned}
\end{equation}

with this DGP, we have 
\begin{align*}
    \E(Y\mid A,M,L,X)&=1+L+M_1+M_2+X+\E(U\mid A,X)=2+L+M_1+M_2+2X+A.
\end{align*}

For the scenario where $Y$ is not in the district of $A$, as illustrated by Figure~\ref{fig:graphs_primal}(d), the DGP for variables $X,A$, and $M$ is the same as in \eqref{sim1:dgp:YinL}. The DGP for the remaining variables is given by \eqref{sim1:dgp:YnotL}.
\begin{equation}\label{sim1:dgp:YnotL}
    \begin{aligned}
    &U_1\sim \N(1+A+X,1),\ U_2\sim \N(1+M_1+M_2+A+X,1),
    \\
    &L\sim\N(1+M_1+M_2+X+U_1,1),\ Y\sim\N(1+L+A+X+U_2,1),
\end{aligned}
\end{equation}
with this DGP, we have 
\begin{align*}
    \E(L\mid A,M,X)&=1+M_1+M_2+X+\E(U_1\mid A,X)=2+M_1+M_2+2X+A
    \\
    \E(Y\mid A,M,L,X)&=1+L+X+A+\E(U_2\mid A,M,X)=2+L+M_1+M_2+2X+2A.
\end{align*}

\begin{figure}[t]
    \centering
    \includegraphics[width=1\textwidth]{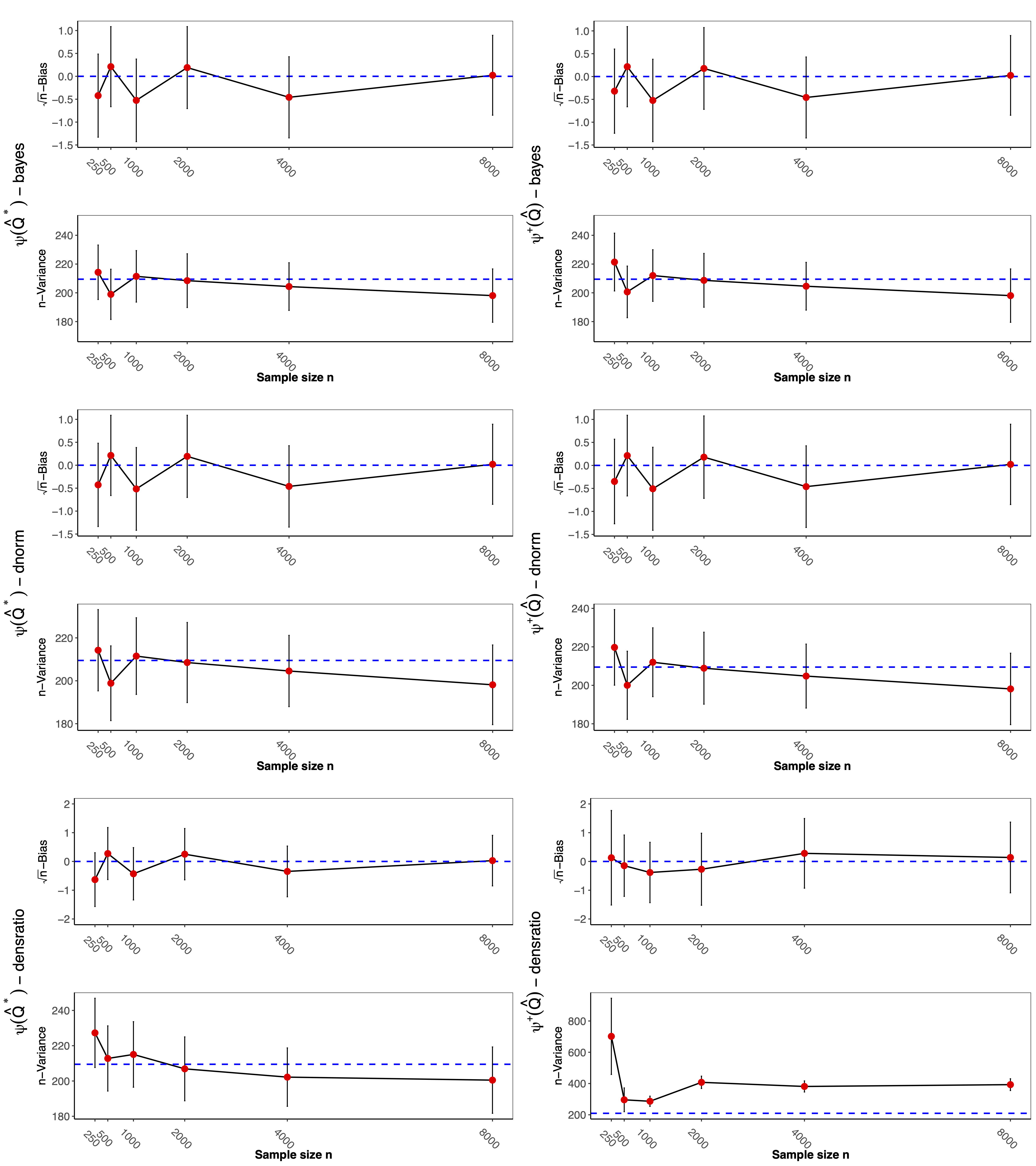}
    \caption{\texttt{[Simulation~1]} Simulation results validating the $n^{1/2}$-consistency behaviors when the outcome is in the district of the treatment. The left column is for TMLE and the right column is for the one-step estimator counterpart. 
    }
    \label{fig:YinL}
\end{figure}

\begin{figure}[t]
    \centering
    \includegraphics[width=1\textwidth]{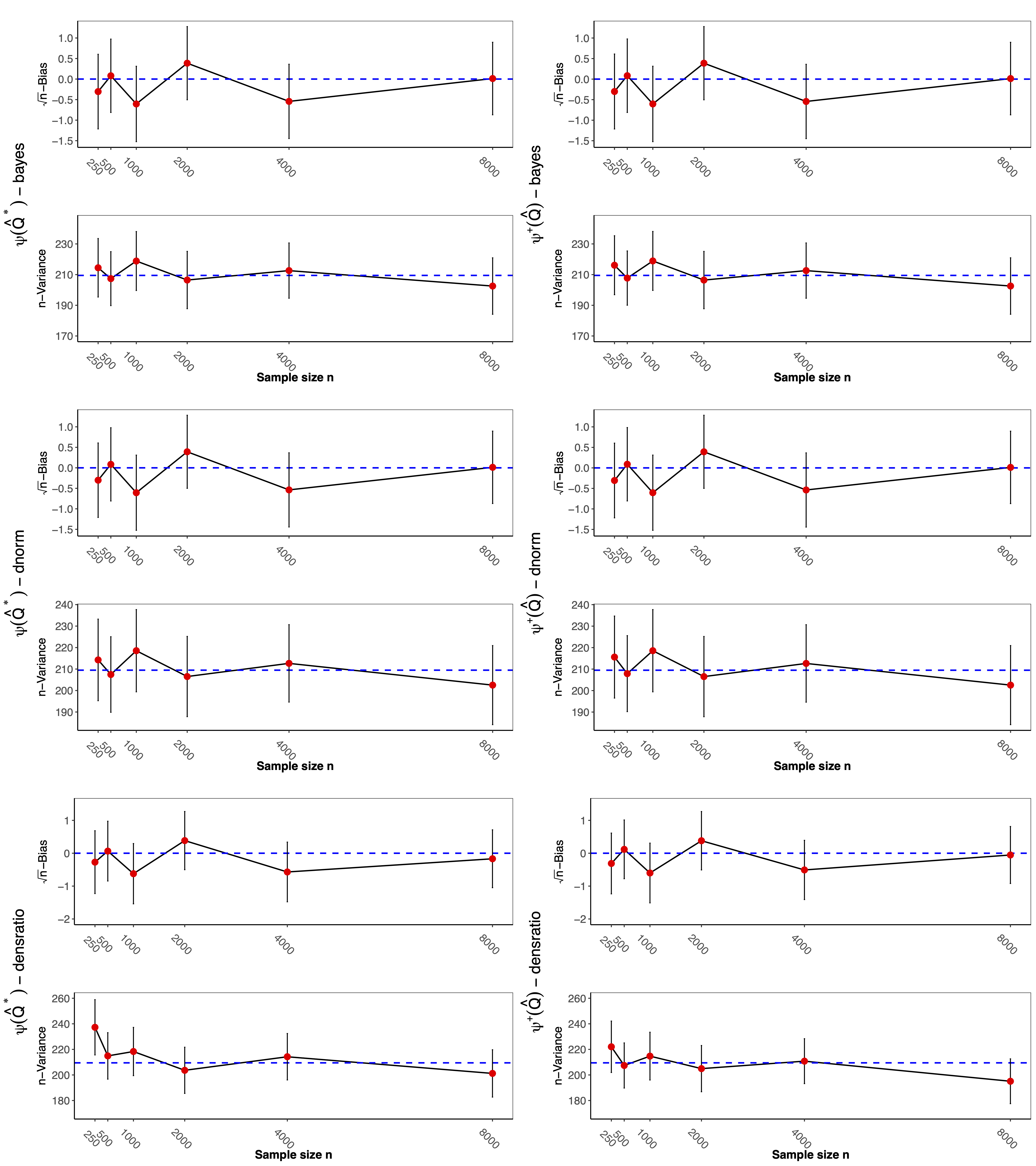}
    \caption{\texttt{[Simulation~1]} Simulation results validating the $n^{1/2}$-consistency behaviors when the outcome is not in the district of the treatment. The left column is for TMLE and the right column is for the one-step estimator counterpart. 
    }
    \label{fig:YnotL}
\end{figure}

The results under this simulation are provided in Figures~\ref{fig:YinL} and~\ref{fig:YnotL}. 

Table~\ref{table:runtime-yinl} reports the combined runtime, in seconds, for computing each one-step estimator and its corresponding TMLE under the DGP displayed in \eqref{sim1:dgp:YinL}. Runtimes were averaged over 1000 replications at each sample size, ranging from 250 to 8000, with each replicate allocated one CPU core and 1 GB of RAM. The Bayes-rule strategy had the shortest average runtime at every sample size, followed by the parametric density strategy and the kernel density-ratio strategy. These timings apply to the nuisance-estimation methods used in this study; computational costs may differ with other learners or tuning procedures.

\begin{table}[!ht]
\centering
\caption{
\blue{\texttt{[Simulation~1]} Average combined runtime (in seconds) for computing each one-step estimator and its corresponding TMLE, averaged over 1000 replications at each sample size under the Simulation~1 DGP, described via \eqref{sim1:dgp:YinL}.}
}
\label{table:runtime-yinl}
\begin{tabular}{lrrrrrr}
\hline
Estimators/Sample size & 250 & 500 & 1000 & 2000 & 4000 & 8000
\\
\hline
bayes & 0.112 & 0.111 & 0.127 & 0.208 & 0.267 & 0.652
\\
densratio & 5.519 & 13.902 & 30.689 & 63.905 & 147.113 & 364.589
\\
dnorm & 0.347 & 0.558 & 1.112 & 2.043 & 4.305 & 7.974
\\
\hline
\end{tabular}
\end{table}

\clearpage
\subsection{Simulation 2: TMLEs versus one-step estimators under weak overlap}
\label{app:sims:overlap}

We generated the treatment variable according to $\bin(\expit(1+5X))$, while the rest of the DGPs, as
specified in displays \eqref{sim1:dgp:YinL} and \eqref{sim1:dgp:YnotL}, remain unchanged.

\subsection{Simulation 3: Misspecified parametric models vs. flexible nuisance estimation} 
\label{app:sims:misspecification}

DGPs for simulation 3 is described below. For the scenario where $Y$ is in the district of $A$, the DGP is shown in \eqref{sim3:dgp:YinL}.

\begin{equation}\label{sim3:dgp:YinL}
    \begin{aligned}
    &X\sim \uniform(0,1),\ A\sim \bin(\expit(1+X)), \ U\sim \N(1+A+X+AX,1)
    \\
    &M=\left[
    \begin{matrix}
    M_1 \\
    M_2
    \end{matrix}\right]
     \stackrel{\operatorname{dim}=2}{\sim} \mathcal{N}\left(\left[\begin{array}{c}
    1+A+X+AX \\
    -1-0.5 A+2 X-AX
    \end{array}\right],\left[\begin{array}{ll}
    2 & 1 \\
    1 & 3
    \end{array}\right]\right),
    \\
    & L\sim\N(1+A+M_1+M_2+X+AX+M_1X+M_2X,1),
    \\
    &Y\sim\N(1+L+M_1+M_2+X+U+LX,1),
\end{aligned}
\end{equation}
with this DGP, we have 
\begin{align*}
    \E(Y\mid A,M,L,X)&=1+L+M_1+M_2+X+\E(U\mid A,X)+M_1X+M_2X
    \\
    &=2+L+M_1+M_2+2X+A+AX+M_1X+M_2X.
\end{align*}

For the scenario where $Y$ is not in the district of $A$, the DGP for variables $X,A$, and $M$ is the same as the display in \eqref{sim1:dgp:YinL}. The DGP for the remaining variables is given by the display in \eqref{sim3:dgp:YnotL}.
\begin{equation}\label{sim3:dgp:YnotL}
    \begin{aligned}
    &U_1\sim \N(1+A+X+AX,1),\ U_2\sim \N(1+M_1+M_2+A+X+AX,1),
    \\
    &L\sim\N(1+M_1+M_2+X+U_1+M_1X+M_2X,1),\ Y\sim\N(1+L+A+X+U_2+LX,1),
\end{aligned}
\end{equation}
with this DGP, we have 
\begin{align*}
    \E(L\mid A,M,X)&=1+M_1+M_2+X+M_1X+M_2X+\E(U_1\mid A,X)
    \\
    &=2+M_1+M_2+2X+A+AX+M_1X+M_2X
    \\
    \E(Y\mid A,M,L,X)&=1+L+X+A+LX+\E(U_2\mid A,M,X)
    \\
    &=2+L+M_1+M_2+2X+2A+AX+LX.
\end{align*}

\subsection{Simulation 4: Impact of cross-fitting} \label{app:sims:cross-fit}

We generated data according to \eqref{dgp:cross_fititng_YinL} for the scenario where $Y$ is in the district of $A$.

\begin{equation}\label{dgp:cross_fititng_YinL}
\begin{aligned}
    &X_i \sim \operatorname{Uniform}(0,1), \ i\in\{1,\ldots,10\}, 
    \\
    &A \sim \mathrm{Binomial}(\mathrm{expit}(V_A \ [1 \ X \ X^2]^T)), \ U \sim \N\left(1+A+X_1, 1\right), 
    \\
    &M=\left[
    \begin{matrix}
    M_1 \\
    M_2
    \end{matrix}\right]
     \stackrel{\operatorname{dim}=2}{\sim} \mathcal{N}\left(V_M\ [1\ A\ X\ AX_{1-5}\ X^2_{6-10}]^T,\left[\begin{array}{ll}
    2 & 1 \\
    1 & 3
    \end{array}\right]\right),
    \\
     &L \sim\N\left(V_L \ [1 \ A \ M_1 \ M_2 \ X \ X^2_{6-10}]^T, 1 \right), 
     \\
    &Y \sim\N\left(V_Y \ [1 \ L \ M_1 \ M_2 \ X \ X_{6-10}^2 \ U]^T, 1 \right). 
\end{aligned}
\end{equation}%
where 
{\small 
\begin{align*}
    &V_A=0.1\times[0.48, 0.07, 1, -1, -0.34, -0.12, 0.3, -0.35, 1, -0.1, 0.46,0.33, 0, 0.45, 0.1, -0.32, -0.08, -0.2, 0.5, 0.5, -0.03]\\
    &V_M=0.025\times\left[\begin{array}{c}
    V_{M_1} \\
    V_{M_2}
    \end{array}\right]\\
    &V_{M_1}=[3.0, 1.5, -1.5, -1.5, -1, -2, -3, -3.0, -1.5, 2.0, 1.5, 3, 1.5, 2.0, 0.5, 0.5, 3.0, -0.2, -0.33, 0.5, 0.3, -0.5]\\
    &V_{M_2}=[1.5, -1.5, -3.0, 2.0, -2, 3, -3, 1.5, -1.5, -1.5, 1.5, -1, -1.5, 0.3, 3.0, -0.33, 0.5, 0.5, 0.50, -0.2, 0.1, 0.2]\\
    &V_L=0.025\ [-3.0, -2.0, -1.5, 1.5, -1.5, -1.0, 0.5, -1.0, 0.3, 3.0, 0.5, 1.5, 0.5, -1.5, -3.0, -0.5, 0.5, 3.0, 1.5]\\
    &V_Y=[1.0, -2.0, -3.0, -1.5, 1.0, 0.5, -2.0, 1.5, -2.0, -3.0, -3.0, -1.5, -1.0, 0.5, 3.0, 1.0, 1.5, -2.0, 3.0, -1.0]\\
    &X=[X_1,X_2,X_3,X_4,X_5,X_6,X_7,X_8,X_9,X_{10}]\\
    &X_{1-5}=[X_1,X_2,X_3,X_4,X_5]\\
    &X_{6-10}=[X_6,X_7,X_8,X_9,X_{10}] \ . 
\end{align*}
}

\subsection{Simulation 6: Parameter-space preservation }\label{app:sims:oob} 

Detailed descriptions of the DGPs used in Simulation~6 are provided below, where $\beta^V_i$ denotes the $i$-th coefficient for variable $V$. This simulation considers four settings with different degrees of weak overlap severity, corresponding to the four parameter configurations presented in \eqref{eq:four-weak-overlap-setting}. For both one-step estimator and TMLE, we estimated the sequential regressions and the density ratios using correctly specified parametric models.
\begin{align*}
    &U\sim\uniform(0,1),
    \quad X\sim\bin(0.5), 
    \quad A\sim\bin\left(\beta^A_1+\beta^A_2X+\beta^A_3U\right),\\
    &M\sim\bin\left\{\expit\left(\beta^M_1+\beta^M_2A+\beta^M_3X\right)\right\},\quad 
    L\sim\bin\left\{\expit\left(\beta^L_1+\beta^L_2M+\beta^L_3A+\beta^L_4X\right)\right\},\\ 
    &Y\sim\bin\left(\beta^Y_1+\beta^Y_2L+\beta^Y_3M+\beta^Y_4X+\beta^Y_5U\right),
\end{align*}
The parameter values under the four simulation scenarios are
{\small
\begin{equation}\label{eq:four-weak-overlap-setting}
    \begin{aligned}
    \text{Favorable:}\qquad
    &\beta^A=(0.44,0.04,0.08)^\top,\quad 
    \beta^M=(-0.50,0.50,0.30)^\top,
    \\
    &\beta^L=(-0.75,0.40,0.50,0.30)^\top,\quad 
    \beta^Y=(0.65,0.05,0.05,0.05,0.05)^\top;
    \\[1mm]
    \text{Moderate:}\qquad
    &\beta^A=(0.04,0.02,0.04)^\top,\quad
    \beta^M=(-1.50,3.25,0.50)^\top,
    \\
    &\beta^L=(-1.75,0.60,3.25,0.50)^\top,\quad 
    \beta^Y=(0.85,0.02,0.02,0.02,0.02)^\top;
    \\[1mm]
    \text{Severe:}\qquad
    &\beta^A=(0.015,0.01,0.02)^\top,\quad
    \beta^M=(-1.75,4.00,0.50)^\top,
    \\
    &\beta^L=(-2.00,0.60,4.00,0.50)^\top,\quad 
    \beta^Y=(0.88,0.015,0.015,0.015,0.015)^\top;
    \\[1mm]
    \text{Extreme:}\qquad
    &\beta^A=(0.005,0.005,0.01)^\top,\quad 
    \beta^M=(-2.00,4.50,0.50)^\top,
    \\
    &\beta^L=(-2.25,0.60,4.50,0.50)^\top,\quad
    \beta^Y=(0.89,0.0125,0.0125,0.0125,0.0125)^\top.
\end{aligned}
\end{equation}
}

\begin{table}[!h]
\centering
\caption{\texttt{[Simulation~6]} Ranges of density ratios and corresponding denominator probabilities across weak overlap scenarios.}
\label{table:weak-overlap-severity}
\renewcommand{\arraystretch}{1.}
\setlength{\tabcolsep}{4pt}
\begin{tabular}{lcccccc}
\toprule
Scenario & $f^r_A$ & $p(A=1\mid X)$ & $f^r_M$ & $p(M\mid A=0,X)$ & $f^r_L$ & $p(L\mid M,A=0,X)$ \\
\midrule
Favorable & $[0.92,1.08]$ & $[0.48,0.52]$ & $[0.77,1.32]$ & $[0.38,0.62]$ & $[0.76,1.37]$ & $[0.32,0.68]$ \\
Moderate  & $[11.50,15.67]$ & $[0.06,0.08]$ & $[0.13,4.67]$ & $[0.18,0.82]$ & $[0.11,5.52]$ & $[0.15,0.85]$ \\
Severe    & $[27.57,39.00]$ & $[0.03,0.04]$ & $[0.08,6.11]$ & $[0.15,0.85]$ & $[0.06,7.39]$ & $[0.12,0.88]$ \\
Extreme   & $[65.67,99.00]$ & $[0.01,0.02]$ & $[0.06,7.75]$ & $[0.12,0.88]$ & $[0.05,9.49]$ & $[0.10,0.91]$ \\
\bottomrule
\end{tabular}
\end{table}

\clearpage
\subsection{Simulation 7: Asymptotic behavior under unequal nuisance convergence rates}
\label{app:sims:rate}

To interpret the results of Simulation~7, write
\begin{equation}
\psi_{a_0}(\widehat Q)-\psi_{a_0}(Q)=P_n\Phi_{a_0}(Q)+R_2(\widehat Q,Q)+D_n,
\label{eq:simulation-rate-expansion}
\end{equation}
where $\psi_{a_0}(\widehat Q)$ denotes either the one-step estimator or the TMLE, and $D_n$ collects the centered empirical-process term and any TMLE targeting error. Under cross-fitting and $L_2(P)$-consistency of the estimated influence function, the empirical-process term is $o_P(n^{-1/2})$; assuming the same order for the targeting error gives $D_n=o_P(n^{-1/2})$. Expressions involving cross-fitted nuisance estimates are interpreted foldwise and averaged over folds.

Let $\sigma^2=P\{\Phi_{a_0}(Q)^2\}$, with $0<\sigma^2<\infty$. Because $P\Phi_{a_0}(Q)=0$, the central limit theorem gives
\begin{equation}
\begin{aligned}
    \sqrt{n}\{\psi_{a_0}(\widehat Q)-\psi_{a_0}(Q)\}=\sqrt{n}\,P_n\Phi_{a_0}(Q)+\sqrt{n}\,R_2(\widehat Q,Q)+o_P(1),\qquad \sqrt{n}\,P_n\Phi_{a_0}(Q)\rightarrow_d \N(0,\sigma^2).
\end{aligned}
\label{eq:simulation-root-n-expansion}
\end{equation}
The product-rate conditions, characterized in Section~\ref{sec:inference} of the main text, ensure that $R_2(\widehat Q,Q)=o_P(n^{-1/2})$, yielding the centered normal limit and, with a consistent variance estimator, asymptotically nominal confidence interval coverage. If only $R_2(\widehat Q,Q)=o_P(n^{-\gamma})$ for $0<\gamma<1/2$ is established, the estimator remains consistent, but this upper bound alone does not determine its root-$n$ limiting behavior.

\paragraph{Data generating mechanism.}
We generated data according to the ADMG in Figure~\ref{fig:graphs_primal}(b), with the corresponding DGP specified in \eqref{eq:sim7_dgp}.
\begin{equation}\label{eq:sim7_dgp}
\begin{aligned}
X &\sim \operatorname{Uniform}(0,1), \\ 
U &\sim \operatorname{Uniform}(-2,2), \\
A\mid X,U &\sim \operatorname{Bernoulli}
 \left\{f(X)+f(X)\bigl(1-f(X)\bigr)\frac{U}{2}\right\}, \quad \text{where} \ f(X) = \operatorname{expit}(1+X), \\
\begin{pmatrix}M_1\\M_2\end{pmatrix}\Bigm| A,X
&\sim \mathcal N\left[
\begin{pmatrix}1+A+X\\-1-0.5A+2X\end{pmatrix},
\begin{pmatrix}1&0.5\\0.5&1.5\end{pmatrix}
\right],\\
L\mid A,M,X
&\sim\mathcal N(1+0.5A+0.5M_1+0.5M_2+X,\,1),\\
Y\mid U,A,M,L,X
&\sim\mathcal N(1+0.5L+0.5M_1+0.5M_2+X+U,\,1).
\end{aligned}
\end{equation}

The treatment probability in the Bernoulli equation is valid for every $u\in[-2,2]$. Since $U\perp X$ and $E(U)=0$, the observed treatment propensity is
\begin{align*}
\pi(1\mid x)
&=\E\left[
f(x)+f(x)\{1-f(x)\}\frac{U}{2}
\,\middle|\, X=x
\right] 
=f(x)=\operatorname{expit}(1+x).
\end{align*}
For the observed outcome regression, Bayes' rule gives
\begin{align*}
    \E(U\mid A=a,X=x)=\frac{2}{3}\{a-f(x)\}.
\end{align*}
Because the conditional distributions of $M$ and $L$ do not depend on $U$ given their observed parents, $U\perp(M,L)\mid(A,X)$. Therefore,
\begin{align*}
\mu(x,a,m,l)
&=\E(Y\mid X=x,A=a,M=m,L=l)\\
&=1+0.5l+0.5m_1+0.5m_2+x
  +E(U\mid A=a,X=x)\\
&=1+0.5l+0.5m_1+0.5m_2+x
  +\frac{2}{3}\left\{a-\operatorname{expit}(1+x)\right\}.
\end{align*}

For both estimators, we fitted the sequential regressions parametrically and estimated the density ratios using the Bayes' rule reparameterization, with correctly specified parametric models for $h_L$, $h_M$, and $g_M$. 

\paragraph{Simulation design.} 
Motivated by the controlled nuisance perturbations in the simulations of \citet{kennedy2023towards}, we perturbed each selected nuisance estimates by
\begin{align*}
    \delta_{k,n}=c\epsilon X\frac{n^{-1/k}}{(\log n)^{1/30}}=o_p(n^{-1/k}),
\end{align*}
where $c>0$ is a real-valued constant, $\epsilon\sim \N(1,1)$, and $X$ is the covariate variable. The Gaussian perturbation $\epsilon$ is mutually independent for different nuisance estimates. Since $c\epsilon X=O_p(1)$, the term $c\epsilon Xn^{-1/k}$ is $O_p(n^{-1/k})$. Dividing by $(\log n)^{1/30}$ ensures that the perturbation is $o_p(n^{-1/k})$.

We considered five simulation scenarios: two fast-rate scenarios, with perturbation pairs $(\delta_{2.5,n}, \delta_{5,n})$ and $(\delta_{3,n}, \delta_{6,n})$, and three slow-rate scenarios, with perturbation pairs $(\delta_{6,n}, \delta_{8,n})$, $(\delta_{9,n}, \delta_{10,n})$, and $(\delta_{10,n}, \delta_{15,n})$. In each pair, the first and second terms correspond to perturbations of the sequential regressions and density ratios, respectively. The perturbations are additive for the sequential outcome regressions and are applied on the logit or log scale for the propensity score and density ratios, respectively, ensuring that the perturbed nuisance estimates remain within their theoretical ranges. The corresponding perturbation products are
{\small\begin{align*}
    &\delta_{2.5,n}\delta_{5,n}=O_p\big(\frac{n^{-3/5}}{(\log n)^{1/15}}\big)=o_p(n^{-3/5}),\quad \delta_{3,n}\delta_{6,n}=O_p\big(\frac{n^{-1/2}}{(\log n)^{1/15}}\big)=o_p(n^{-1/2}),
    \\
    &\delta_{6,n}\delta_{8,n}=O_p\big(\frac{n^{-7/24}}{(\log n)^{1/15}}\big)=o_p(n^{-7/24}),\quad \delta_{9,n}\delta_{10,n}=O_p\big(\frac{n^{-19/90}}{(\log n)^{1/15}}\big)=o_p(n^{-19/90}), \\
    &\delta_{10,n}\delta_{15,n}=O_p\big(\frac{n^{-1/6}}{(\log n)^{1/15}}\big)=o_p(n^{-1/6}),
\end{align*}}
the last three terms decaying more slowly than $o_p(n^{-1/2})$.

\paragraph{Moment conditions and an illustrative remainder calculation.}
Convergence in probability alone does not justify convergence of repeated-sampling bias or variance. To explain the simulation design, we consider the product remainder under suitable regularity and moment conditions. As an illustrative derivation, we analyze the product between $f^r_L(X,M,L,a_0)$ and $\mu(X,M,L,a_1)$, with $f_L^r$ and $f_M^r$ both defined as the conditional density under $a_0$ divided by that under $a_1$. The density ratio $f^r_L$ is perturbed on the log scale, while $\mu$ receives an additive perturbation. 

We first consider the one-step estimator. Let $\hat f_L^{r,0}$ and $\hat\mu^{\,0}$ denote the initial parametric fits, and let $\hat f_L^r$ and $\hat\mu$ denote their perturbed versions. The perturbations are applied as
\begin{align*}
    \log\hat f_L^r
    =\log\hat f_L^{r,0}+cr_{n,k}X\epsilon_{rL},
    \qquad
    \hat\mu=\hat\mu^{\,0}+cr_{n,k'}X\epsilon_Y,
\end{align*}
where $r_{n,k}=n^{-1/k}/(\log n)^{1/30}$. In particular,
\begin{align*}
    \hat f_L^r
    =\hat f_L^{r,0}\exp\{cr_{n,k}X\epsilon_{rL}\}
    =\hat f_L^{r,0}
     \{1+cr_{n,k}X\epsilon_{rL}+o_p(r_{n,k})\}.
\end{align*}
Under correctly specified parametric nuisance models and suitable
regularity and moment conditions, the errors of the initial fits are
$o_p(r_{n,k})$ and $o_p(r_{n,k'})$, respectively. Hence
\begin{align*}
    \hat f_L^r-f_L^r
=cr_{n,k}f_L^r X\epsilon_{rL}+o_p(r_{n,k}),
\qquad
\hat\mu-\mu
=cr_{n,k'}X\epsilon_Y+o_p(r_{n,k'}).
\end{align*}

Substituting these expansions gives
\begin{align*}
&\E\left[f^r_M(X,M)\I(A=a_1)\{\hat f^r_L(X,M,L)-f^r_L(X,M,L)\}\{\mu(X,M,L,a_1)-\hat\mu(X,M,L,a_1)\}\right]\\
&\quad=-c^2r_{n,k}r_{n,k'}\epsilon_{rL}\epsilon_Y\E\left[\I(A=a_1)f^r_M(X,M)f^r_L(X,M,L)X^2\right]+o_p(r_{n,k}r_{n,k'})\\
&\quad=-c^2r_{n,k}r_{n,k'}\epsilon_{rL}\epsilon_Y\E\left[\pi(a_1\mid X)X^2\right]+o_p(r_{n,k}r_{n,k'}).
\end{align*}
Since $\epsilon_{rL}$ and $\epsilon_Y$ are independent $\N(1,1)$ random variables, we have $\operatorname{Var}(\epsilon_{rL}\epsilon_Y)=3$. Under the additional moment condition that the expected squared approximation error is $o(r_{n,k}^2r_{n,k'}^2)$, the scaled variance of the selected expectation term satisfies
\begin{align*}
&n\operatorname{Var}\left(\E\left[f^r_M(X,M)\I(A=a_1)\{\hat f^r_L(X,M,L)-f^r_L(X,M,L)\}\{\mu(X,M,L,a_1)-\hat\mu(X,M,L,a_1)\}\right]\right)\\
&\quad=3c^4nr_{n,k}^2r_{n,k'}^2\E^2\left[\pi(a_1\mid X)X^2\right]+o(nr_{n,k}^2r_{n,k'}^2).
\end{align*}
In the second scenario, where $k=6$ and $k'=3$,
\begin{align*}
nr_{n,6}^2 r_{n,3}^2 &=\frac{1}{(\log n)^{2/15}}\longrightarrow0.
\end{align*}
Therefore, the scaled variance of the selected expectation term converges to zero.

In the third scenario, where $k=8$ and $k'=6$,
\begin{align*}
nr_{n,6}^2r_{n,8}^2&=\frac{n^{5/12}}{(\log n)^{2/15}}\longrightarrow\infty.
\end{align*}
Since $\E^2[\pi(a_1\mid X)X^2]>0$, the scaled variance of the selected expectation term diverges. Similar arguments apply to the remaining scenarios.

Analogous calculations can be applied to the remaining terms in $R_2$ to show that their scaled variances also converge to zero under the root-$n$ rate scenario and diverge under the slow-rate scenario, provided the required moment conditions hold.

We now examine the behavior of TMLE, for which the targeting procedure may lead to finite-sample performance different from that of the corresponding one-step estimator. To illustrate the potential effect of targeting, we consider a single targeting step in the following derivation.

\begin{figure}[!t]
    \centering
    \includegraphics[width=1\linewidth]{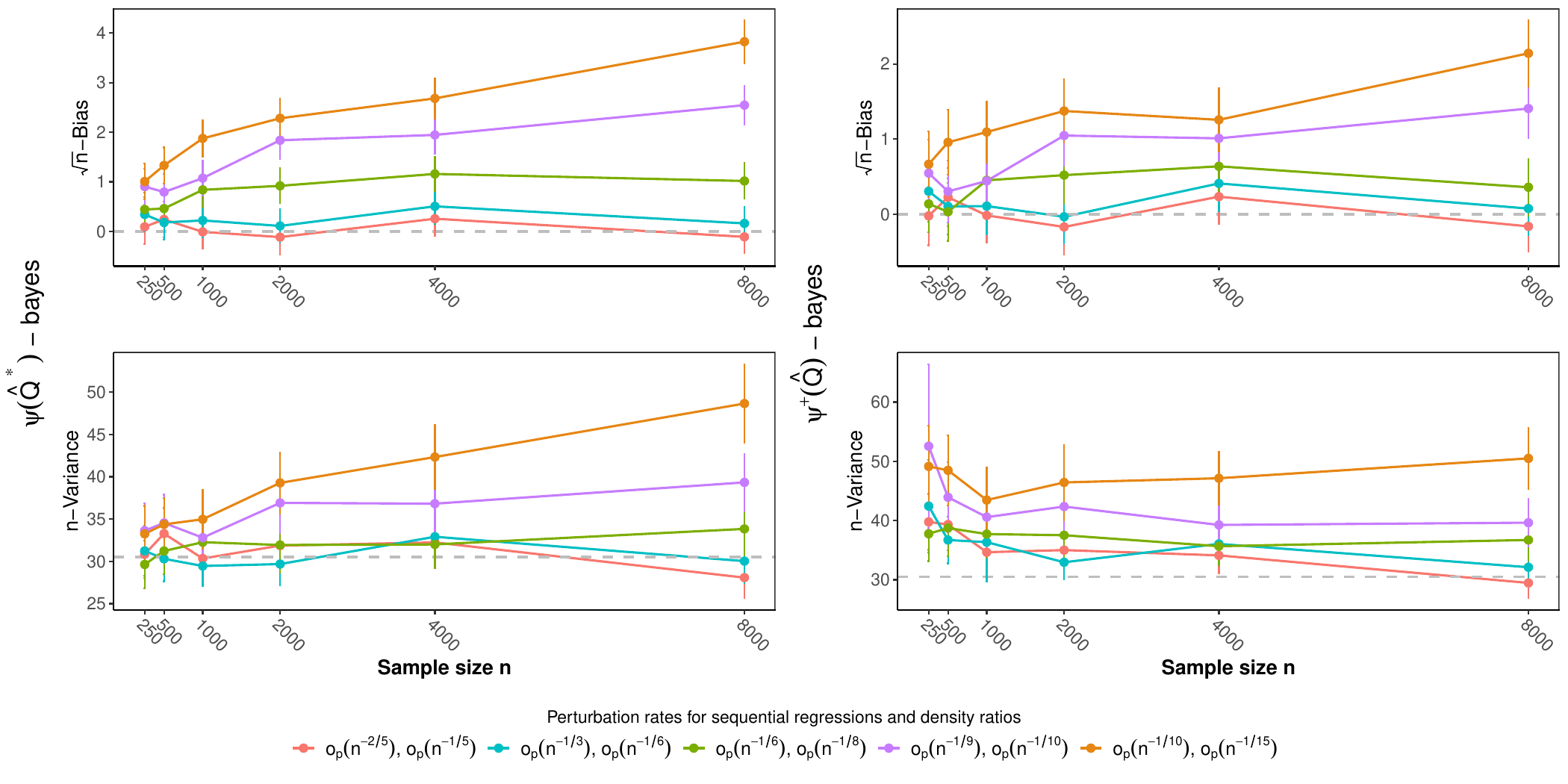}
    \caption{\texttt{[Simulation~7]} Simulation results illustrating the asymptotic behavior of the one-step estimator $\onebayes$ and TMLE $\tmlebayes$ under five perturbation scenarios. The sequential regressions and density ratios are perturbed by magnitudes $(\delta_{2.5,n}, \delta_{5,n})$, $(\delta_{3,n}, \delta_{6,n})$, $(\delta_{6,n}, \delta_{8,n})$, $(\delta_{9,n}, \delta_{10,n})$, and $(\delta_{10,n}, \delta_{15,n})$ in the first through fifth scenarios, represented by red, blue, green, purple, and orange lines, respectively, where $\delta_{k,n}=n^{-1/k}/(\log n)^{1/30}$.}
    \label{fig:rates}
\end{figure}

Define
\begin{align*}
\bar x&=\frac{\E\{\I(A=a_1)f^r_M(X,M)f^r_L(X,M,L)X\}}{\E\{\I(A=a_1)f^r_M(X,M)f^r_L(X,M,L)\}}=\frac{\E\{\pi(a_1\mid X)X\}}{\E\{\pi(a_1\mid X)\}}.
\end{align*}
The weighted-intercept fluctuation of $\mu$ removes the weighted mean of its initial perturbation. Under suitable regularity conditions, the targeting coefficient satisfies $\varepsilon_Y=-cr_{n,k'}\bar x\epsilon_Y+o_p(r_{n,k'}).$ Consequently,
\begin{align*}
\hat\mu^*-\mu&=cr_{n,k'}\epsilon_Y(X-\bar x)+o_p(r_{n,k'}).
\end{align*}
Since $f^r_L$ is not targeted, its leading perturbation error remains
\begin{align*}
\tilde f^r_L-f^r_L&=cr_{n,k}f^r_LX\epsilon_{rL}+o_p(r_{n,k}).
\end{align*}
Substituting these expansions into the selected expectation term gives
\begin{align*}
&\E\left[f^r_M(X,M)\I(A=a_1)\{\tilde f^r_L(X,M,L)-f^r_L(X,M,L)\}\{\mu(X,M,L,a_1)-\widehat\mu^*(X,M,L,a_1)\}\right]\\
&\quad=-c^2r_{n,k}r_{n,k'}\epsilon_{rL}\epsilon_Y\E\left[\I(A=a_1)f^r_M(X,M)f^r_L(X,M,L)X(X-\bar x)\right]+o_p(r_{n,k}r_{n,k'})\\
&\quad=-c^2r_{n,k}r_{n,k'}\epsilon_{rL}\epsilon_Y\E\left[\pi(a_1\mid X)(X-\bar x)^2\right]+o_p(r_{n,k}r_{n,k'}).
\end{align*}
Assuming the same second-moment condition as above, the scaled variance of the selected expectation term satisfies
\begin{align*}
&n\operatorname{Var}\left(\E\left[f^r_M(X,M)\I(A=a_1)\{\tilde f^r_L(X,M,L)-f^r_L(X,M,L)\}\{\mu(X,M,L,a_1)-\widehat\mu^*(X,M,L,a_1)\}\right]\right)\\
&\quad=3c^4nr_{n,k}^2r_{n,k'}^2\E^2\left[\pi(a_1\mid X)(X-\bar x)^2\right]+o(nr_{n,k}^2r_{n,k'}^2).
\end{align*}
It follows that the scaled variance of the selected expectation term under TMLE exhibits the same asymptotic behavior as its one-step counterpart, but with a different leading coefficient. Specifically, the one-step coefficient involves the weighted uncentered second moment of $X$, whereas the TMLE coefficient involves its weighted centered second moment. Under the considered DGP, the latter is smaller. To facilitate comparison of the behavior of the two estimators at the same sample sizes, we set $c=1.2$ for TMLEs and $c=0.43$ for one-step estimators to account for differences in their leading coefficients. The results are shown in Figure~\ref{fig:rates}. 

\clearpage
\subsection{Simulation 8: Leveraging ordinary independence restrictions for efficiency gains}
\label{app:sims:eff}

Detailed descriptions of the DGP used in Simulation~8 are provided below. Data were generated according to a modification of the ADMG in Figure~\ref{fig:graphs_primal}(b), obtained by removing the direct edge from $A$ to $L$. We implemented both one-step estimators and TMLEs based on either the nonparametric EIF or the semiparametric EIF. Because the DGP primarily involves binary variables, we fitted the sequential regressions parametrically and estimated the density ratios by fitting the corresponding conditional densities with parametric models.
\begin{equation}
\begin{aligned}
    &U\sim\uniform(0,1),\qquad  
    X\sim\bin(0.5),\\
    &A\sim\bin\left\{\expit(1+X-0.5U)\right\},\qquad
    M\sim\bin\left\{\expit(1+A+X)\right\},\\
    &L\sim\bin\left\{\expit(1+M+X)\right\},\qquad 
    Y\sim\text{Normal}(1+4L+M+X+U,1).
\end{aligned}
\end{equation}

\end{document}